\documentclass[letterpaper,english,12pt,aps,letter,superscriptaddress,floatfix,pre,nofootinbib]{revtex4}
\pdfoutput=1
\usepackage[T1]{fontenc}
\usepackage[latin1]{inputenc}
\usepackage{babel}
\usepackage{array}
\usepackage{verbatim}
\usepackage{multirow}
\usepackage{amsmath}
\usepackage{amssymb}
\usepackage{graphicx}
\usepackage[unicode=true,pdfusetitle,
 bookmarks=true,bookmarksnumbered=false,bookmarksopen=false,
 breaklinks=false,pdfborder={0 0 1},backref=false,colorlinks=false]
 {hyperref}

\makeatletter

\pdfpageheight\paperheight
\pdfpagewidth\paperwidth

\providecommand{\tabularnewline}{\\}

\@ifundefined{textcolor}{}
{%
 \definecolor{BLACK}{gray}{0}
 \definecolor{WHITE}{gray}{1}
 \definecolor{RED}{rgb}{1,0,0}
 \definecolor{GREEN}{rgb}{0,1,0}
 \definecolor{BLUE}{rgb}{0,0,1}
 \definecolor{CYAN}{cmyk}{1,0,0,0}
 \definecolor{MAGENTA}{cmyk}{0,1,0,0}
 \definecolor{YELLOW}{cmyk}{0,0,1,0}
}

\usepackage{ae,aecompl}
\usepackage{multirow}

\renewcommand{\citet}[1]{\cite{#1}}

\makeatother

\begin{document}

\title{Hydrodynamics of Suspensions of Passive and Active Rigid Particles:\\
A Rigid Multiblob Approach}

\author{Florencio Balboa Usabiaga}

\affiliation{Courant Institute of Mathematical Sciences, New York University,
New York, NY 10012}

\author{Bakytzhan Kallemov}

\affiliation{Courant Institute of Mathematical Sciences, New York University,
New York, NY 10012}

\affiliation{Energy Geosciences Division, Lawrence Berkeley National Laboratory,
Berkeley, CA, 94720}

\author{Blaise Delmotte}

\affiliation{Courant Institute of Mathematical Sciences, New York University,
New York, NY 10012}

\author{Amneet Pal Singh Bhalla}

\affiliation{Department of Mathematics, University of North Carolina, Chapel Hill,
NC 27599}

\author{Boyce E. Griffith}

\affiliation{Department of Mathematics, University of North Carolina, Chapel Hill,
NC 27599}

\affiliation{Department of Biomedical Engineering, University of North Carolina,
Chapel Hill, NC 27599}

\author{Aleksandar Donev}

\email{donev@courant.nyu.edu}

\selectlanguage{english}%

\affiliation{Courant Institute of Mathematical Sciences, New York University,
New York, NY 10012}
\begin{abstract}
We develop a rigid multiblob method for numerically solving the mobility
problem for suspensions of passive and active rigid particles of complex
shape in Stokes flow in unconfined, partially confined, and fully
confined geometries. As in a number of existing methods, we discretize
rigid bodies using a collection of minimally-resolved spherical blobs
constrained to move as a rigid body, to arrive at a potentially large
linear system of equations for the unknown Lagrange multipliers and
rigid-body motions. Here we develop a block-diagonal preconditioner
for this linear system and show that a standard Krylov solver converges
in a modest number of iterations that is essentially independent of
the number of particles. Key to the efficiency of the method is a
technique for fast computation of the product of the blob-blob mobility
matrix and a vector. For unbounded suspensions, we rely on existing
analytical expressions for the Rotne-Prager-Yamakawa tensor combined
with a fast multipole method (FMM) to obtain linear scaling in the
number of particles. For suspensions sedimented against a single no-slip
boundary, we use a direct summation on a Graphical Processing Unit
(GPU), which gives quadratic asymptotic scaling with the number of
particles. For fully confined domains, such as periodic suspensions
or suspensions confined in slit and square channels, we extend a recently-developed
rigid-body immersed boundary method {[}``An immersed boundary method
for rigid bodies'', B. Kallemov, A. Pal Singh Bhalla, B. E. Griffith,
and A. Donev, Communications in Applied Mathematics and Computational
Science, 11-1, 79-141, 2016{]} to suspensions of freely-moving passive
or active rigid particles at zero Reynolds number. We demonstrate
that the iterative solver for the coupled fluid and rigid body equations
converges in a bounded number of iterations regardless of the system
size. In our approach, each iteration only requires a few cycles of
a geometric multigrid solver for the Poisson equation, and an application
of the block-diagonal preconditioner, leading to linear scaling with
the number of particles. We optimize a number of parameters in the
iterative solvers and apply our method to a variety of benchmark problems
to carefully assess the accuracy of the rigid multiblob approach as
a function of the resolution. We also model the dynamics of colloidal
particles studied in recent experiments, such as passive boomerangs
in a slit channel, as well as a pair of non-Brownian active nanorods
sedimented against a wall.
\end{abstract}
\maketitle
\global\long\def\V#1{\boldsymbol{#1}}
\global\long\def\M#1{\boldsymbol{#1}}
\global\long\def\Set#1{\mathbb{#1}}

\global\long\def\D#1{\Delta#1}
\global\long\def\d#1{\delta#1}

\global\long\def\norm#1{\left\Vert #1\right\Vert }
\global\long\def\abs#1{\left|#1\right|}

\global\long\def\av#1{\langle#1\rangle}

\global\long\def\grad{\boldsymbol{\nabla}}
\global\long\def\div{\boldsymbol{\nabla\cdot}}
\global\long\def\curl{\grad\times}

\global\long\def\Re{\text{Re}}

\global\long\def\slip{\breve{\V u}}

\global\long\def\sM#1{\M{\mathcal{#1}}}

\global\long\def\Id{\sM I}

\global\long\def\J{\sM J}

\global\long\def\S{\sM S}

\global\long\def\A{\sM A}

\global\long\def\Div{\sM D}

\global\long\def\Grad{\sM G}

\global\long\def\Lap{\sM L_{v}}

\global\long\def\L{\sM L}

\global\long\def\Lp{\sM L_{p}}

\global\long\def\Mob{\sM M}

\global\long\def\BMob{\sM N}

\global\long\def\K{\sM K}

\global\long\def\Greens{\M{\Set G}}

\global\long\def\Oseen{\M{\Set O}}

\section{Introduction}

The study of the hydrodynamics of colloidal suspensions of passive
particles is an old yet still active subject in soft condensed matter
physics and chemical engineering. In recent years there has been a
growing interest in suspensions of active colloids \citet{ActiveSuspensions},
which exhibit rich collective behaviors quite distinct from those
of passive suspensions. There is a growing number of computational
methods for modeling active suspensions \citet{IrreducibleActiveFlows_PRL,MicroSwimmers_Lushi,SquirmersFCM,StokesianDynamics_Rigid,ActiveFilaments_Adhikari,ActiveFilaments_RPY,BoundaryIntegralGalerkin,Galerkin_Wall_Spheres},
which are typically built upon well-developed techniques for passive
suspensions in steady Stokes flow, i.e., at zero Reynolds number.
Since active particles typically have metallic subcomponents, they
are often significantly denser than the solvent and sediment toward
the bottom wall, making it necessary to address confinement and implement
non-periodic boundary conditions in any method aimed at simulating
experimentally-relevant configurations. Furthermore, since collective
motions seen in active suspensions involve large numbers of particles,
and since hydrodynamic interactions among particles decay slowly like
the inverse of the distance, it is crucial to develop methods that
can capture long-ranged hydrodynamic effects, yet still scale to tens
or hundreds of thousands of particles.

For suspensions of passive particles the methods of Brownian \citet{BrownianDynamics_DNA,BrownianDynamics_OrderN}
and Stokesian dynamics \citet{BrownianDynamics_OrderNlogN,StokesianDynamics_Wall}
have dominated in chemical engineering, and related techniques have
been used in biochemical engineering \citet{HYDROLIB,SphereConglomerate,HYDROPRO,HYDROPRO_Globular,Multiblob_RPY_Rotation}.
These methods simulate the overdamped (diffusive) dynamics of the
particles by using Green's functions for steady Stokes flow to capture
the effect of the fluid. While this sort of implicit solvent approach
works very well in many situations, it has several notable technical
difficulties: achieving near linear scaling for many-particle systems
is technically challenging, handling non-trivial boundary conditions
(bounded systems) is complicated and has to be done on a case-by-case
basis \citet{BrownianDynamics_DNA2,StokesianDynamics_Wall,StokesianDynamics_Slit,StokesianDynamics_Confined,BD_LB_Comparison,RegularizedStokeslets_Walls,RegularizedStokeslets_Periodic,SpectralEwald_Stokes,BoundaryIntegral_Wall,BoundaryIntegral_Periodic3D,BrownianDynamics_OrderN2},
generalizations to non-spherical (and in particular complex) particle
shapes is difficult, and including thermal fluctuations is non-trivial
due to the need to obtain stochastic increments with the desired covariance.
In this work we develop relatively low-accuracy but flexible and simple
rigid multiblob methods that address these difficulties. Our approach
builds heavily on a number of existing techniques, combining elements
from several distinct but related methods. We extensively test the
proposed methods and study their accuracy and performance on a number
of test problems.

The continuum formulation of the Stokes equations with suitable boundary
conditions on the surfaces of a collection of rigid particles is well-known
and summarized in more detail in Appendix \ref{sec:ContinuumFormulation}.
Due to the linearity of the Stokes equations, there is an affine mapping
from the applied forces $\V f$ and torques $\V{\tau}$ and any specified
\emph{apparent slip} velocity due to active boundary layers $\breve{\V u},$
to the resulting particle motion given by the linear velocities $\V u$
and the angular velocities $\V{\omega}$. Specifically, 
\begin{equation}
\left[\begin{array}{c}
\V u\\
\V{\omega}
\end{array}\right]=\M{\mathcal{N}}\left[\begin{array}{c}
\V f\\
\V{\tau}
\end{array}\right]-\breve{\M{\mathcal{N}}}\breve{\V u},\label{eq:N_def_cont}
\end{equation}
where $\M{\mathcal{N}}$ is the \emph{mobility matrix}, and $\breve{\M{\mathcal{N}}}$
is an \emph{active mobility} linear operator. The \emph{mobility problem}
consists of computing the rigid-body motion given the applied forces
and torques and apparent slip. The inverse of this problem is the
\emph{resistance problem}, which computes the forces and torques on
the body given a specified motion of the body and active slip. Solving
the mobility problem is a key component of any numerical method for
modeling the deterministic or fluctuating (Brownian) dynamics of the
particles.

In this paper we develop a \emph{mobility solver} for suspensions
of rigid particles immersed in viscous fluid, specifically, we develop
novel preconditioners for iterative solvers for the unknown motions
of the rigid bodies, given the applied external forces and torques
as well as active apparent slip on the surface of the particles. As
we discuss in more detail in the body of the paper, our formulation
can readily solve the resistance problem; however, our iterative solvers
will prove to be more scalable for mobility computations (which are
of primary interest) than for resistance computations. Key to the
success of our iterative solvers is the idea that instead of eliminating
variables using \emph{exact} Schur complements and solving a \emph{reduced}
system iteratively, as done in the majority of existing methods \citet{RigidMultiblobs_Swan,StokesianDynamics_Rigid,FluctuatingFCM_DC},
one should instead iteratively solve an \emph{extended} system that
includes all of the variables. This has the key advantage that the
matrix-vector product becomes an efficient direct calculation, and
the Schur complement can be computed only \emph{approximately} and
used to construct an effective preconditioner. 

Like many other authors, we construct rigid bodies of essentially
arbitrary shape as a collection of rigidly-connected collection of
``blobs'' or ``beads'' forming a composite object \citet{RigidMultiblobs_Swan}
that we will refer to as a \emph{rigid multiblob}. The hydrodynamic
interactions between blobs are represented using a Rotne-Prager tensor
generalized to the desired domain geometry (boundary conditions) \citet{RPY_Shear_Wall},
specifically, we use the the Rotne-Prager-Yamakawa (RPY) tensor \citet{RotnePrager}
for an unbounded domain, and the Rotne-Prager-Blake (RPB) tensor \citet{StokesianDynamics_Wall}
for a half-space domain. In Section \ref{sec:RigidMultiblobs} we
describe how to obtain the hydrodynamic coupling between a large collection
of rigid multiblobs by solving a large linear system for Lagrange
multipliers enforcing the rigidity. A key contribution of our work
is to develop an indefinite saddle-point preconditioner for iterative
solution of the resulting linear system. This preconditioner is based
on a block-diagonal approximation of the blob-blob mobility matrix,
in which all hydrodynamic interactions among distinct bodies (more
precisely, among blobs on distinct bodies) are neglected. The only
system-specific component is the implementation of a fast matrix-vector
multiplication routine, which in turn requires a scalable method for
multiplying the RPY mobility matrix by a vector.

For simple geometries such as an unbounded domain or particles sedimented
next to a no-slip boundary, simple analytical formulas for the RPY
tensor are well-known \citet{StokesianDynamics_Wall,RPY_Shear_Wall},
and can be used to construct an efficient matrix-vector multiplication
routine, for example, using fast multipole methods (FMMs) \citet{RPY_FMM,OseenBlake_FMM},
or even direct summation on a GPU. We numerically study the performance
and accuracy of the rigid multiblob methods for suspensions in an
unbounded domain in Section \ref{sec:ResultsUnbounded}, and study
particles sedimented near a no-slip boundary in Section \ref{sec:ResultsWall}.
We find that resolving spherical particles with twelve blobs placed
on the vertices of an icosahedron \citet{MultiblobSprings} is notably
more accurate than the FTS (force-torque-stresslet plus degenerate
quadrupole) truncation typically employed in Stokesian dynamics simulations,
provided that the effective hydrodynamic radius of the rigid multiblob
is adjusted to account for the finite size of the blobs. We also find
that a small number of iterations of a Krylov method are required
to solve the required linear system, and importantly, the number of
iterations is constant \emph{independent} of the the number of rigid
bodies, making it possible to develop a linear or near-linear scaling
algorithm. For \emph{resistance problems}, however, we observe a number
of iterations growing at least as fast as the linear dimensions of
the system. This is consistent with similar studies of iterative solvers
for Stokesian dynamics by Ichiki \citet{libStokes}.

For confined systems, however, even in the simplest case of a periodic
system, the Green's function for Stokes flow and the associated RPY
tensor is difficult to obtain in closed form, and when it is possible
to write an analytical result, the resulting formulas are typically
based on infinite series that are expensive to evaluate. For periodic
systems this is commonly addressed by using Ewald summation \citet{RotnePrager_Periodic}
based on the fast Fourier transform (FFT) \citet{RigidMultiblobs_Swan};
the present state-of-the-art for Stokes flow is the spectral Ewald
method \citet{SpectralEwald_Stokes}, which has recently been used
for Stokesian dynamics simulations of periodic suspensions \citet{SD_SpectralEwald}.
A key deficiency of most existing methods is that they rely critically
on having triply periodic domains and the use of the FFT. Generalizing
these methods to non-periodic domains while keeping their linear scaling
requires a large development effort and typically a new implementation
for every different geometry \citet{BrownianDynamics_OrderN2,BoundaryIntegral_Wall}.
Furthermore, in a number of applications involving active particles
\citet{ActiveDimers_EHD,Hematites_Science}, there is a surface slip
(e.g., electrohydrodynamic or osmophoretic flow) induced on the bottom
boundary due to the gradients created by the particles, and this slip
drives or at least strongly affects the motion of the particles. Accounting
for this slip requires solving an additional equation such as a Poisson
or Laplace equation for the electric potential or concentration of
chemical fuel with nontrivial boundary conditions on the particle
and wall surfaces. The solution of this additional equation provides
the slip boundary condition for the Stokes equations, which must be
solved to find the resulting fluid flow and active particle motion.
Such nontrivial multi-physics coupling is quite hard to accomplish
in existing methods.

To address these difficulties, in Section \ref{sec:RigidIBAMR} we
develop a method for general cuboidal confined domains which does
not require analytical Green's functions. This relies on an immersed
boundary (IB) method for obtaining an approximation to the RPY tensor
in confined geometries, as recently developed by some of us \citet{BrownianBlobs}.
This technique has been combined with the concept of multiblob representation
of rigid bodies in a follow-up work \citet{MultiblobSprings}, but
in this work stiff elastic springs were used to enforce the rigidity.
By contrast, we ensure the rigidity of the multiblobs via Lagrange
multipliers which are solved concurrently with solving for the fluid
pressure and velocity. Our key novel contribution is an effective
preconditioner for the rigidly-constrained Stokes problem in periodic
and non-periodic domains, obtained by combining our recently-developed
preconditioner for a rigid-body IB method \citet{RigidIBM} with a
block-diagonal preconditioner for the mobility subproblem.

In the IB method developed in Section \ref{sec:RigidIBAMR} and studied
numerically in Section \ref{sec:ResultsConfined}, analytical Green's
functions are replaced by an ``on the fly'' computation carried
out by a standard finite-volume fluid solver. This Stokes solver can
readily handle nontrivial boundary conditions, for example, slip along
the walls \citet{ActiveDimers_EHD,Hematites_Science} can easily be
accounted for. Furthermore, suspensions at small but nonzero Reynolds
numbers can be handled with little extra work \citet{ISIBM,RigidIBM}.
Additionally, we avoid uncontrolled approximations relying on truncations
of multipole expansions to a fixed order \citet{BrownianDynamics_OrderNlogN,ForceCoupling_Stokes,ISIBM,IrreducibleActiveFlows_PRL},
and we can seamlessly handle arbitrary body shapes and deformation
kinematics. Lastly, and importantly, in the spirit of fluctuating
hydrodynamics \citet{BrownianBlobs,ForceCoupling_Fluctuations,SELM},
it is straightforward to generate the stochastic increments required
to simulate the Brownian motion of small rigid particles suspended
in a fluid by including a fluctuating stress in the fluid equations,
as we will discuss in more detail in future work; here we focus on
the deterministic mobility and resistance problems. At the same time,
our method also has some disadvantages compared to methods such as
boundary integral or boundary element methods. Notably, it requires
filling the domain with a dense uniform fluid grid, which is expensive
at low densities. It is also a low-order method that cannot compute
solutions as accurately as spectral boundary integral formulations.
We do believe, nevertheless, that the method developed here offers
a good compromise between accuracy, efficiency, scalabilty, flexibility
and extensibility, compared to other more specialized formulations.

We apply our methods to a number of test problems for which analytical
solutions are known, but also study a few nontrivial problems that
have not been properly addressed in the literature. In Section \ref{sub:CylinderWall}
we study the mobility of a cylinder of finite aspect ratio that is
parallel to a no-slip boundary and compare to experimental measurements
and asymptotic theory based on a slender-body approximation. In Section
\ref{sub:ActivePair} we study the formation of a stable rotating
pair of active ``extensor'' or ``pusher'' nanorods next to a no-slip
boundary, and confirm the direction of rotation observed in recent
experiments \citet{TripleNanorods_Megan}. In Section \ref{sub:Boomerang}
we compute the effective diffusion coefficient of a boomerang-shaped
colloid in a slit channel, and compare to recent experimental measurements
\citet{BoomerangDiffusion,AsymmetricBoomerangs}. In Section \ref{sub:BinarySedimentation}
we study the mean and variance of the sedimentation velocity in a
binary suspension of spheres of size ratio two, and compare to recent
Stokesian dynamics simulations \citet{BinarySuspension_SD,SD_SpectralEwald}.

\section{\label{sec:RigidMultiblobs}Rigid Multiblob Models of Colloidal Suspensions}

In this section we develop the rigid multiblob model of colloidal
particles at zero Reynolds number. The kind of models we use here
are not new, but we present the method in detail instead of relying
on previous presentations, the most relevant of which are those of
Swan \emph{et al.} \citet{StokesianDynamics_Rigid,RigidMultiblobs_Swan}.
This is in part to present the formulation in our notation, and in
part to explain the differences with other closely-related methods.
Our key novel contribution in this section is the preconditioned iterative
solver described in Section \ref{sub:Preconditioner}; the performance
and scaling of our mobility solver is studied numerically for unbounded
domains in Section \ref{sub:ConvergenceFMM}, and for particles confined
near a single wall in Section \ref{sub:ConvergenceWall}.

The modeling of suspensions of rigid spheres at small Reynolds numbers
is a well-developed field with a long history. A powerful class of
methods are related to Brownian Dynamics with Hydrodynamic Interactions
(BDHI) \citet{BrownianDynamics_DNA,BrownianDynamics_OrderN,LBM_vs_BD_Burkhard,BD_LB_Ladd}
and Stokesian Dynamics (SD) \citet{BrownianDynamics_OrderNlogN,StokesianDynamics_Wall,HYDROLIB,StokesianDynamics_Slit,StokesianDynamics_Brownian,SD_SpectralEwald}
(note that these terms are used differently in different communities).
The difference between these two (as we define them here) is that
BDHI uses what we call a minimally-resolved model \citet{BrownianBlobs}
in which each colloid (for colloidal suspensions) or polymer bead
(for polymeric suspensions) is only resolved at the monopole level,
more precisely, at the Rotne-Prager level \citet{RigidMultiblobs_Swan}.
By contrast, in SD the next level in a multipole expansion is taken
into account and torques and stresslets are also accounted for. It
has been shown recently that yet one more order needs to be kept in
the multipole expansion to model suspensions of active spheres \citet{IrreducibleActiveFlows_PRL,BoundaryIntegralGalerkin},
and a suitable Galerkin truncation of the multipole hierarchy has
been developed for active spheres in unbounded domains \citet{BoundaryIntegralGalerkin},
as well as for active spheres confined near a no-slip boundary \citet{Galerkin_Wall_Spheres}.
It is also possible to account for higher-order multipoles \citet{HYDROMULTIPOLE,HYDROMULTIPOLE_Wall,BoundaryIntegralGalerkin,HydroMultipole_Ladd,HydroMultipole_Ladd_Lubrication},
leading to more complicated (and computationally expensive) but also
more accurate models. It has also been shown that multipole expansions
converge very poorly for nearly touching spheres due to the divergence
of the lubrication forces, and in most methods for dense colloidal
suspensions of hard spheres pairwise lubrication corrections are added
in a somewhat \emph{ad hoc} manner; we will refer to this approach
as SD with lubrication.

Given the well-developed tools for modeling sphere suspensions, it
is natural to leverage them when modeling suspensions of particles
of more complex shapes. Here we describe a technique capable of, in
principle, modeling passive rigid particles of arbitrary shape. The
method can also be used to model, without any extra effort, active
particles with active slip layers, i.e., particles which are phoretic
(e.g., osmo-phoretic, electro-phoretic, chemo-phoretic, etc.) due
to an apparent slip at their surface. For the purposes of hydrodynamic
calculations, we discretize rigid bodies by constructing them out
of multiple rigidly-connected spherical ``blobs'' or beads of hydrodynamic
radius $a$. These blobs can be thought of as hydrodynamically minimally-resolved
spheres forming a rigid conglomerate that approximates the hydrodynamics
of the actual rigid object being studied. We prefer the word ``blob''
over ``sphere'' or ``point'' or ``monopole'' because blobs are
not spheres as they do not have a well-defined surface like spheres
do, they have a finite size associated with them (the hydrodynamic
blob radius $a$) unlike points, and they account for a degenerate
quadrupole associated to the Faxen corrections in addition to a force
monopole. The word ``bead'' is also appropriate, but we prefer to
reserve that for polymer models (bead-spring or bead-link models).

\begin{figure*}[tbph]
\begin{centering}
\includegraphics[width=0.25\textwidth]{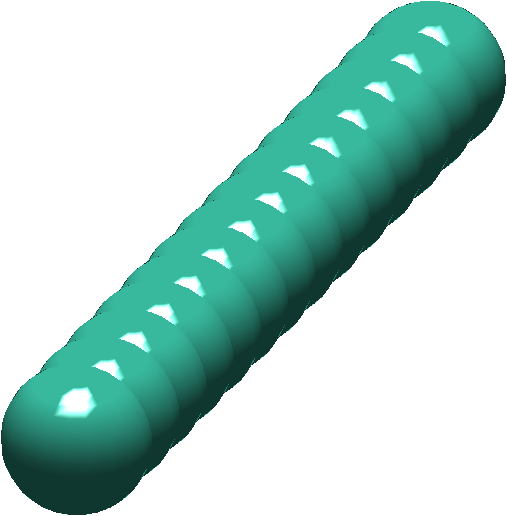}\includegraphics[width=0.25\textwidth]{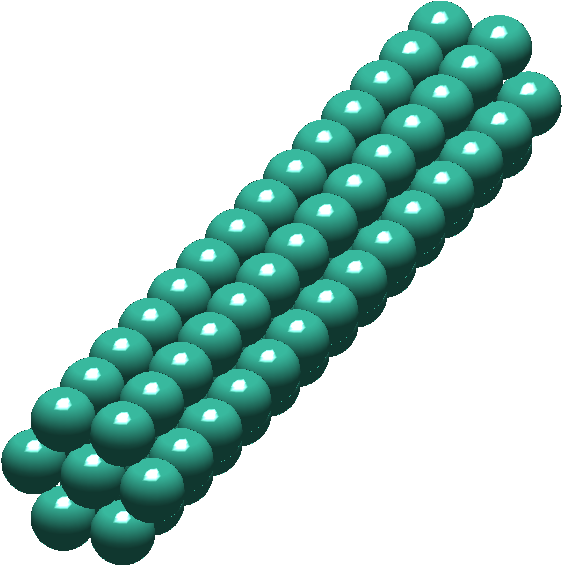}\includegraphics[width=0.25\textwidth]{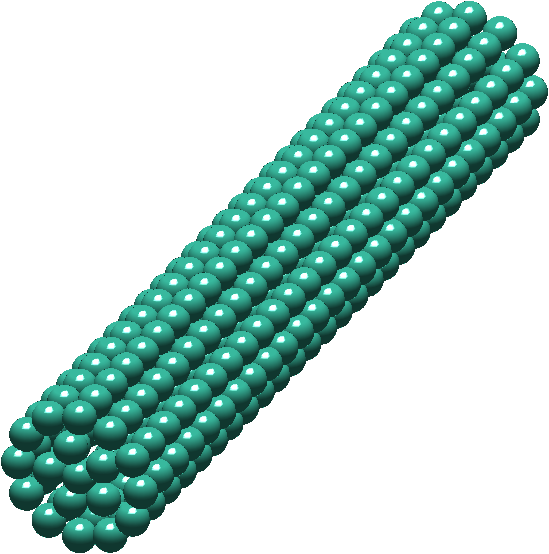}\includegraphics[clip,width=0.25\textwidth]{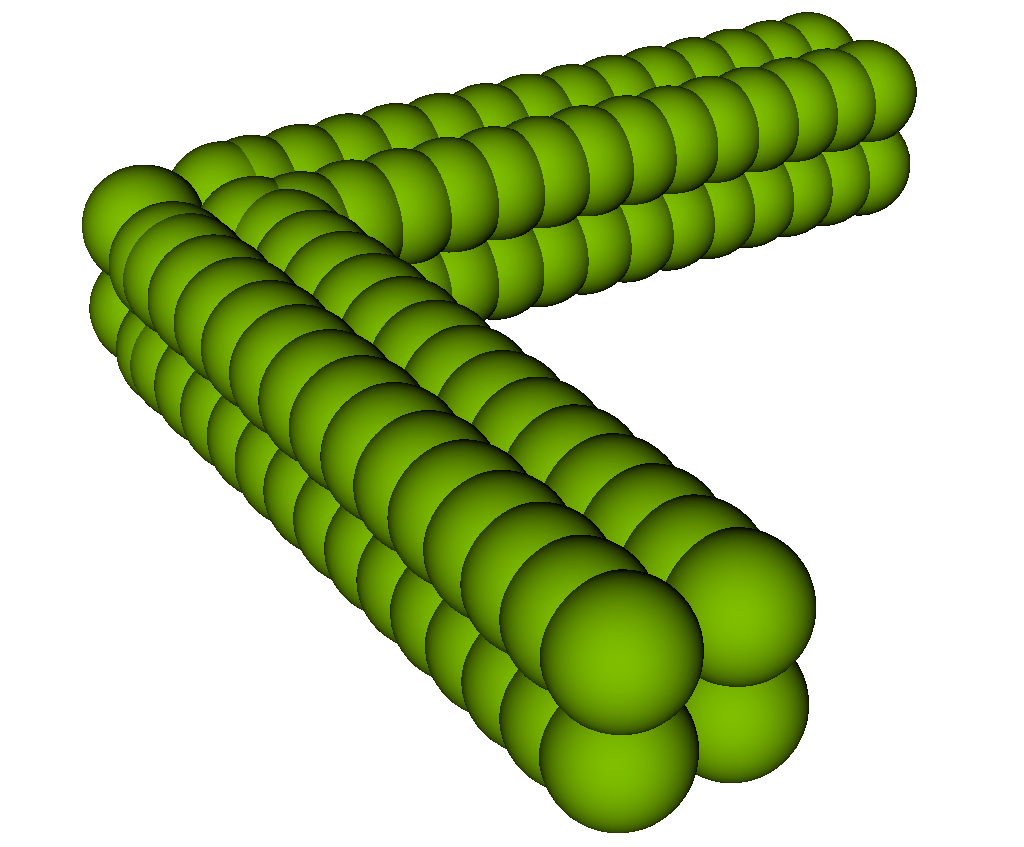} 
\par\end{centering}

\caption{\label{fig:BlobModels}Rigid multiblob models of colloidal particles
manufactured in recent experimental work. (Left three panels) A cylinder
of aspect ratio of about six, similar to the active nanorods studied
experimentally in \citet{FlippingNanorods,TripleNanorods_Megan},
for three different resolutions; from left to right: minimally-resolved
model with 14 blobs, marginally-resolved model with 86 blobs, and
well-resolved model with 324 blobs. (Rightmost panel) A 120-blob model
of a boomerang with square cross-section, as studied experimentally
in \citet{BoomerangDiffusion}.}
\end{figure*}

Examples of ``multiblob'' \citet{MultiblobSprings} models of two
types of colloidal particles are illustrated in Fig. \ref{fig:BlobModels}.
In the leftmost panel, we show a minimally-resolved model of a rigid
rod, with dimensions similar to active metallic ``nanorods'' used
in recent experiments \citet{FlippingNanorods,TripleNanorods_Megan}.
In this minimally-resolved model the blobs, shown as spheres with
radius equal to $a$, are placed in a row along the axes of the cylinder.
Such minimally-resolved models are particularly suited for cylinders
of large but finite aspect ratio; for very thin rods such as actin
filaments boundary integral methods based on slender-body theory \citet{Actin_BI_FMM}
will be more effective. In the more resolved model illustrated in
the second panel from the left, a hexagon of blobs is placed around
the circumference of the cylinder to better resolve it. A yet more
resolved model with a dodecagon of blobs around the cylinder circumference
is shown in the third panel from the left. In the rightmost panel
of Fig. \ref{fig:BlobModels} we show a blob model of a colloidal
boomerang with a square cross-section, as manufactured using lithography
and studied in \citet{BoomerangDiffusion}. Similar ``bead'' or
``raspberry'' models appear in a number of studies of hydrodynamics
of particle suspensions \citet{HYDROPRO,HYDROPRO_Globular,RotationalBD_Torre,Raspberry_MPCD,Raspberry_LBM,SPM_Rigid,StokesianDynamics_Rigid,HYDROLIB,SphereConglomerate,RigidBody_SD,IBM_Sphere,MultiblobSprings,ActiveFilaments_Adhikari,MultibeadRods_Channel}.

In many studies, stiff elastic springs between the blobs are used
to keep the structure rigid; in some models the fluid or particle
inertia is included also. Here, we keep the structures \emph{strictly
rigid} and refer to the resulting structures as \emph{rigid multiblob}
models. Such rigid multiblob models have been used in a number of
prior studies \citet{HYDROPRO,HYDROPRO_Globular,RotationalBD_Torre,StokesianDynamics_Rigid,HYDROLIB,SphereConglomerate,RigidBody_SD,RegularizedStokeslets},
but we refer to \citet{StokesianDynamics_Rigid} for a detailed exposition.
Our primary focus in this section will be to develop algorithmic techniques
that allow suspensions of tens or even hundreds of thousands of rigid
multiblob particles to be simulated efficiently. This is in many ways
primarily an exercise in numerical linear algebra, but one that is
\emph{necessary} to make the rigid multiblob approach useful for simulating
moderately dense suspensions. A second goal, which will be realized
in the results sections of this paper, will be to carefully assess
the accuracy of rigid multiblob models as a function of their resolution
(number of blobs per body).

\subsection{Hydrodynamics of rigid multiblobs}

We now summarize the main equations used to solve the mobility and
resistance problems for a collection of rigid multiblobs immersed
in a viscous fluid. We first discuss the hydrodynamic interaction
between blobs, and then discuss the hydrodynamic interactions between
rigid bodies. 

In the notation used below, we will use the Latin indices $i,j,k,l$
for individual blobs, and reserve Latin indices $p,q,r,s$ for bodies.
We will denote with $\mathcal{B}_{p}$ the set of blobs comprising
body $p$. We will consider a suspension of $N$ rigid bodies with
a chosen reference \emph{tracking point} on body $p$ having position
$\V q_{p}$, and the orientation of body $p$ relative to a \emph{reference
configuration} represented by the quaternion $\V{\theta}_{p}$ \citet{BrownianMultiBlobs}.
The linear velocity of (the chosen tracking point on) body $p$ will
be denoted with $\V u_{p}$, and its angular velocity will be denoted
with $\V{\omega}_{p}$. The total force applied on body $p$ is $\V f_{p}$,
and the total torque is $\V{\tau}_{p}$. The composite configuration
vector of position and orientation of body $p$ will be denoted with
$\V Q_{p}=\left\{ \V q_{p},\,\V{\theta}_{p}\right\} $, the composite
vector of linear and angular velocity will be denoted with $\V U_{p}=\left\{ \V u_{p},\,\V{\omega}_{p}\right\} $,
and the composite vector of forces and torques with $\V F_{p}=\left\{ \V f_{p},\,\V{\tau}_{p}\right\} $.
The position of blob $i\in\mathcal{B}_{p}$ will be denoted with $\V r_{i}$,
and its velocity will be denoted with $\dot{\V r}_{i}$. When not
subscripted, vectors will refer to the composite vector formed by
all bodies or all blobs on all bodies. For example, $\V U$ will denote
the linear and angular velocities of all bodies, and $\V r$ will
denote the positions of all of the blobs. We will use a superscript
to denote portions of composite vectors for all blobs belonging to
one body, for example, $\V r^{(p)}=\left\{ \V r_{i}\;|\; i\in\mathcal{B}_{p}\right\} $
will denote the vector of positions of all blobs belonging to body
$p$. 

The fact that the multiblob $p$ is rigid is expressed by the ``no-slip''
kinematic condition,
\begin{equation}
\dot{\V r}_{i}=\V u_{p}+\V{\omega}_{p}\times\left(\V r_{i}-\V q_{p}\right),\quad\forall i\in\mathcal{B}_{p}.\label{eq:noslip_rigid}
\end{equation}
This no-slip condition can be written for all bodies succinctly as
\begin{equation}
\dot{\V r}=\M{\mathcal{K}}\V U,\label{eq:no_slip}
\end{equation}
where $\M{\mathcal{K}}\left(\V Q\right)$ is a simple geometric matrix
\citet{RigidMultiblobs_Swan}. We will denote the apparent velocity
of the fluid at point $\V r_{i}$ with $\V w_{i}\approx\V v\left(\V r_{i}\right)$.
For a \emph{passive blob}, i.e., a blob that represents a passive
part of the rigid particle, the \emph{no-slip} boundary condition
requires that $\V w_{i}=\dot{\V r}_{i}$. However, for \emph{active
blobs} an additional apparent slip of the fluid relative to the surface
of the body can be imposed, resulting in a nonzero \emph{slip} $\slip_{i}=\V w_{i}-\dot{\V r}_{i}$.
This kind of active propulsion is termed ``implicit swimming gait''
by Swan and Brady \citet{StokesianDynamics_Wall}. An ``explicit
swimming gait'' \citet{StokesianDynamics_Wall} can be taken into
account without any modifications to the formulation or algorithm
by simply replacing (\ref{eq:noslip_rigid}) with 
\begin{equation}
\V w_{i}=\dot{\V r}_{i}=\V u_{p}+\V{\omega}_{p}\times(\V r_{i}-\V q_{p})+\slip_{i}.\label{eq:noslip_explicit}
\end{equation}
That is, the only difference between ``slip'' and ``deformation''
is whether the blobs move relative to the rigid body frame dragging
the fluid along, or stay fixed in the body frame while the fluid passes
by them. One can of course even combine the two and have the blobs
move relative to the rigid body while also pushing flow, for example,
this can be used to model an active filament where there is slip along
the filament but the filament itself is moving. In the end, the only
thing that matters to the formulation is the velocity difference
\begin{equation}
\slip_{i}\approx\V v\left(\V r_{i}\right)-\left(\V u_{p}+\V{\omega}_{p}\times\left(\V r_{i}-\V q_{p}\right)\right).\label{eq:noslip_i}
\end{equation}
In Appendix \ref{app:Permeable} we explain how to model permeable
(porous) bodies by making the apparent slip proportional to the fluid-blob
force $\V{\lambda}$.

The fundamental problem tackled in this paper is the solution of the
\emph{mobility problem}, that is, the computation of the motion of
the bodies given the applied forces and torques on the bodies and
the slip velocity. Because of the linearity of the Stokes equations
and the boundary conditions, there exists an affine linear mapping
\[
\V U=\M{\mathcal{N}}\V F-\breve{\M{\mathcal{M}}}\slip,
\]
where the \emph{body mobility matrix} $\M{\mathcal{N}}\left(\V Q\right)$
depends on the configuration and is the central object of the computation.
The \emph{active mobility matrix} $\breve{\M{\mathcal{M}}}$ is a
discretization of the active mobility operator $\breve{\M{\mathcal{N}}}$,
and gives the active motion of force- and torque-free particles. Note
that $\breve{\M{\mathcal{M}}}$ is related to, but different from,
the propulsion matrix introduced in \citet{BoundaryIntegralGalerkin}.
The propulsion matrix is essentially a finite-dimensional projection
of the operator $\breve{\M{\mathcal{N}}}$ that only depends on the
choice of basis functions used to express the surface slip velocity
$\breve{\V u}$, and does not depend on the specific discretization
of the body or quadrature rules, as does $\breve{\M{\mathcal{M}}}$.

In the remainder of this section we develop a method for computing
$\V U$ given $\V F$ and $\slip$, i.e., a method for computing the
combined action of $\M{\mathcal{N}}$ and $\breve{\M{\mathcal{M}}}$,
for large collections of non-overlapping rigid particles. We will
also briefly discuss the \emph{resistance problem}, in which we are
given the motion of the bodies as a specified kinematics, and seek
the resulting drag forces and torques, which have the form
\[
\V F=\M{\mathcal{R}}\V U+\breve{\M{\mathcal{R}}}\slip,
\]
where the \emph{body resistance matrix} $\M{\mathcal{R}}=\M{\mathcal{N}}^{-1}$
and $\breve{\M{\mathcal{R}}}=\M{\mathcal{N}}^{-1}\breve{\M{\mathcal{M}}}$
is the \emph{active resistance matrix}.

\subsubsection{Blob mobility matrix}

The blob-blob translational mobility matrix $\M{\mathcal{M}}$ describes
the hydrodynamic interactions between the $N_{b}$ blobs, accounting
for the influence of the boundaries. Specifically, if the blobs are
free to move (i.e., not constrained rigidly) with the fluid under
the action of set of translational forces $\V{\lambda}_{i}$, the
translational velocities of the blobs will be
\begin{equation}
\V w=\dot{\V r}+\slip=\M{\mathcal{M}}\V{\lambda}.\label{eq:w_Mlambda}
\end{equation}
The mobility matrix $\M{\mathcal{M}}$ is a block matrix of dimension
$\left(dN_{b}\right)\times\left(dN_{b}\right)$, where $d$ is the
dimensionality. The $d\times d$ block $\M{\mathcal{M}}_{ij}$ computes
the velocity of blob $i$ given the force on blob $j$, neglecting
the presence of the other blobs in a \emph{pairwise} approximation.

To construct a suitable $\M{\mathcal{M}}$, we can think of blobs
as spheres of hydrodynamic radius $a$. For two well-separated spheres
$i$ and $j$ of radius $a$ we have the far-field approximation\textbf{
}\citet{BD_LB_Ladd,StokesianDynamics_Wall,RPY_Shear_Wall}
\begin{equation}
\Mob_{ij}\approx\eta^{-1}\left(\M I+\frac{a^{2}}{6}\grad_{\V r^{\prime}}^{2}\right)\left(\M I+\frac{a^{2}}{6}\grad_{\V r^{\prime\prime}}^{2}\right)\Greens(\V r^{\prime},\V r^{\prime\prime})\big|_{\V r^{\prime\prime}=\V r_{i},}^{\V r^{\prime}=\V r_{j},}\label{eq:MobilityFaxen}
\end{equation}
where $\eta$ is the fluid viscosity and $\Greens$ is the Green's
function for the steady Stokes problem with unit viscosity, with the
appropriate boundary conditions such as no-slip on the boundaries
of the domain. The differential operator $\M I+\left(a^{2}/6\right)\grad^{2}$
is called the Faxen operator \citet{BD_LB_Ladd}. Note that the form
of (\ref{eq:MobilityFaxen}) guarantees that the mobility matrix is
symmetric positive semidefinite (SPD) by construction since $\Greens$
is an SPD kernel.

For a three dimensional unbounded domain with fluid at rest at infinity,
the Green's function is isotropic and given by the Oseen tensor,
\begin{equation}
\Greens(\V r^{\prime},\V r^{\prime\prime})\equiv\Oseen(\V r=\V r^{\prime}-\V r^{\prime\prime})=\frac{1}{8\pi r}\left(\M I+\frac{\V r\otimes\V r}{r^{2}}\right).\label{eq:OseenTensor}
\end{equation}
Using this expression in (\ref{eq:MobilityFaxen}) yields the far-field
component of the Rotne-Prager-Yamakawa (RPY) tensor \citet{RotnePrager},
commonly used in BDHI. A correction needs to be introduced when particles
are close to each other to ensure an SPD mobility matrix \citet{RotnePrager},
which can be derived by using an integral form of the RPY tensor valid
even for overlapping particles \citet{RPY_Shear_Wall}, to give 
\begin{equation}
\Mob_{ij}=\frac{1}{6\pi\eta a}\begin{cases}
C_{1}(r_{ij})\M I+C_{2}(r_{ij})\frac{\V r_{ij}\otimes\V r_{ij}}{r_{ij}^{2}}, & \quad r_{ij}>2a\\
C_{3}(r_{ij})\M I+C_{4}(r_{ij})\frac{\V r_{ij}\otimes\V r_{ij}}{r_{ij}^{2}}, & \quad r_{ij}\leq2a
\end{cases}\label{eq:RPYTensor}
\end{equation}
where $\V r_{ij}=\V r_{i}-\V r_{j}$, and
\begin{eqnarray*}
C_{1}(r)=\frac{3a}{4r}+\frac{a^{3}}{2r^{3}}, & \quad & C_{2}(r)=\frac{3a}{4r}-\frac{3a^{3}}{2r^{3}},\\
C_{3}(r)=1-\frac{9r}{32a}, & \quad & C_{4}(r)=\frac{3r}{32a}.
\end{eqnarray*}
The diagonal blocks of the mobility matrix, i.e., the self-mobility
can be obtained by setting $r_{ij}=0$ to obtain $\Mob_{ii}=\left(6\pi\eta a\right)^{-1}\M I$,
which matches the Stokes solution for the drag on a translating sphere;
this is an important continuity property of the RPY tensor \citet{DDFT_Hydro}.
We will use the RPY tensor (\ref{eq:RPYTensor}) for simulations of
rigid-particle suspensions in unbounded domains in Section \ref{sec:ResultsUnbounded}.

In principle, it is possible to generalize the RPY tensor to any flow
geometry, i.e., to any boundary conditions (and imposed external flow)
\citet{RPY_Shear_Wall}, including periodic domains \citet{RotnePrager_Periodic,RPY_Periodic_Shear},
as well as confined domains \citet{StokesianDynamics_Wall,StokesianDynamics_Slit}.
However, we are not aware of any tractable analytical expressions
for the complete RPY tensor (including near-field corrections) even
for the simplest confined geometry of particles near a single no-slip
boundary. In the presence of a single no-slip wall, an analytic approximation
to $\Mob_{ij}$ is given by Swan and Brady \citet{StokesianDynamics_Wall}
(and re-derived later in \citet{ArtificialCilia_Stark}) as a generalization
of the Rotne-Prager (RP) tensor \citet{RotnePrager} to account for
the no-slip boundary using Blake's image construction \citet{blake1971note}.
As shown in Ref. \citet{RPY_Shear_Wall}, the corrections to the Rotne-Prager
tensor (\ref{eq:MobilityFaxen}) for particles that overlap each other
but not the wall are independent of the boundary conditions, and are
thus given by the standard RPY expressions (\ref{eq:RPYTensor}) for
unbounded domains. Therefore, in Section \ref{sec:ResultsWall} we
compute $\Mob$ by adding to the RPY tensor (\ref{eq:RPYTensor})
wall corrections corresponding to the translation-translation part
of the Rotne-Prager-Blake mobility given by Eqs. (B1) and (C2) in
\citet{StokesianDynamics_Wall}, ignoring the higher order torque
and stresslet terms in the spirit of the minimally-resolved blob model.
The expressions derived by Swan and Brady \citet{StokesianDynamics_Wall}
assume that neither particle overlaps the wall and the resulting expressions
are not guaranteed to lead to an SPD $\Mob$ if one or more blobs
overlap the wall, as we discuss in more detail in the Conclusions.

For more complicated geometries, such as a slit or a square (duct)
channel, analytical computations of the Green's function become quite
complicated and tedious, and numerical computations typically require
pre-tabulations \citet{BD_LB_Ladd,StokesianDynamics_Slit,HE_Spheres_TwoWalls}.
In Section \ref{sec:ResultsConfined} we explain how a grid-based
finite volume Stokes solver can be used to obtain the action of the
Green's function and thus compute the action of the mobility matrix
for confined domains, for essentially arbitrary combinations of periodic,
free-slip, no-slip, or stress boundary conditions.

\subsubsection{Body mobility matrix}

After discretizing the rigid bodies as rigid multiblobs, we can write
down a system of equations that constrain the blobs to move rigidly
in a straightforward manner. Letting $\V{\lambda}$ be a vector of
forces (Lagrange multipliers) that acts on each blob to enforce the
rigidity of the body, we have the following linear system for $\V{\lambda}$,
$\V u$, and $\V{\omega}$ for all bodies $p$,
\begin{align}
\sum_{j}\Mob_{ij}\V{\lambda}_{j}= & \V u_{p}+\V{\omega}_{p}\times\left(\V r_{i}-\V q_{p}\right)+\slip_{i},\quad\forall i\in\mathcal{B}_{p},\label{eq:rigidSystem}\\
\sum_{i\in\mathcal{B}_{p}}\V{\lambda}_{i}= & \V f_{p},\nonumber \\
\sum_{i\in\mathcal{B}_{p}}(\V r_{i}-\V q_{p})\times\V{\lambda}_{i}= & \V{\tau}_{p}.\nonumber 
\end{align}
The first equation is the no-slip condition obtained by combining
(\ref{eq:w_Mlambda}) and (\ref{eq:noslip_rigid}). The second and
third equations are the force and torque balance conditions for body
$p$. Note that the physical interpretation of $\V{\lambda}$ is that
of a total force on the portion of the surface of the body associated
with a given blob. If one wants to think of (\ref{eq:rigidSystem})
as a regularized discretization of the first-kind integral equation
(\ref{eq:first_kind}) and obtain a pointwise value of the traction
force\emph{ density}, one should divide $\V{\lambda}_{j}$ by the
surface area $\D A_{j}$ associated with blob $j$, which plays the
role of a quadrature weight \citet{RegularizedStokeslets}; we will
discuss more sophisticated quadrature rules \citet{RigidRegularizedStokeslets,RegularizedStokesletsPhoretic}
in the Conclusions.

We can write the \emph{mobility problem} (\ref{eq:rigidSystem}) in
compact matrix notation as a \emph{saddle-point }linear system of
equations for the rigidity forces $\V{\lambda}$ and unknown motion
$\V U$,
\begin{equation}
\left[\begin{array}{cc}
\M{\mathcal{M}} & -\M{\mathcal{K}}\\
-\M{\mathcal{K}}^{T} & \M 0
\end{array}\right]\left[\begin{array}{c}
\V{\lambda}\\
\V U
\end{array}\right]=\left[\begin{array}{c}
\slip\\
-\V F
\end{array}\right].\label{eq:saddle_M}
\end{equation}
Forming the Schur complement by eliminating $\V{\lambda}$ we get
(see also Eq. (1) in \citet{StokesianDynamics_Rigid} or Eq. (32)
in \citet{RigidMultiblobs_Swan}) 
\[
\V U=\M{\mathcal{N}}\V F-\left(\M{\mathcal{N}}\M{\mathcal{K}}^{T}\M{\mathcal{M}}^{-1}\right)\slip=\M{\mathcal{N}}\V F-\breve{\M{\mathcal{M}}}\slip,
\]
where the body mobility matrix $\M{\mathcal{N}}$ is
\begin{align}
\M{\mathcal{N}} & =\left(\M{\mathcal{K}}^{T}\M{\mathcal{M}}^{-1}\M{\mathcal{K}}\right)^{-1},\label{eq:BodyMob}
\end{align}
and is evidently SPD since $\M{\mathcal{M}}$ is. Although written
in this form using the inverse of $\M{\mathcal{M}}$, unlike in a
number of prior works \citet{HYDROPRO,RotationalBD_Torre,HYDROLIB,SphereConglomerate,RigidBody_SD},
we obtain $\V U$ by solving (\ref{eq:saddle_M}) directly using an
iterative solver, as we explain in more detail in Section \ref{sub:Preconditioner}.
We note that one can compute a fluid velocity field $\V v\left(\V r\right)$
from $\V{\lambda}$ using a procedure we describe in Appendix \ref{app:RenderFlow}.

The \emph{resistance problem}, on the other hand, consists of solving
for $\V{\lambda}$ in 
\begin{equation}
\M{\mathcal{M}}\V{\lambda}=\M{\mathcal{K}}\V U+\slip,\label{eq:resistance_problem}
\end{equation}
and then computing $\V F=\M{\mathcal{K}}^{T}\V{\lambda}$, giving
\[
\V F=\left(\M{\mathcal{K}}^{T}\M{\mathcal{M}}^{-1}\M{\mathcal{K}}\right)\V U+\left(\M{\mathcal{K}}^{T}\M{\mathcal{M}}^{-1}\right)\slip=\M{\mathcal{R}}\V U+\breve{\M{\mathcal{R}}}\slip.
\]
At first glance, it appears that solving the resistance system (\ref{eq:resistance_problem})
is easier than solving the saddle-point problem (\ref{eq:saddle_M});
however, as we explain in more detail in Section \ref{sub:ConvergenceFMM},
the mobility problem is significantly easier to solve using iterative
methods than the resistance problem, consistent with similar observations
in the context of Stokesian Dynamics \citet{libStokes}. Observe that
the saddle-point formulation (\ref{eq:saddle_M}) applies more broadly
to \emph{mixed} mobility/resistance problems, where some of the rigid
body degrees of freedom are constrained but some are free \citet{BeadModels_Blaise}.
An example is a suspension of spheres being rotated by a magnetic
field at a specified angular velocity but free to move translationally,
or a suspension of colloids fixed in space by strong laser tweezers
but otherwise free to rotate, or even a hinged body that can only
move in a partially-constrained manner. In cases such as these we
simply redefine $\V U$ to contain the free kinematic degrees of freedom
and modify the definition of the kinematic matrix $\M{\mathcal{K}}$.
Much of what we say below continues to apply, but with the caveat
that the expected speed of convergence of iterative methods is expected
to depend on the nature of the imposed constraints, as we discuss
in Section \ref{sub:ConvergenceFMM}.

Note that the formula (\ref{eq:BodyMob}) is somewhat formal, and
in practice all inverses should be replaced by pseudo-inverses. For
instance, in the limit when infinitely many blobs cover the surface
of a body, the mobility matrix $\M{\mathcal{M}}$ is not invertible
since making $\V{\lambda}$ perpendicular to the surface will not
yield any flow because it will try to compress the (fictitious) incompressible
fluid inside the body. Note that this nontrivial null space of the
mobility poses no problem when using an iterative method to solve
(\ref{eq:saddle_M}) because the right hand side is in the proper
range due to the imposition of the volume-preservation constraint
(\ref{eq:slip_solvability}). It is also possible that the matrix
$\M{\mathcal{K}}^{T}\M{\mathcal{M}}^{-1}\M{\mathcal{K}}$ is not invertible.
A typical example for this is the minimally-resolved cylinder shown
in the left-most panel of Fig. \ref{fig:BlobModels}. Because all
of the forces $\V{\lambda}$ are applied exactly on the semi-axes
of the cylinder, they cannot exert a torque around the symmetry axes
of the rod. Again, there is no problem with iterative solvers for
(\ref{eq:saddle_M}) if the applied force is in the appropriate range
(e.g., one should not apply a torque around the semi-axes of a minimally-resolved
cylinder).

\subsection{\label{sub:Preconditioner}Iterative Mobility Solver}

For a small number of blobs, the equation (\ref{eq:saddle_M}) can
be solved by direct inversion of $\M{\mathcal{M}}$, as done in most
prior works. For large systems, which is the focus of our work, iterative
methods are required. A standard approach used in the literature is
to eliminate one of the variables $\V{\lambda}$ or $\V U$. Eliminating
$\V{\lambda}$ leads to the equation
\begin{equation}
\left(\M{\mathcal{K}}^{T}\M{\mathcal{M}}^{-1}\M{\mathcal{K}}\right)\V U=\V F-\M{\mathcal{K}}^{T}\M{\mathcal{M}}^{-1}\slip,\label{eq:symmetric_1}
\end{equation}
which requires the action of $\M{\mathcal{M}}^{-1}$, which must itself
be obtained inside a nested iterative solver, increasing both the
complexity and the cost of the method. Swan and Wang \citet{RigidMultiblobs_Swan}
have recently used the Conjugate Gradient method to solve (\ref{eq:symmetric_1}),
preconditioning using the block-diagonal matrix $\M{\mathcal{P}}=\left(6\pi\eta a\right)\left(\M{\mathcal{K}}^{T}\M{\mathcal{K}}\right)$.

An alternative is to write an equivalent system to (\ref{eq:saddle_M}),
for an arbitrary constant $c\neq0$,
\begin{equation}
\left[\begin{array}{cc}
\M{\mathcal{M}} & -\M{\mathcal{K}}\\
-\M{\mathcal{K}}^{T}\left(\M{\mathcal{I}}+c\M{\mathcal{M}}\right) & c\left(\M{\mathcal{K}}^{T}\M{\mathcal{K}}\right)
\end{array}\right]\left[\begin{array}{c}
\V{\lambda}\\
\V U
\end{array}\right]=\left[\begin{array}{c}
\slip\\
-\left(\V F+c\M{\mathcal{K}}^{T}\slip\right)
\end{array}\right],\label{eq:saddle_M_eq}
\end{equation}
from which we can easily eliminate $\V U$ to obtain an equation for
$\V{\lambda}$ only, in the form\textbf{
\begin{equation}
\left[\M{\mathcal{M}}\left(\M{\mathcal{I}}-\M{\mathcal{K}}(\M{\mathcal{K}}^{T}\M{\mathcal{K}})^{-1}\M{\mathcal{K}}^{T}\right)-c^{-1}\M{\mathcal{K}}(\M{\mathcal{K}}^{T}\M{\mathcal{K}})^{-1}\M{\mathcal{K}}^{T}\right]\V{\lambda}=\mathrm{rhs},\label{eq:symmetric_2}
\end{equation}
}where we omit the full expression for the right hand side for brevity.
The system (\ref{eq:symmetric_2}) can now be solved using (preconditioned)
conjugate gradients, and only requires the inverse of the simpler
matrix $\M{\mathcal{K}}^{T}\M{\mathcal{K}}$. Note that, although
not presented in this way, this is the essence of the approach that
is followed and recommended by Swan \emph{et al}. \citet{StokesianDynamics_Rigid}
(see Appendix \citet{StokesianDynamics_Rigid} and note that $c$
is denoted by $\lambda$ in that paper); they recommend computing
the action of $\left(\M{\mathcal{K}}^{T}\M{\mathcal{K}}\right)^{-1}$
by an iterative method preconditioned by an incomplete Cholesky factorization.
A similar approach is followed in boundary integral formulations (which
are usually formulated using a double layer density), where a continuum
operator related to $\M{\mathcal{K}}(\M{\mathcal{K}}^{T}\M{\mathcal{K}})^{-1}\M{\mathcal{K}}^{T}$
is computed and then discretized using a quadrature rule \citet{BoundaryIntegral_Pozrikidis,BoundaryIntegral_Periodic3D}.

In contrast to the approaches taken by Swan \emph{et al}. \citet{StokesianDynamics_Rigid,RigidMultiblobs_Swan},
we have found that numerically the best approach to solving for the
unknown rigid-body motions of the particles is to solve the extended
saddle-point problem (\ref{eq:saddle_M}) for \emph{both} $\V U$
and $\V{\lambda}$ \emph{directly}, using a preconditioned iterative
Krylov method. In fact, as we will demonstrate in the results section
of this paper, such an approach has computational complexity that
is essentially linear in the number of blobs because the number of
iterations required to solve (\ref{eq:saddle_M}) is quite modest
when an appropriate preconditioner, described below, is used. This
approach does not require computing (the action of) $\left(\M{\mathcal{K}}^{T}\M{\mathcal{K}}\right)^{-1}$
and leads to a very simple implementation.

\subsubsection{Matrix-Vector Product}

A Krylov solver for (\ref{eq:saddle_M}) requires two components:
\begin{enumerate}
\item An efficient algorithm for performing the matrix-vector product, which
in our case amounts to a fast method to multiply the dense but low-rank
mobility matrix $\M{\mathcal{M}}$ by a vector of blob forces $\V{\lambda}$.
\item A suitable preconditioner, which is an approximate solver for (\ref{eq:saddle_M}).
\end{enumerate}
How to efficiently compute $\M{\mathcal{M}}\V{\lambda}$ depends very
much on the boundary conditions and thus the form of the Green's function
used to construct $\M{\mathcal{M}}$. For unbounded domains, in this
work we use the Fast Multipole Method (FMM) developed specifically
for the RPY tensor in \citet{RPY_FMM}; alternative kernel-independent
FMMs could also be used, and have also been generalized to periodic
domains \citet{PeriodicFMM_Zorin}. The FMM method has an essentially
linear computational cost of $O\left(N_{b}\log N_{b}\right)$ for
a single matrix-vector multiplication. In the simulations presented
here we use a fixed and rather tight relative tolerance for the FMM
$\sim10^{-9}$ throughout the iterative solution process. Krylov methods,
however, allow one to \emph{lower} the accuracy of the matrix-vector
product as the residual is reduced \citet{InexactKrylov_Relaxation};
this has recently been used to lower the cost of FMM-based boundary
integral methods \citet{InexactFMM_Krylov}. We will explore such
optimizations in future work.

For rigid particles sedimented near a single no-slip wall, we have
implemented a Graphics Processing Unit (GPU) based direct summation
matrix-vector product based on the Rotne-Prager-Blake tensor derived
by Swan and Brady \citet{StokesianDynamics_Wall}. This has, asymptotically,
a quadratic computational cost of $O\left(N_{b}^{2}\right)$; however,
the computation is trivially parallel so the multiplication is remarkably
fast even for one million blobs because of the very large number of
threads available on modern GPUs. Gimbutas \emph{et al}. have recently
developed an FMM method for the Blake tensor by using a simple image
construction (image Stokeslet plus a harmonic scalar correction) and
applying an infinite-space FMM method to the extended system of singularities
\citet{OseenBlake_FMM}. However, this construction has not yet been
generalized to the Rotne-Prager-Blake tensor, and, furthermore, the
FMM will not be more efficient than the direct product on GPUs in
practice unless a large number of blobs is considered. For fully confined
domains, we will adopt an extended saddle-point formulation that will
be described in Section \ref{sec:ResultsConfined}.

\subsubsection{Preconditioner}

In this work we demonstrate that a very efficient yet simple preconditioner
for (\ref{eq:saddle_M}) is obtained by neglecting hydrodynamic interactions
between different bodies, that is, setting the elements of $\M{\mathcal{M}}$
corresponding to pairs of blobs on \emph{distinct} bodies to zero
in the preconditioner. This amounts to making a block-diagonal approximation
of the mobility $\widetilde{\M{\mathcal{M}}}$ defined by only keeping
the diagonal blocks corresponding to a single body interacting \emph{only}
with the boundaries of the domain,
\begin{equation}
\widetilde{\M{\mathcal{M}}}^{(pq)}=\delta_{pq}\M{\mathcal{M}}^{(pp)}.\label{eq:M_tilde_block_diag}
\end{equation}
We will demonstrate here that the \emph{indefinite block-diagonal}
preconditioner,
\begin{equation}
\M{\mathcal{P}}=\left[\begin{array}{cc}
\widetilde{\M{\mathcal{M}}} & -\M{\mathcal{K}}\\
-\M{\mathcal{K}}^{T} & \M 0
\end{array}\right],\label{eq:indef_block_P}
\end{equation}
is a very effective preconditioner for solving (\ref{eq:saddle_M}).

Applying the preconditioner (\ref{eq:indef_block_P}) amounts to solving
the linear system
\begin{equation}
\left[\begin{array}{cc}
\widetilde{\M{\mathcal{M}}} & -\M{\mathcal{K}}\\
-\M{\mathcal{K}}^{T} & \M 0
\end{array}\right]\left[\begin{array}{c}
\V{\lambda}\\
\V U
\end{array}\right]=\left[\begin{array}{c}
\slip\\
-\V F
\end{array}\right],\label{eq:saddle_M_precon}
\end{equation}
which is quite easy to do since the approximate body mobility matrix
(Schur complement),
\[
\widetilde{\M{\mathcal{N}}}=\left(\M{\mathcal{K}}^{T}\widetilde{\M{\mathcal{M}}}^{-1}\M{\mathcal{K}}\right)^{-1},
\]
is itself a block-diagonal matrix where each block on the diagonal
refers to a single body neglecting all hydrodynamic interactions with
other bodies,
\[
\widetilde{\M{\mathcal{N}}}_{pq}=\delta_{pq}\left(\left(\M{\mathcal{K}}^{(p)}\right)^{T}\left(\M{\mathcal{M}}^{(pp)}\right)^{-1}\M{\mathcal{K}}^{(p)}\right)^{-1}.
\]
Computing $\widetilde{\M{\mathcal{N}}}_{pq}$ requires a dense matrix
inversion (e.g., Cholesky factorization) of the much smaller mobility
matrix $\M{\mathcal{M}}^{(pp)}$, whose size is $\left(dN_{b}^{(p)}\right)\times\left(dN_{b}^{(p)}\right)$,
where $N_{b}^{(p)}$ is the number of blobs on body $p$. In the case
of an infinite domain, the factorization of $\M{\mathcal{M}}^{(pp)}$
can be precomputed once at the beginning of a dynamic simulation and
reused during the simulation due to the rotational and translational
invariance of the RPY tensor; one only needs to apply rotation matrices
to the right-hand side and the result to convert between the original
reference configuration of the body and the current configuration.
Furthermore, particles of the same shape and size discretized with
the same number of blobs as body $p$ can share a single factorization
of $\M{\mathcal{M}}^{(pp)}$ and $\widetilde{\M{\mathcal{N}}}_{pp}$.
In cases where $\M{\mathcal{M}}^{(pp)}$ depends in a nontrivial way
on the position of the body, as for (partially) confined domains,
one needs to factorize $\M{\mathcal{M}}^{(pp)}$ for all bodies $p$
at every time step; this factorization can still be reused during
the iterative solve in each application of the preconditioner.

Because our preconditioner is indefinite, one cannot use the preconditioned
Conjugate Gradient (PCG) Krylov method to solve (\ref{eq:saddle_M})
without modification. One of the most robust iterative methods, which
we use in this work, is the Generalized Minimum Residual Method (GMRES).
The key advantage of GMRES is that it is guaranteed to reduce the
residual from iteration to iteration. Its main downside is that it
requires storing a large number of intermediate vectors (i.e., the
history of the iterates). GMRES also can stall, although this can
be corrected to some extent by restarts. An alternative to GMRES is
the (stabilized) Bi-Conjugate Gradient (BiCG(Stab)) method, which
works for non-symmetric matrices as well. In our implementation we
have relied on the PETSc library \citet{PETSc} for iterative solvers;
this library makes it very easy to experiment with different iterative
solvers.

\section{\label{sec:RigidIBAMR}Rigid Multiblobs in Confined Domains}

The rigid multiblob method described in Section \ref{sec:RigidMultiblobs}
requires a technique for multiplying the blob-blob mobility matrix
with a vector. Therefore, this approach, like all other Green's function
based methods \citet{HYDROMULTIPOLE,HydroMultipole_Ladd,HydroMultipole_Ladd_Lubrication,HYDROMULTIPOLE_Wall,StokesianDynamics_Slit,StokesianDynamics_Wall,BoundaryIntegralGalerkin,BoundaryIntegral_Periodic3D,BoundaryIntegral_Wall,SpectralEwald_Stokes,RegularizedStokeslets_Periodic,RegularizedStokeslets_Walls,BD_LB_Ladd,BrownianDynamics_OrderN2},
is very geometry-specific and does not generalize easily to more complicated
boundary conditions. To handle geometries for which there is no simple
analytical expression for the Green's function, such as slit or square
channels, pre-tabulation of the Green's function is necessary, and
ensuring a positive semi-definite mobility matrix is in general difficult.
Another difficulty with Green's function based methods is that including
a ``background'' flow is only simple when this flow can be computed
easily analytically, such as simple shear flows. But for more complicated
geometries, such as Poiseuille flow through a square channel, computing
the base flow is itself not trivial or requires evaluating expensive
infinite-series solutions.

An alternative approach is to use a traditional Stokes solver to solve
the fluid equations numerically \citet{BrownianDynamics_OrderN}.
This requires filling the domain with a grid, which can increase the
number of degrees of freedom considerably over just discretizing the
surface of the immersed bodies. However, the number of fluid degrees
of freedom can be held approximately constant as more bodies are included,
so that the methods typically scale very well with the number of particles
and are well-suited to dense particle suspensions. Previous work \citet{BrownianBlobs,SELM,SIBM_Brownian}
has shown how to use an immersed boundary (IB) method \citet{IBM_PeskinReview}
to obtain the action of the Green's function in complex geometries.
In this approach, spherical particles are minimally resolved using
only a single blob per particle. In subsequent work this approach
was extended to multiblob models \citet{MultiblobSprings}, but the
rigidity constraint was imposed only approximately using stiff springs,
leading to numerical stiffness. A class of related minimally-resolved
methods based on the Force Coupling Method (FCM) \citet{ForceCoupling_Monopole,ForceCoupling_Stokes,ForceCoupling_Fluctuations,FluctuatingFCM_DC}
can include also torques and stresslets, as well as particle activity
\citet{SquirmersFCM}, but a number of these methods have relied strongly
on periodic boundaries since they use the Fast Fourier Transform (FFT)
to solve the (fluctuating) Stokes equations.

In recent work \citet{RigidIBM}, some of us have developed an IB
method for rigid bodies. This method applies to a broad range of Reynolds
numbers. In the case of zero Reynolds number it becomes equivalent
to the rigid multiblob method presented in Section \ref{sec:RigidMultiblobs},
but with a blob-blob mobility that is computed by the fluid solver.
In Ref. \citet{RigidIBM} only rigid bodies with specified motion
(kinematics) were considered; here we extend the method to handle
freely-moving rigid bodies in Stokes flow. We will present here the
key ideas and focus on the new components necessary to solve for the
unknown motion of the particles; we refer the reader interested in
more technical details to Refs. \citet{BrownianBlobs,RigidIBM}. The
key novel contribution of our work is the preconditioner described
in Section \ref{sub:Preconditionining-Algorithm}; the performance
and scalability of our preconditioned iterative solvers is studied
numerically in Section \ref{sub:ConvergenceIBAMR}. To begin, we present
a semi-continuum formulation where the relation to Section \ref{sec:RigidMultiblobs}
is most obvious, and then we discuss the fully discrete formulation
used in the actual implementation. In Appendix \ref{app:Permeable}
we demonstrate how to handle permeable bodies using a small modification
of the formulation. Numerical results obtained using the method described
here are given in Section \ref{sec:ResultsConfined}.

\subsection{\label{sub:ContinuumIB}Semi-Continuum Formulation}

We consider here a semi-discrete model in which the rigid body has
already been discretized using blobs but a continuum description is
used for the fluid, that is, we consider a rigid multiblob model immersed
in a continuum Stokesian fluid. In the IB literature blobs are referred
to as markers, and are often thought of as ``points'' or ``discrete
delta functions''. We use the term ``blob,'' however, to connect
to Section \ref{sec:RigidMultiblobs} and to emphasize that the blobs
have a finite physical and hydrodynamic extent.

In the IB method \citet{IBM_PeskinReview} (and also the force coupling
method \citet{ForceCoupling_Monopole}), the shape of the blob and
its effective interaction with the fluid is captured through a smooth
kernel function $\delta_{a}\left(\V r\right)$ that integrates to
unity and whose support is localized in a region of size comparable
to the blob radius $a$. In our rigid multiblob IB method, to obtain
the fluid-blob interaction forces $\V{\lambda}\left(t\right)$ that
constrain the unknown rigid motion of the $N_{b}$ blobs, we need
to solve a constrained Stokes problem \citet{RigidIBM} for the fluid
velocity field $\V v\left(\V r,t\right)$, the fluid pressure field
$\pi\left(\V r,t\right)$, the blob constraint forces $\V{\lambda}\left(t\right)$,
and the unknown rigid-body motions $\V u\left(t\right)$ and $\V{\omega}\left(t\right)$,
\begin{align}
\grad\pi & =\eta\grad^{2}\V v+\sum_{i=1}^{N_{b}}\V{\lambda}_{i}\delta_{a}\left(\V r_{i}-\V r\right),\nonumber \\
\div\V v & =0,\nonumber \\
\int\delta_{a}\left(\V r_{i}-\V r^{\prime}\right)\V v\left(\V r^{\prime},t\right)\, d\V r^{\prime} & =\V u_{p}+\V{\omega}_{p}\times\left(\V r_{i}-\V q_{p}\right)+\slip_{i},\quad\forall i\in\mathcal{B}_{p},\label{eq:semi_continuum}\\
\sum_{i\in\mathcal{B}_{p}}\V{\lambda}_{i} & =\V f_{p},\quad\forall p,\nonumber \\
\sum_{i\in\mathcal{B}_{p}}(\V r_{i}-\V q_{p})\times\V{\lambda}_{i} & =\V{\tau}_{p},\quad\forall p.\nonumber 
\end{align}
Note that here the velocity and pressure fields contain both the ``background''
and the ``perturbational'' contributions to the flow. In the first
equation in (\ref{eq:semi_continuum}), the kernel function is used
to transfer (spread) the force exerted on the blob to the fluid, and
in the third equation the same kernel is used to average the fluid
velocity in the region covered by the blob and constrain it to follow
the imposed rigid body motion plus additional slip or body deformation.
The handling of the spreading of constraint forces and averaging of
the fluid velocity near physical boundaries is discussed in Appendix
D in \citet{RigidIBM}. We have implicitly assumed that appropriate
boundary conditions are specified for the fluid velocity and pressure.
Notably, we will apply the above formulation to cases where periodic
or no-slip boundary conditions are applied along the boundaries of
a cubic prism (recall that periodic boundaries are not actual physical
boundaries). This includes, for example, a slit channel, a square
channel, or a cubical container. It is also relatively straightforward
to handle stress-based boundary conditions such as free-slip or pressure
valves \citet{NonProjection_Griffith}.

It is not difficult to show that (\ref{eq:semi_continuum}) is equivalent
to the system (\ref{eq:rigidSystem}) with the mobility matrix between
two blobs $i$ and $j$ identified with \citet{SIBM_Brownian,BrownianBlobs,RigidIBM,ForceCoupling_Fluctuations,ForceCoupling_Stokes,ForceCoupling_Monopole}
\begin{equation}
\Mob_{ij}\left(\V r_{i},\,\V r_{j}\right)=\eta^{-1}\int\delta_{a}(\V r_{i}-\V r^{\prime})\Greens(\V r^{\prime},\V r^{\prime\prime})\delta_{a}(\V r_{j}-\V r^{\prime\prime})\ d\V r^{\prime}d\V r^{\prime\prime}\label{eq:GreensMobility}
\end{equation}
where we recall that $\Greens$ is the Green's function for the Stokes
problem with unit viscosity and the specified boundary conditions.
This expression can directly be compared to (\ref{eq:MobilityFaxen})
after realizing that for a smooth velocity field \citet{ForceCoupling_Stokes,ForceCoupling_Monopole},
\[
\int\delta_{a}(\V r_{i}-\V r)\V v(\V r)d\V r\approx\left[\M I+\left(\int\frac{x^{2}}{2}\delta_{a}\left(x\right)dx\right)\grad^{2}\right]\V v\left(\V r\right)\big|_{\V r=\V r_{i}}=\left(\M I+\frac{a_{F}^{2}}{6}\grad^{2}\right)\V v\left(\V r\right)\big|_{\V r=\V r_{i}},
\]
where we assumed a spherical blob, $\delta_{a}\left(\V r\right)\equiv\delta_{a}\left(r\right)$.
We have defined here the ``Faxen'' radius of the blob $a_{F}\equiv\left(3\int x^{2}\delta_{a}(x)\, dx\right)^{1/2}$
through the second moment of the kernel function.

In multipole expansion based methods, the self-mobility of a body
is treated separately by solving the single-body problem exactly (this
is only possible for simple particle shapes). However, in the type
of approach followed here the self-mobility $\Mob_{ii}$ is also given
by the same formula (\ref{eq:GreensMobility}) with $i=j$ and does
not need to be treated separately. In fact, the self-mobility of a
particle in an unbounded three-dimensional domain \emph{defines} the
effective hydrodynamic radius $a$ of a blob,
\[
\Mob_{ii}=\frac{1}{6\pi\eta a}\M I=\eta^{-1}\int\delta_{a}(\V r^{\prime})\Oseen(\V r^{\prime}-\V r^{\prime\prime})\delta_{a}(\V r^{\prime\prime})\ d\V r^{\prime}d\V r^{\prime\prime},
\]
where the Oseen tensor $\Oseen$ is given in (\ref{eq:OseenTensor}).
In general, $a_{F}\neq a$, but for a suitable choice of the kernel
one can accomplish $a_{F}\approx a$ (for example, for a Gaussian
$a/a_{F}=\sqrt{3/\pi}$ \citet{ForceCoupling_Monopole}) and thus
accurately obtain the Faxen correction for a rigid sphere \citet{BrownianBlobs}.

For an isotropic or tensor product kernel $\delta_{a}$ and an unbounded
domain, the pairwise blob-blob mobility (\ref{eq:GreensMobility})
will take the form
\begin{equation}
\Mob_{ij}=f\left(r_{ij}\right)\Id+g\left(r_{ij}\right)\hat{\V r}_{ij}\otimes\hat{\V r}_{ij},\label{eq:M_tilde_ij}
\end{equation}
where $\V r_{ij}=\V r_{i}-\V r_{j}$, and hat denotes a unit vector.
The functions of distance $f(r)$ and $g(r)$ depend on the specific
kernel (and in the fully discrete setting on the spatial discretization
of the Stokes equations) and will be different from those appearing
in the RPY tensor (\ref{eq:RPYTensor}). Nevertheless, as we will
show numerically in Section \ref{sub:TransInv}, the functions $f$
and $g$ for our IB method are quite close in form to those appearing
in the RPY tensor. We note that the RPY tensor itself can be seen
as a realization of (\ref{eq:GreensMobility}) with the kernel being
a surface delta function over a sphere of radius $a$ \citet{RPY_Shear_Wall}.

We have demonstrated above that solving (\ref{eq:semi_continuum})
is a way to apply the blob-blob mobility for a confined domain. In
the method of regularized Stokeslets \citet{RegularizedStokeslets_Walls,RegularizedStokeslets_Periodic,RegularizedStokeslets,RegularizedStokeslets_2D}
the mobility is obtained \emph{analytically} by averaging the analytical
Green's function with a kernel or envelope function specifically chosen
to make the resulting integrals analytical. Note however that in that
method the kernel $\delta_{a}$ appears only once inside the integral
in (\ref{eq:GreensMobility}) because only the force spreading is
regularized but not the interpolation of the velocity; this leads
to non-symmetric mobility matrix inconsistent with the Faxen formula
(\ref{eq:MobilityFaxen}). By contrast, our approach is guaranteed
to lead to a symmetric positive semidefinite (SPD) mobility matrix
$\Mob$, which is crucial when including thermal fluctuations \citet{BrownianBlobs,SIBM_Brownian,ForceCoupling_Fluctuations}.

\subsection{Fully Discrete Formulation}

To obtain a fully discrete formulation of the linear system (\ref{eq:semi_continuum})
we need to spatially discretize the Stokes equations on a grid. The
spatial discretization of the fluid equation used in this work uses
a uniform Cartesian grid with grid spacing $h$, and is based on a
second-order accurate staggered-grid finite volume (equivalently,
finite difference) discretization, in which vector-valued quantities
such as velocity, are represented on the faces of the Cartesian grid
cells, while scalar-valued quantities such as pressure are represented
at the centers of the grid cells \citet{IBAMR,NonProjection_Griffith,ISIBM,RigidIBM}.
The viscous terms are discretized using a standard $7$-point Laplacian
(in three dimensions), accounting for boundary conditions using ghost
cell extrapolation \citet{NonProjection_Griffith,RigidIBM}.

\subsubsection{Spreading and interpolation}

In the fully discrete formulation of the fluid-body coupling, we replace
spatial integrals in the semi-continuum formulation (\ref{eq:semi_continuum})
by sums over fluid grid points. The regularized delta function kernel
is discretized using a tensor product of one-dimensional immersed
boundary kernels $\phi_{a}\left(x\right)$ of compact support, following
Peskin \citet{IBM_PeskinReview}. To maximize translational and rotational
invariance (i.e., improve grid-invariance) we use the smooth (three-times
differentiable) six-point kernel recently described by Bao \emph{et
al.} \citet{New6ptKernel}. This kernel is more expensive than the
traditional four-point kernel \citet{IBM_PeskinReview} because it
increases the support of the kernel to $6^{3}=216$ grid points in
three dimensions; however, this cost is justified because the new
six-point kernel improves the translational invariance by orders of
magnitude compared to other standard IB kernel functions \citet{New6ptKernel}.

The interaction between the fluid and the rigid body is mediated through
two crucial operations. The discrete velocity-interpolation operator
$\J$ averages velocities on the staggered grid in the neighborhood
of blob $i$ via 
\[
\left(\J\V v\right)_{i}^{\alpha}=\sum_{k}v_{k}^{\alpha}\;\phi_{a}\left(\V r_{i}-\V r_{k}^{\alpha}\right),
\]
where the sum is taken over faces $k$ of the grid, $\alpha$ indexes
coordinate directions ($x,y,z$) as a superscript, and $\V r_{k}^{\alpha}$
is the position of the center of the grid face $k$ in the direction
$\alpha$. The discrete force-spreading operator $\S$ spreads forces
from the blobs to the faces of the staggered grid via
\begin{equation}
\left(\S\V{\lambda}\right)_{k}^{\alpha}=\D V^{-1}\sum_{i}\V{\lambda}_{i}^{\alpha}\;\phi_{a}\left(\V r_{i}-\V r_{k}^{\alpha}\right),\label{eq:S_unweighted}
\end{equation}
where now the sum is over the blobs and $\D V=h^{3}$ is the volume
of a grid cell. These operators are adjoint with respect to a suitably-defined
inner product, and the discrete matrices satisfy $\J=\D V\,\S^{T}$,
which ensures conservation of energy \citet{IBM_PeskinReview}. Extensions
of the basic interpolation and spreading operators to account for
the presence of physical boundary conditions are described in Appendix
D in \citet{RigidIBM}.

We note that it is possible to change the effective hydrodynamic and
Faxen radii of a blob by changing the kernel $\delta_{a}$. Such flexibility
in the kernel can be accomplished without compromising the required
kernel properties postulated by Peskin \citet{IBM_PeskinReview} by
using shifted or \emph{split kernels} \citet{ISIBM},
\[
\phi_{a,s}\left(\V q-\V r_{k}\right)=\frac{1}{2^{d}}\,\prod_{\alpha=1}^{d}\left\{ \phi_{a}\left[q_{\alpha}-\left(r_{k}\right)_{\alpha}-\frac{s}{2}\right]+\phi_{a}\left[q_{\alpha}-\left(r_{k}\right)_{\alpha}+\frac{s}{2}\right]\right\} ,
\]
where $s$ denotes a shift that parametrizes the kernel. By varying
$s$ in a certain range, for example, $0\leq s\leq h$, one can smoothly
increase the support of the kernel and thus increase the hydrodynamic
radius of the blob by as much as a factor of two. We do not use split
kernels in this work but have found them to work as well as the unshifted
kernels, while allowing increased flexibility in varying the grid
spacing relative to the hydrodynamic radius of the particles.

\subsubsection{Discrete constrained Stokes equations}

Following spatial discretization, we obtain a finite-dimensional linear
system of equations for the discrete velocities and pressures and
the blob and body degrees of freedom. For the resistance problem,
we obtain the following rigidly constrained discrete Stokes system
\citet{RigidIBM},

\begin{equation}
\left[\begin{array}{ccc}
\A & \Grad & -\S\\
-\Div & \M 0 & \M 0\\
-\J & \M 0 & -\M{\Omega}
\end{array}\right]\left[\begin{array}{c}
\V v\\
\pi\\
\V{\lambda}
\end{array}\right]=\left[\begin{array}{c}
\V g=\V 0\\
\V h=\V 0\\
\V w=-\slip
\end{array}\right],\label{eq:constrained_Stokes}
\end{equation}
where $\Grad$ is the discrete (vector) gradient operator, $\Div=-\Grad^{T}$
is the discrete (vector) divergence operator, and $\A=-\eta\Lap$
where $\Lap$ is a discrete (vector) Laplacian; these finite-difference
operators take into account the specified boundary conditions \citet{NonProjection_Griffith}.
For impermeable bodies $\M{\Omega}=\M 0$, which makes the linear
system (\ref{eq:constrained_Stokes}) a nested saddle-point problem
in both Lagrange multipliers $\pi$ and $\V{\lambda}$. As explained
in Appendix \ref{app:Permeable}, for permeable bodies $\M{\Omega}$
is a diagonal matrix with $\Omega_{ii}=\kappa_{p}/\left(\eta\D V_{i}\right)$
for blob $i\in\mathcal{B}_{p}$, where $\kappa_{p}$ is the permeability
of body $p$ and $\D V_{i}$ is a volume associated with blob $i$.
The right-hand side could include any external fluid forcing terms,
slip, inhomogeneous boundary conditions, etc. The system (\ref{eq:constrained_Stokes})
can be made symmetric by excluding the volume weighting $\D V^{-1}$
in the spreading operator (\ref{eq:S_unweighted}); this makes $\V{\lambda}$
have units of force density rather than total force.

This nested saddle-point structure continues if one considers impermeable
rigid bodies that are free to move, leading to the \emph{discrete
mobility problem} %
\footnote{Note that in actual codes it is better to use an increment formulation
of the linear system where the unknowns are the changes of the unknowns
from their values at the previous time step; this is particularly
important when there is a non trivial background flow to ensure that
the (small) perturbative flows are resolved accurately.%
}
\begin{equation}
\left[\begin{array}{cccc}
\A & \Grad & -\S & \M 0\\
-\Div & \M 0 & \M 0 & \M 0\\
-\J & \M 0 & \M 0 & \K\\
\M 0 & \M 0 & \K^{T} & \M 0
\end{array}\right]\left[\begin{array}{c}
\V v\\
\pi\\
\V{\lambda}\\
\V U
\end{array}\right]=\left[\begin{array}{c}
\V g\\
\V h=\V 0\\
\V w=-\slip\\
\V z=\V F
\end{array}\right].\label{eq:free_kinematics_Stokes}
\end{equation}
After eliminating the velocity and pressure from this system, we obtain
the saddle-point system (\ref{eq:saddle_M}) with the identification
of the mobility with its discrete approximation
\begin{equation}
\Mob=\J\L^{-1}\S=\D V\,\S^{T}\L^{-1}\S,\label{eq:mob_def}
\end{equation}
which is SPD. Here $\L^{-1}$ is a discrete Stokes solution operator,
\begin{equation}
\L^{-1}=\A^{-1}-\A^{-1}\Grad\left(\Div\A^{-1}\Grad\right)^{-1}\Div\A^{-1},\label{eq:L_inv_general}
\end{equation}
where we have assumed for now that $\A^{-1}$ is invertible; see \citet{RigidIBM}
for the handling of periodic systems, for which the Laplacian is not
invertible. Unlike for Green's function based methods, we never explicitly
compute or form $\L^{-1}$ or $\Mob$; rather, we solve the Stokes
velocity-pressure subsystems iteratively using the preconditioners
described in \citet{NonProjection_Griffith,StokesKrylov}.

\subsection{\label{sub:Preconditionining-Algorithm}Preconditioning Algorithm}

In this section we describe how to solve the system (\ref{eq:free_kinematics_Stokes})
using an iterative solver, as we have implemented in the Immersed
Boundary Adaptive Mesh Refinement software framework (IBAMR) \citet{IBAMR}.
Our codes are integrated into the public release of the IBAMR library.
Note that the matrix-vector product is a straightforward and inexpensive
application of finite-difference stencils on the fluid grid and summations
over blobs. The key to an effective solver is the design of a good
preconditioner, i.e., a good approximate solver for (\ref{eq:free_kinematics_Stokes}).
The basic idea is to combine a preconditioner for the Stokes problem
\citet{Elman_FEM_Book,NonProjection_Griffith,StokesKrylov} with the
indefinite preconditioner (\ref{eq:indef_block_P}) with a block-diagonal
approximation of the mobility $\widetilde{\M{\mathcal{M}}}$ constructed
based on empirical fits of the blob-blob mobility, as we know explain
in detail.

\subsubsection{\label{sub:ApproximateBlobMob}Approximate blob-blob mobility matrix}

A preconditioner for solving the resistance problem (\ref{eq:constrained_Stokes})
was developed by some of us in \citet{RigidIBM}; readers interested
in additional details should refer to this work. The preconditioner
is based on approximating the blob-blob mobility with the functional
form (\ref{eq:M_tilde_ij}), where the functions $f(r)$ and $g(r)$
are obtained by fitting numerical data for the blob-blob mobility
in an \emph{unbounded} system (in practice, a large periodic system).
This involves two important approximations, the validity of which
only affects the \emph{efficiency} of the linear solver but does \emph{not}
affect the \emph{accuracy} of the method since the Krylov method will
correct for the approximations. The first approximation comes from
the fact that the true blob-blob mobility for the immersed boundary
method is not perfectly translationally and rotationally invariant,
so that the form (\ref{eq:M_tilde_ij}) does not hold exactly. The
second approximation is that the boundary conditions are not correctly
taken into account when constructing the approximation of the mobility
$\widetilde{\M{\mathcal{M}}}$. This approximation is crucial to the
feasibility of our method and is much more severe, but, as we will
demonstrate numerically in Section \ref{sec:ResultsConfined}, the
Krylov solver converges in a reasonable number of iterations, correctly
incorporating the boundary conditions in the solution.

The empirical fits of $f(r)$ and $g(r)$ are described in Appendix
A of \citet{RigidIBM}, and code to evaluate the empirical fits is
publicly available for a number of kernels constructed by Peskin and
coworkers (three-, four-, and six-point) at \url{http://cims.nyu.edu/~donev/src/MobilityFunctions.c}.
As we show in Section \ref{sub:TransInv}, these functions are quite
similar to those appearing in the RPY tensor (\ref{eq:RPYTensor}),
and, in fact, it is possible to use the RPY functions $f_{RPY}(r)$
and $g_{RPY}(r)$ in the preconditioner, with a value of the effective
hydrodynamic radius $a$ that depends on the choice of the kernel.
Nevertheless, somewhat better performance is achieved by using the
empirical fits for $f(r)$ and $g(r)$ developed in \citet{RigidIBM}.

In \citet{RigidIBM}, we considered general fluid-structure interaction
problems over a range of Reynolds numbers, and constructed $\widetilde{\M{\mathcal{M}}}$
as a dense matrix of size $\left(dN_{b}\right)\times\left(dN_{b}\right)$,
which was then factorized using dense linear algebra. This is infeasible
for suspensions of many rigid bodies. In this work, we use the block-diagonal
approximation (\ref{eq:M_tilde_block_diag}) to the blob-blob mobility
matrices, in which there is one block per rigid particle. Once $\widetilde{\M{\mathcal{M}}}$
is constructed and its diagonal blocks factorized, the corresponding
approximate body mobility matrix $\widetilde{\M{\mathcal{N}}}$ is
easy to form, as discussed in more detail in Section \ref{sub:Preconditioner}.
Note that these matrices and their factorizations need to be constructed
only once at the beginning of the simulation, and can be reused throughout
the simulation.

\subsubsection{\label{sub:FluidSolver}Fluid solver}

A key component of solving the constrained Stokes problems (\ref{eq:constrained_Stokes})
or (\ref{eq:free_kinematics_Stokes}) is an iterative solver for the
unconstrained discrete Stokes sub-problem,
\[
\left[\begin{array}{cc}
\A & \Grad\\
-\Div & \M 0
\end{array}\right]\left[\begin{array}{c}
\V v\\
\pi
\end{array}\right]=\left[\begin{array}{c}
\V g\\
\V h
\end{array}\right],
\]
for which a number of techniques have been developed in the finite-element
context \citet{Elman_FEM_Book}. To solve this system, we can use
GMRES with a preconditioner $\sM P_{S}^{-1}$ that assumes periodic
boundary conditions so that the various finite-difference operators
commute \citet{ApproximateCommutators}. Specifically, the preconditioner
for the Stokes system that we use in this work is based on a projection
preconditioner developed by Griffith \citet{NonProjection_Griffith,StokesKrylov},
\begin{equation}
\sM P_{S}^{-1}=\left(\begin{array}{cc}
\Id & \; h^{2}\Grad\widetilde{\Lp}^{-1}\\
\V 0 & \eta\Id
\end{array}\right)\left(\begin{array}{cc}
\Id & \V 0\\
-\Div & -\Id
\end{array}\right)\left(\begin{array}{cc}
\eta^{-1}\widetilde{\Lap}^{-1} & \V 0\\
\V 0 & \Id
\end{array}\right),\label{P_Stokes}
\end{equation}
where $\Lp=h^{2}\left(\Div\Grad\right)$ is the dimensionless pressure
(scalar) Laplacian, and $\widetilde{\Lap}^{-1}\approx\left(\Lap\right)^{-1}$
and $\widetilde{\Lp}^{-1}\approx\left(\Lp\right)^{-1}$ denote approximate
solvers obtained by a \emph{single} V-cycle of a geometric multigrid
method, as performed using the \emph{hypre} library \citet{hypre}
in our IBAMR implementation. In this paper we will primarily report
the options we have found to be best without listing all of the different
combinations we have tried. For completeness, we note that we have
tried the better-known lower and upper triangular preconditioners
\citet{Elman_FEM_Book,StokesKrylov} for the Stokes problem. While
these simpler preconditioners are better when solving pure Stokes
problems than the projection preconditioner (\ref{P_Stokes}) since
they avoid the pressure multigrid application $\widetilde{\Lp}^{-1}$,
we have found them to perform much worse in the context of suspensions
of rigid bodies. A possible explanation is that the projection preconditioner
$\sM P_{S}^{-1}$ is the only one that is exact for periodic systems
if exact subsolvers for the velocity and pressure subproblems are
used.

Observe that one application of $\sM P_{S}^{-1}$ is relatively inexpensive
and involves only $(d+1)$ scalar multigrid V-cycles. The number of
iterations required for convergence depends strongly on the boundary
conditions; fast convergence is obtained within 10-20 iterations for
periodic systems, but as many as a hundred GMRES iterations may be
required for highly confined systems \citet{StokesKrylov}. We emphasize
that the performance of this preconditioner is highly dependent on
the details of the staggered geometric multigrid method, which is
not highly optimized in the \emph{hypre} library, especially for domains
of high aspect ratios such as narrow slit channels. For periodic boundary
conditions, one can use FFTs to solve the Stokes problem, and this
is likely to be more efficient than geometric multigrid especially
because FFTs have been highly optimized for common hardware architectures.
However, such an approach would require 3 scalar FFTs for \emph{each}
iteration of the iterative solver for the constrained Stokes problem
(\ref{eq:constrained_Stokes}) or (\ref{eq:free_kinematics_Stokes}),
and this will in general be substantially more expensive than using
only a few cycles of geometric multigrid as an \emph{approximate}
Stokes solver.

The use of an approximate Stokes solver instead of an exact one is
an important difference between implementing the rigid multiblob method
for periodic systems using the spectral Ewald method \citet{SpectralEwald_Stokes,SD_SpectralEwald}
and our approach. The product of the blob-blob mobility with a vector
can be computed more accurately and faster using the spectral Ewald
method, in particular because one can adjust the cutoff for splitting
the computation between real and Fourier space arbitrarily, unlike
in our method where the grid spacing is tied to the particle radius.
However, for rigid multiblobs, one must solve the system (\ref{eq:saddle_M}),
which requires potentially many matrix-vector products, i.e., many
FFTs in the spectral Ewald approach. By contrast, in our method we
solve the extended problem (\ref{eq:free_kinematics_Stokes}), and
only solve the Stokes problems approximately using a few cycles of
multigrid in each iteration. This will require more iterations but
each iteration can be substantially cheaper than performing three
FFTs each Krylov iteration. For non-periodic systems, there is no
equivalent of the spectral Ewald method, but see \citet{BrownianDynamics_OrderN,BrownianDynamics_OrderN2}
for some steps in this direction. Our method computes the hydrodynamic
interactions in a confined geometry ``on the fly'' without ever
actually computing the action of the Green's function exactly, rather,
it is computed only approximately and the outer Krylov solver corrects
for any approximations made in the preconditioner.

\subsubsection{Preconditioning Algorithm}

We now have the necessary ingredients to compose a preconditioner
for solving (\ref{eq:free_kinematics_Stokes}), i.e., to construct
an approximate solver for this linear system. Each application of
our preconditioner involves the following steps:
\begin{enumerate}
\item Approximately solve the fluid sub-problem,
\[
\left[\begin{array}{cc}
\A & \Grad\\
-\Div & \M 0
\end{array}\right]\left[\begin{array}{c}
\tilde{\V v}\\
\tilde{\pi}
\end{array}\right]=\left[\begin{array}{c}
\V g\\
\V h
\end{array}\right],
\]
using $N_{s}^{(1)}$ iterations of an iterative method with the preconditioner
(\ref{P_Stokes}).
\item Interpolate $\tilde{\V v}$ to get the relative slip at each of the
blobs, $\tilde{\V w}=\M{\mathcal{J}}\tilde{\V v}+\V w$, and rotate
the corresponding component from the current frame to the reference
frame of each body.
\item Approximately compute the unknown body kinematics $\V U$:

\begin{enumerate}
\item Calculate $\tilde{\V{\lambda}}=\widetilde{\M{\mathcal{M}}}^{-1}\tilde{\V w}$
and rotate the result back to the fixed frame of reference. Here $\widetilde{\M{\mathcal{M}}}$
is a block-diagonal approximation to the blob-blob mobility matrix
in the reference frame, as described in Section \ref{sub:ApproximateBlobMob};
the factorization of the blocks of $\widetilde{\M{\mathcal{M}}}$
is performed once at the beginning of the simulation.
\item Calculate $\widetilde{\V{\mathcal{F}}}=\V{\mathcal{F}}+\K\tilde{\V{\lambda}}$
and transform (rotate) $\widetilde{\V{\mathcal{F}}}$ to the body
frame of reference.
\item Compute $\V U=\widetilde{\M{\mathcal{N}}}\widetilde{\V{\mathcal{F}}}$
and transform it back to the fixed frame of reference, where $\widetilde{\M{\mathcal{N}}}=\left(\K\widetilde{\M{\mathcal{M}}}^{-1}\K^{T}\right)^{-1}$.
\end{enumerate}
\item Calculate the updated relative slip velocity at each of the blobs,
\[
\D{\V U}=\K^{T}\V U-\tilde{\V w},
\]
and transform (rotate) it to reference body frame.
\item Compute $\V{\lambda}=\widetilde{\M{\mathcal{M}}}^{-1}\D{\V U}$ and
transform $\V{\lambda}$ back to the fixed frame of reference if necessary.
\item Solve the corrected fluid subproblem to obtain the fluid velocity
and pressure:
\[
\left[\begin{array}{cc}
\A & \Grad\\
-\Div & \M 0
\end{array}\right]\left[\begin{array}{c}
\V v\\
\pi
\end{array}\right]=\left[\begin{array}{c}
\V g+\M{\mathcal{S}}\V{\lambda}\\
\V h
\end{array}\right],
\]
using $N_{s}^{(2)}$ iterations of an iterative method with the preconditioner
(\ref{P_Stokes}).
\end{enumerate}

A few comments are in order. The above preconditioner is not SPD so
the outer Krylov solver should be a method such as GMRES of BiCGStab
\citet{Iterative_Saad_Book}. We prefer to use right-preconditioned
Krylov solvers because in this case the residual computed by the iterative
solver is the true residual (as opposed to the preconditioned residual
for left preconditioning), and therefore termination criteria ensure
that the original system was solved to the desired target tolerance.
We expect that the long-term recurrence GMRES method will require
a smaller number of iterations than the short-term recurrence used
in BiCGStab (but note that each iteration of BiCGStab requires \emph{two}
applications of the preconditioner). However, observe that GMRES can
require substantially more memory since it requires storing a complete
history of the iterative process %
\footnote{Each vector requires storing a complete velocity and pressure field,
i.e., $4$ floating-point numbers per grid cell, which can make the
memory requirements of a GMRES-based solver with a large restart frequency
quite high for large grid sizes.%
}. This can be ameliorated by restarts at a cost of slowed convergence.
If the iterative solver used for the Stokes solver in steps 1 and
6 is a nonlinear method (most Krylov methods are nonlinear), then
the outer solver must be a flexible method such as FGMRES. This flexibility
typically increases the memory requirements of the iterative method
(for example, it exactly doubles the number of stored intermediate
vectors for FGMRES versus GMRES), and so an alternative is to use
a linear method such as Richardson's method %
\footnote{All of these iterative methods are available in the PETSc library
\citet{PETSc} we use in our IBAMR implementation \citet{IBAMR} of
the above preconditioner, making it simple to try different combinations
and study their effectiveness on any particular problem of interest.%
}. Note that when a preconditioned Krylov method is used for the Stokes
subsolver, one additional application of the preconditioner is required
to convert the system to preconditioned form for both left and right
preconditioning, making the total number of applications of the Stokes
preconditioner (\ref{P_Stokes}) be $N_{s}^{(1)}+N_{s}^{(2)}+2$ per
Krylov iteration. By contrast, if Richardson's method is used in the
Stokes subsolver, the number of preconditioner applications is $N_{s}^{(1)}+N_{s}^{(2)}$.
Since in many practical cases the cost is dominated by the multigrid
cycles, this difference can be important in the overall performance
of the preconditioner. We will explore the performance of the preconditioner
and the effect of the various choices in detail in Section \ref{sub:ConvergenceIBAMR}.

\section{\label{sec:ResultsUnbounded}Results: Unbounded Domain}

In this section we investigate the accuracy of rigid multiblob models
of spheres as a function of the number of blobs. We focus on spheres
in an unbounded domain because of the availability of analytical results
to compare to, and not because the rigid multiblob method is particularly
good for suspensions of spheres, for which there already exist a number
of well-developed multipole expansion approaches. We also investigate
the performance of the preconditioner developed in Sec. \ref{sub:Preconditioner}
for solving (\ref{eq:saddle_M}), for suspensions of spheres in an
unbounded domain (e.g., clusters of colloids formed in a gel). For
unbounded domains, we compute the product of the blob-blob mobility
matrix $\M{\mathcal{M}}$ with a vector using the Fast Multipole Method
(FMM) developed specifically for the RPY tensor in \citet{RPY_FMM};
this software makes four calls to the Poisson FMM implemented in the
FMMLIB3D library (\url{http://www.cims.nyu.edu/cmcl/fmm3dlib/fmm3dlib.html})
per matrix-vector product. As we will demonstrate empirically, the
asymptotic cost of the rigid-multiblob method scales as $N_{b}\,\ln N_{b}$,
where $N_{b}$ is the total number of blobs, with a coefficient that
grows only weakly with density. We note that in this paper we use
relatively tight tolerances ($\sim10^{-9}-10^{-8}$) when computing
the matrix-vector products solving the linear systems in order to
test the robustness of the preconditioners; in practical applications
much lower tolerances ($\sim10^{-5}-10^{-3}$) would typically be
employed, potentially lowering the overall computational effort considerably
over what is reported here.

In this work, each sphere is discretized with $n$ blobs of hydrodynamic
radius $a$ distributed on the surface of a sphere of \emph{geometric}
radius $R_{g}$. We discretize the surface of a sphere as a shell
of blobs constructed by a recursive procedure suggested to us by Charles
Peskin (private communication); the same procedure is used in \citet{RigidMultiblobs_Swan}.
We start with 12 blobs placed at the vertices of an icosahedron \citet{MultiblobSprings},
which gives a uniform triangulation of a sphere by 20 triangular faces.
Then, we place a new blob at the center of each edge and recursively
subdivide each triangle into four smaller triangles, projecting the
vertices back to the surface of the sphere along the way. Each subdivision
approximately quadruples the number of vertices, with the $k$-th
subdivision producing a model with $10\cdot4^{k-1}+2$ blobs, leading
to shells with 12, 42, 162 or 642 blobs, see Fig. 2 in \citet{BrownianMultiBlobs}
for an illustration. In this section we study the optimal choice of
$a$ for a given resolution (number of blobs) and $R_{g}$.

An important concept that will be used heavily in the rest of this
paper is that of an \emph{effective hydrodynamic radius $R_{h}\approx R_{g}+a/2$}
of a blob model of a sphere (more generally, effective hydrodynamic
extent). If we approach the rigid multiblob method from a boundary
integral perspective, we would assign $R_{h}=R_{g}$ as the radius
and treat the additional enlargement of the effective hydrodynamic
radius as a numerical (quadrature+regularization) error. This is more
or less how results are presented in the recent work of Swan and Wang
\citet{RigidMultiblobs_Swan} (see for example their Fig. 8), making
the accuracy appear low even in the far field for small number of
blobs per sphere. However, we instead think of a rigid multiblob as
an effective \emph{model} of a sphere, whose hydrodynamic response
mimics that of an equivalent sphere. A similar effect appears in lattice
Boltzmann simulations, with $a$ being related to the lattice spacing
\citet{VACF_Ladd,FiniteRe_3D_Ladd}. To appreciate why it is imperative
to use an effective radius, observe that even a single blob acts as
an approximation of a sphere with radius $a>0$. Similarly, one should
not treat a line of blobs (see left-most panel in Fig. \ref{fig:BlobModels})
as a zero-thickness object (line); rather, such a line of rigidly-connected
blobs should be considered to model a rigid cylinder with finite thickness
proportional to $a$ \citet{IBM_Sphere}. We compute the effective
hydrodynamic radius of our blob models of spheres next.

\subsection{\label{sub:ConvergenceMoments}Effective hydrodynamic radii of rigid
multiblob spheres}

In this section we consider an isolated rigid multiblob sphere in
an unbounded domain, and compute its response to an applied force
$f_{p}$, an applied torque $\tau_{p}$, and an applied linear shear
flow with strain rate $\gamma$. Each of these defines an effective
hydrodynamic radius by comparing to the analytical results for a sphere,
therefore, each model of a sphere will have three distinct hydrodynamic
radii.

The \emph{translational radius} is measured from (see also \citet{Multiblob_RPY_Rotation})
\[
R_{h}=\frac{f_{p}}{6\pi\eta u_{p}},
\]
where $u_{p}$ is the resulting sphere linear velocity, the \emph{rotational
radius} is (see also \citet{Multiblob_RPY_Rotation})
\[
R_{\tau}=\left(\frac{\tau_{p}}{8\pi\eta\omega_{p}}\right)^{1/3}
\]
where $\omega_{p}$ is the resulting angular velocity, and the effective
stresslet radius is
\[
R_{s}=\left(-\frac{3s_{11}}{20\pi\eta\gamma}\right)^{1/3}.
\]
Here we compute the stresslet $\V s$ induced on the rigid multiblob
under an applied shear by setting an apparent slip $\slip_{i}=-\V v(\V r_{i})=-\gamma(x,-y,0)$
on blob $i$, and then solving the mobility problem to compute the
constraint (rigidity) forces $\V{\lambda}$. The stresslet $\V s$
is the symmetric traceless component of the\textbf{ }first moment
of the constraint forces $\sum_{i\in\mathcal{B}_{p}}\V{\lambda}_{i}\otimes\V r_{i}.$
In this work, we use $R_{h}$ as the effective hydrodynamic radius
when comparing to theory. This is because the translational mobility
is controlled by the most long-ranged $1/r$ hydrodynamic interactions,
and therefore the far-field response of a rigid multiblob is controlled
by $R_{h}$.

Observe that since we only account for translation of the blobs, only
$R_{h}$ is nonzero for a single blob, while $R_{\tau}$ and $R_{s}$
are zero. Therefore, the minimal model of a sphere that allows for
nontrivial rotlet and stresslets is the icosahedral model (12 blobs)
\citet{MultiblobSprings}. Since the rigid multiblob models are able
to exert a stress on the fluid they can change the viscosity of a
suspension \citet{MultiblobSprings}, unlike the single-blob models,
which do not resist shear. It is important to note that the rigid
multiblob models of a sphere are \emph{not} perfectly rotationally
invariant, especially for low resolutions. Therefore, the rigid multiblobs
may exhibit a small translational velocity even in the absence of
an applied force, or they may exhibit a small rotation even in the
absence of an applied torque. In other words, the effective mobility
matrix for a rigid multiblob model of a sphere can exhibit small off-diagonal
components. Similarly, there will in general be small but nonzero
components of the stresslet that would be identically zero for a perfect
sphere. In general, we find these spurious components to be very small
even for the minimally resolved icosahedral rigid multiblob.

A key parameter that we need to choose is how to relate the blob hydrodynamic
radius $a$ with the typical spacing between the blobs. Since our
multiblob models of spheres are regular the minimal spacing between
markers $s$ is well-defined, and we expect that there will be some
optimal ratio $a/s$ that will make the rigid multiblob represent
a true rigid sphere as best as possible. In a number of prior works
the intuitive choice $a/s=1/2$ has been used, since this corresponds
to the idea that the blobs act as a sphere of radius $a$ and we would
like them to touch the other blobs. However, as we explained above,
it is not appropriate to think of blobs as spheres with a well-defined
surface, and it is therefore important to study the optimal spacing
more carefully.

\begin{table}[tbph]
\begin{tabular}{|c|c|c|c|c|c|c|}
\hline 
Number of blobs &
\multicolumn{3}{c|}{$a/s=1/2$} &
\multicolumn{3}{c|}{$a/s=1/4$}\tabularnewline
\cline{2-7} 
 &
$R_{h}/R_{g}$ &
$R_{\tau}/R_{g}$ &
$R_{s}/R_{g}$ &
$R_{h}/R_{g}$ &
$R_{\tau}/R_{g}$ &
$R_{s}/R_{g}$\tabularnewline
\hline 
12 &
1.2625 &
1.2313 &
1.2461 &
1.0154 &
1.0292 &
0.9890\tabularnewline
\hline 
42 &
1.1220 &
1.1019 &
1.1316 &
1.0035 &
1.0147 &
0.9959\tabularnewline
\hline 
162 &
1.0530 &
1.0472 &
1.0567 &
0.9998 &
1.0073 &
0.9968\tabularnewline
\hline 
642 &
1.0239 &
1.0227 &
1.0250 &
0.9992 &
1.0036 &
0.9932\tabularnewline
\hline 
2562 &
1.0113 &
1.0111 &
1.0115 &
0.9994 &
1.0018 &
0.9986\tabularnewline
\hline 
\end{tabular}\hspace{1cm}%
\begin{tabular}{|c|c|c|}
\hline 
\multirow{2}{*}{$\phi$} &
\multicolumn{2}{c|}{iterations}\tabularnewline
\cline{2-3} 
 & 12 blobs &
42 blobs\tabularnewline
\hline 
\hline 
$1.4\cdot10^{-3}$ &
4 &
4\tabularnewline
\hline 
0.011 &
5 &
6\tabularnewline
\hline 
0.09 &
9 &
10\tabularnewline
\hline 
0.18 &
13 &
13\tabularnewline
\hline 
0.36 &
20 &
23\tabularnewline
\hline 
\end{tabular}\caption{\label{tab:Stresslets_3D}(Left) Effective translational, rotational
and stresslet hydrodynamic radii for rigid multiblob models of a sphere,
for two choices of the blob-blob relative spacing $a/s$. (Right)
Iterations to solve the mobility problem with tolerance $10^{-8}$
for 4096 spheres discretized with 12 blobs, or for 512 spheres discretized
with 42 blobs, arranged in a simple cubic lattice at different volume
fractions $\phi$, see Sec. \ref{sub:ConvergenceFMM}.}
\end{table}

In the left half of Table \ref{tab:Stresslets_3D} we present the
effective sphere radii obtained for different resolutions for two
choices of $a/s$ (we have investigated a broad range of spacings,
not shown). The important observation we make is that even when the
radii are far from the geometric, such as for the 12-blob shell, the
different radii are rather consistent with each other. This means
that even low-resolution rigid multiblobs act like spheres as far
as low-order moments (multipoles) are concerned. We note that the
method developed in \citet{Multiblob_RPY_Rotation}, in which rotational
degrees of freedom are added to the blob description, gives $R_{h}\approx R_{\tau}$
to within a fraction of a percent even for only 12 blobs per sphere.
As the resolution is increased all hydrodynamic radii converge to
the geometric radius $R_{g}$ \emph{linearly} with $a/R_{g}$ (data
not shown). The table also suggests that $a/s=1/4$ is better than
$a/s=0.5$ because for $a/s=1/4$ all of the effective hydrodynamic
radii are remarkably close to $R_{g}$ even for the 12-blob model.
However, as we show in the next section, the choice $a/s=1/2$ is
substantially better when looking at how well lubrication forces are
resolved between two spheres.

\subsection{\label{sub:Pair-mobility}Mobility for a pair of spheres}

To determine the best value of $a/s$ in this section we examine the
hydrodynamic interaction between two spheres as they approach each
other. Since pairwise lubrication corrections are not added in our
approach, it is important to investigate how well lubrication is resolved
for different resolutions. To assert the accuracy of the rigid multiblob
models we will examine several non-trivial components of the mobility
between two spheres. Since rigid multiblob models are not rotationally
invariant the exact value for the pair mobility depends on the relative
orientation of the rigid multiblobs; here we report the mean and (twice
the) standard deviation of the particle velocity as error bars, averaged
over a sample of random orientations of the two particles. We note
that we have compared our results to those obtained with the method
developed in \citet{Multiblob_RPY_Rotation}, where rotational DOFs
are included in the blob description, and found only a small difference
(not shown). This means that the inclusion of blob torques does not
lead to an improvement in the accuracy with which pairwise hydrodynamic
interactions are computed.

\begin{figure*}[tbph]
\includegraphics[width=0.49\textwidth]{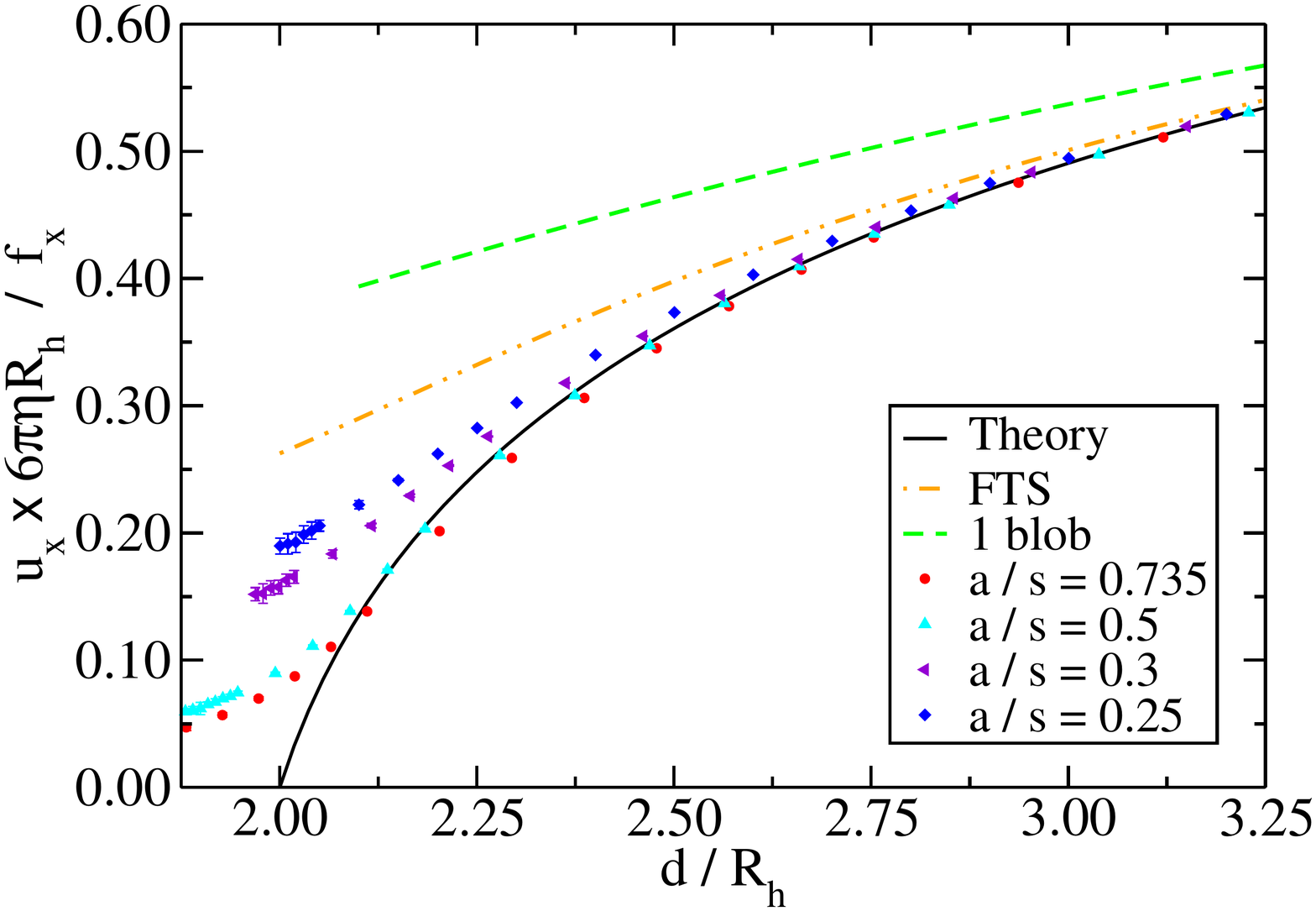}\includegraphics[width=0.49\textwidth]{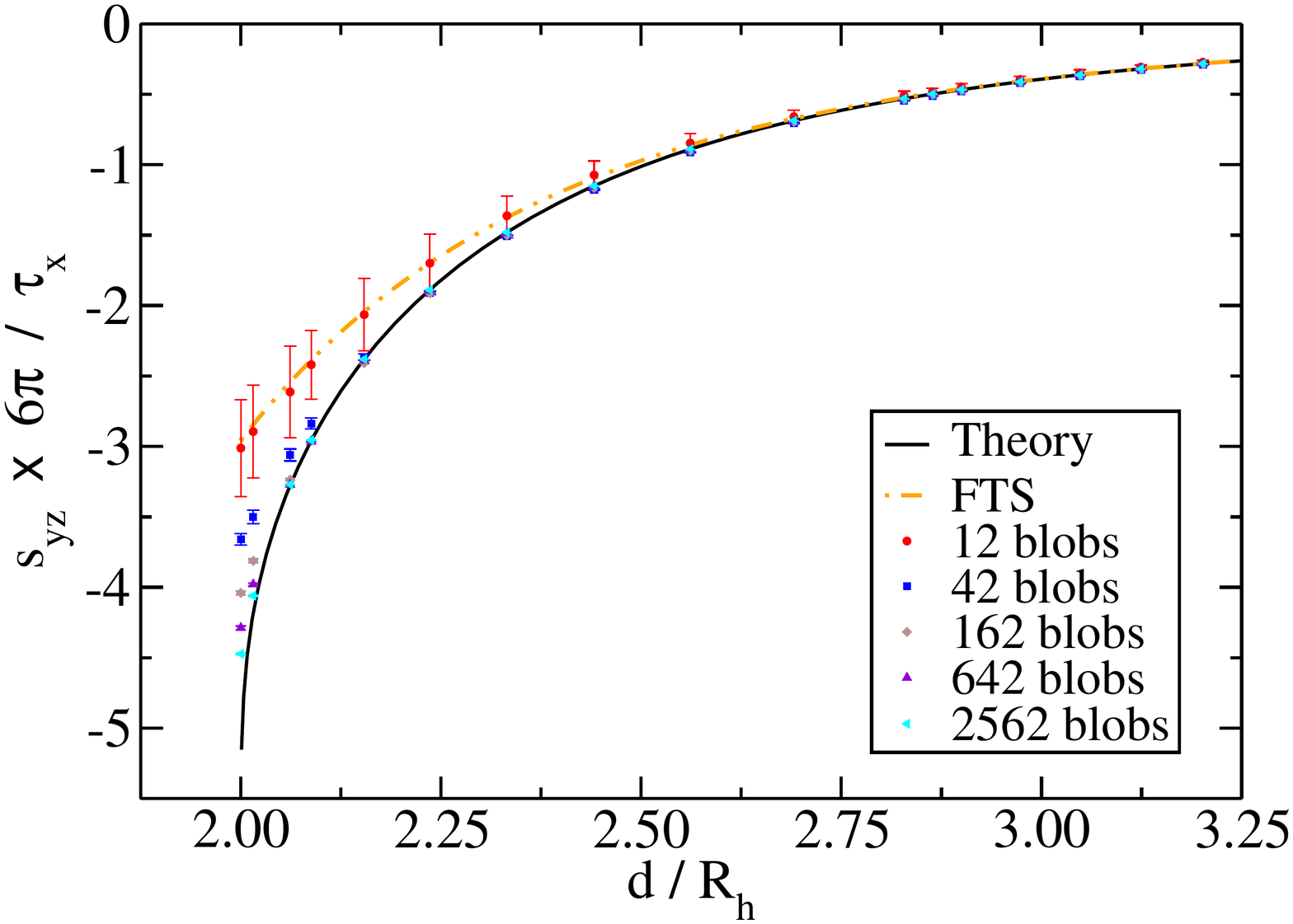}

\includegraphics[width=0.49\textwidth]{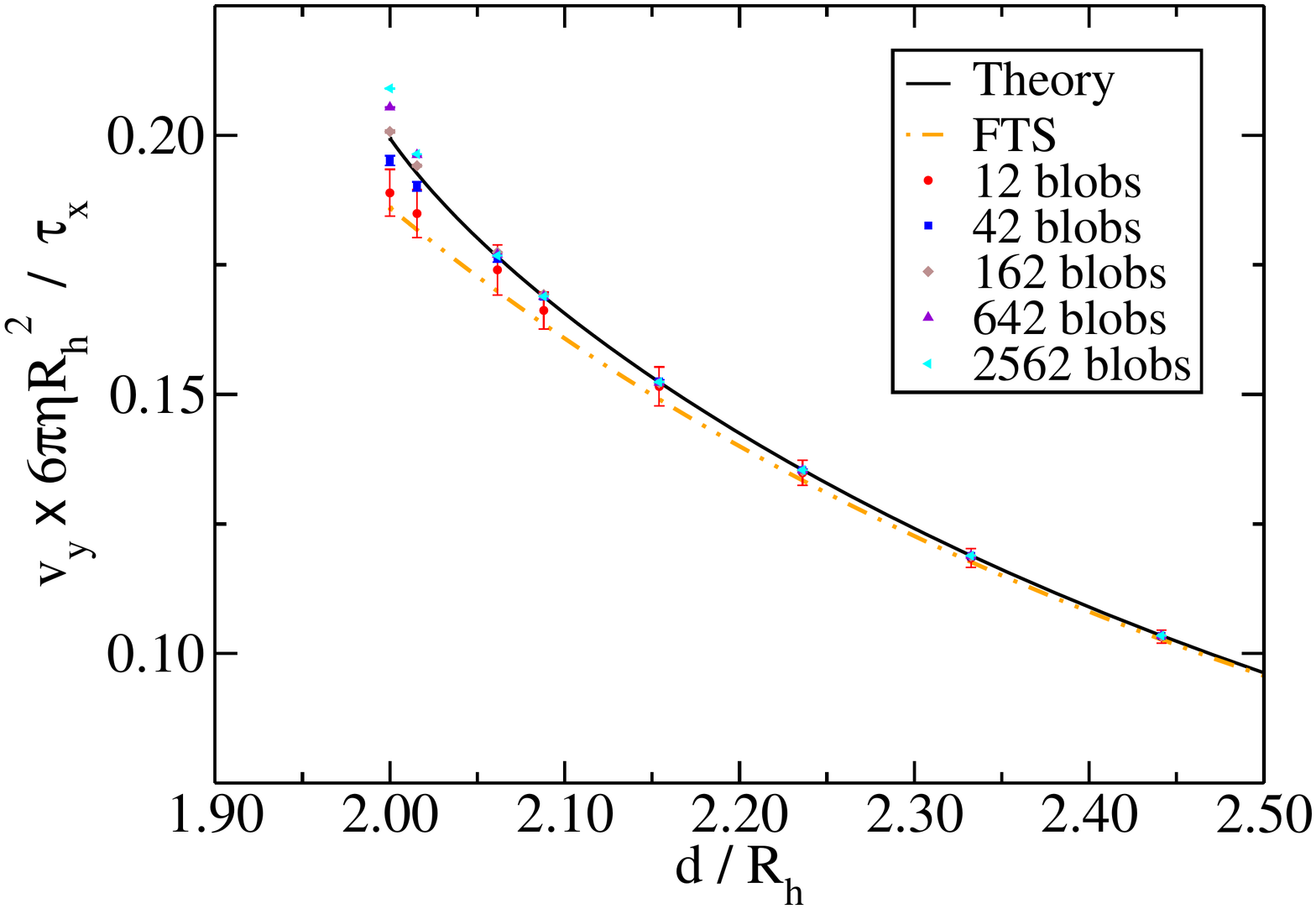}\includegraphics[width=0.49\textwidth]{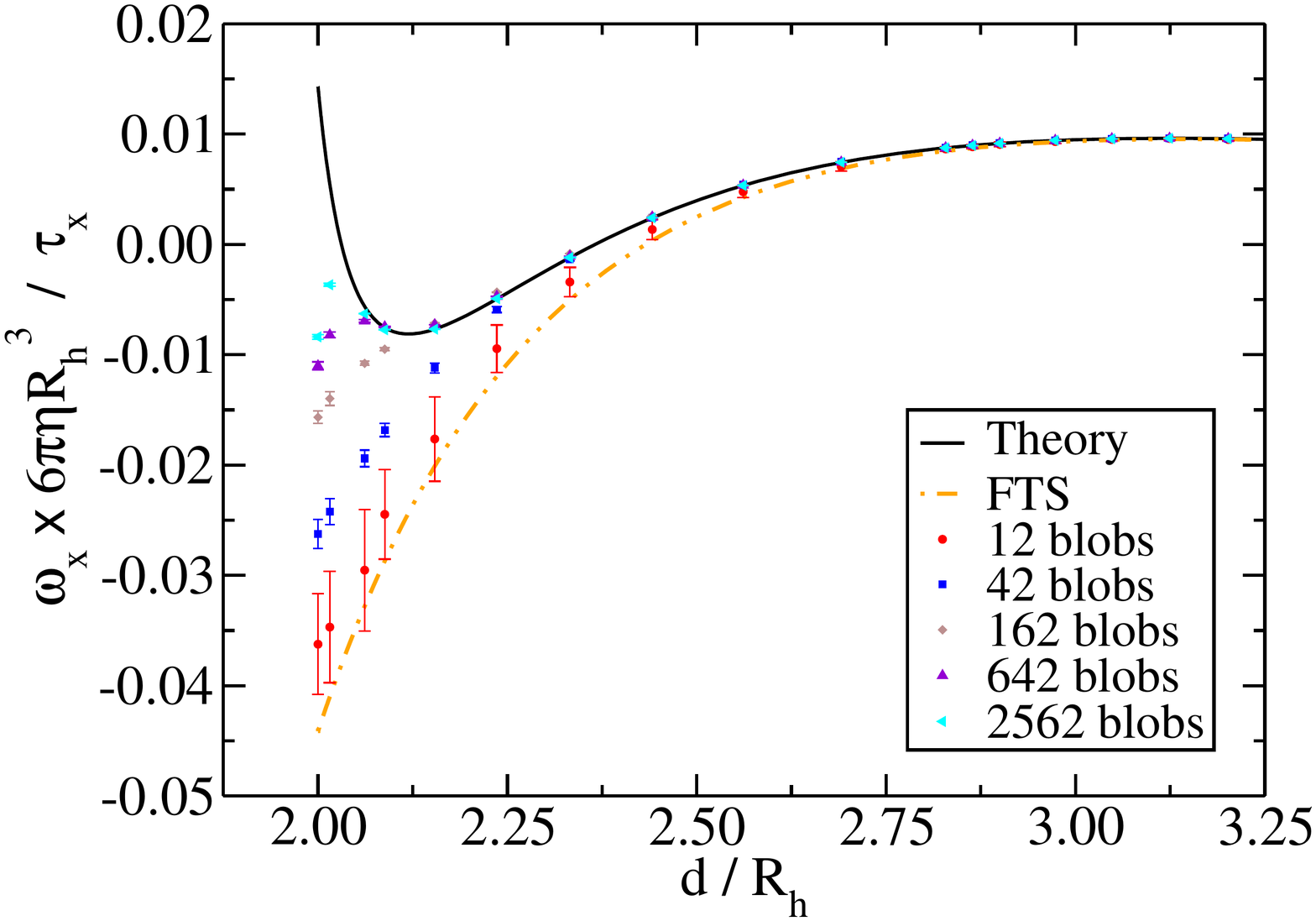}

\caption{\textbf{\label{fig:pairMobility}}Hydrodynamic coupling between two
identical spheres as a function of the center to center distance $d$.
Twice the standard deviation as the two spheres are rotated relative
to one another is shown as an error bar. Comparison is made to Stokesian
dynamics without lubrication corrections, i.e., truncation at the
FTS level, and to ``exact'' theory \citet{Brenner1961,SpheresStokes_JeffreyOnishi},
see legend. The top left panel shows the average sphere velocity under
the action of external unit forces $\V f_{1}=-\V f_{2}=\V f$ directed
along the line of collision, for a resolution of 162 blobs and for
several values of $a/s$. The remaining three panels show non-trivial
components of the pairwise mobility for a fixed $a/s=0.5$ and different
resolutions (number of blobs per sphere, see legend). One sphere,
located at $\left(\left(d{}^{2}-4R_{h}^{2}\right){}^{1/2},0,2R_{h}\right)$,
is subject to an external torque of magnitude one around the $x$-axis.
The response of the second sphere located at the origin is measured:
the top right panel shows the stresslet $\V s$ (i.e., the rotation-stresslet
coupling), the bottom left panel shows the linear velocity (i.e.,
the rotation-translation coupling) $\V v$, and the bottom right panel
shows the angular velocity $\V{\omega}$ (i.e., the rotation-rotation
coupling) of the second sphere.}
\end{figure*}

In our first test, we pull two spheres toward each other with equal
but opposite forces directed along the line of collision. In the top-left
panel of Fig. \ref{fig:pairMobility} we compare the numerical results
for spheres with $162$ markers and different blob radius $a$ with
the exact result derived by Brenner \citet{Brenner1961}. One can
see that for long distances all sensible choices of $a$ provide a
good agreement with the exact theory once the results have been scaled
with the hydrodynamic radius $R_{h}$ computed in Section \ref{sub:ConvergenceMoments}.
However, the choice of $a$ makes a big difference at short distances.
Specifically, for small $a/s$ flow can ``leak'' inbetween the blobs
and the lubrication force is substantially lowered. For large $a/s$,
we expect that the corrugation of the effective hydrodynamic surface
of the sphere will introduce deviations from a spherical shape. As
the figure illustrates, the best agreement between theory and numerical
results is obtained for $a/s=1/2$. This intuitive choice has been
used in other related methods \citet{StokesianDynamics_Rigid,SPM_Rigid,HYDROPRO},
while the IB community has favored shorter distance between markers
(but see discussion in Ref. \citet{RigidIBM}). In the rest of the
tests on this section we use $a/s=1/2$ and in the rest of the paper
we will use $a/s\approx1/2$ unless otherwise noted.

\begin{figure*}[tbph]
\includegraphics[width=0.49\textwidth]{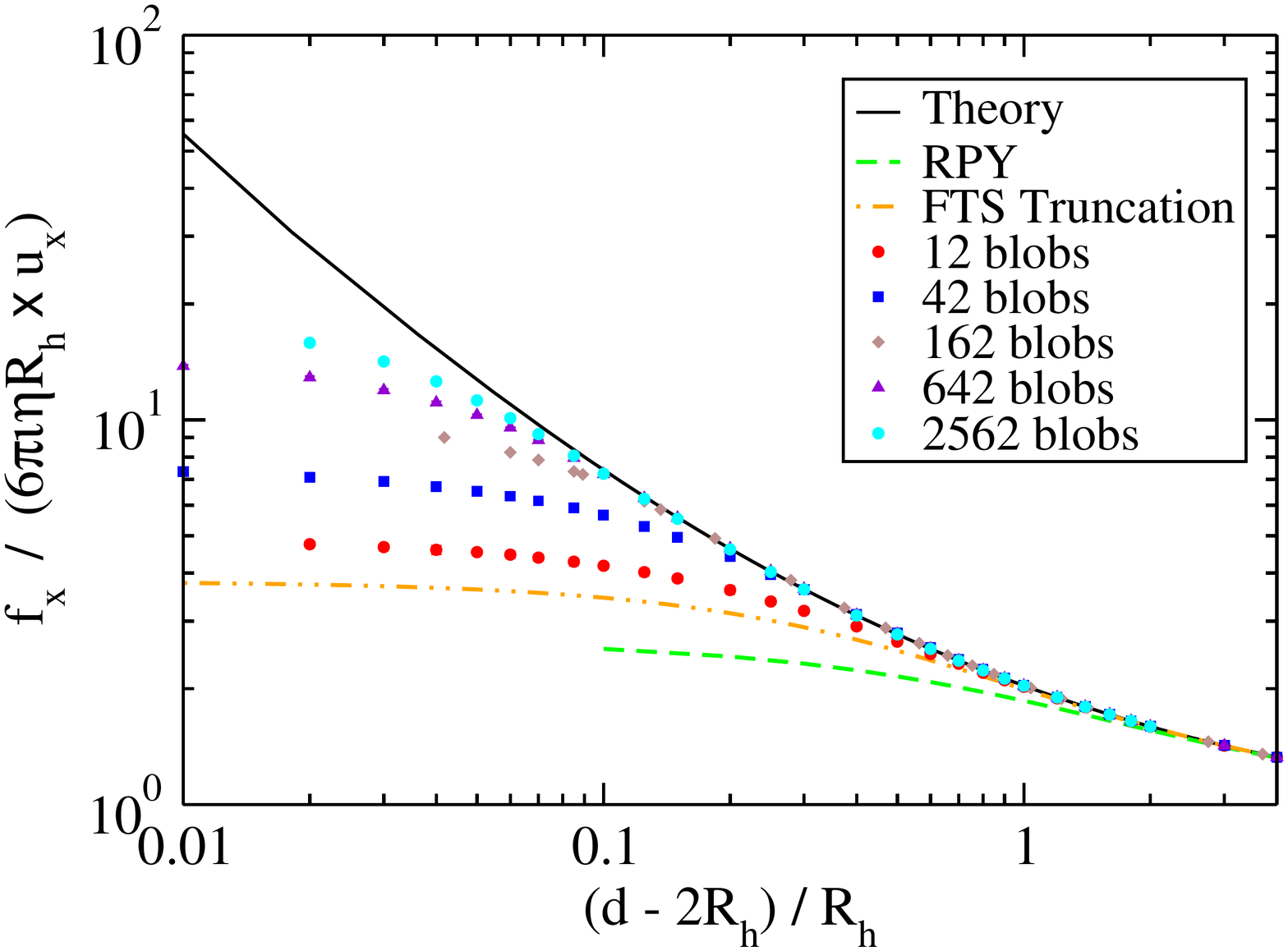}\includegraphics[width=0.49\textwidth]{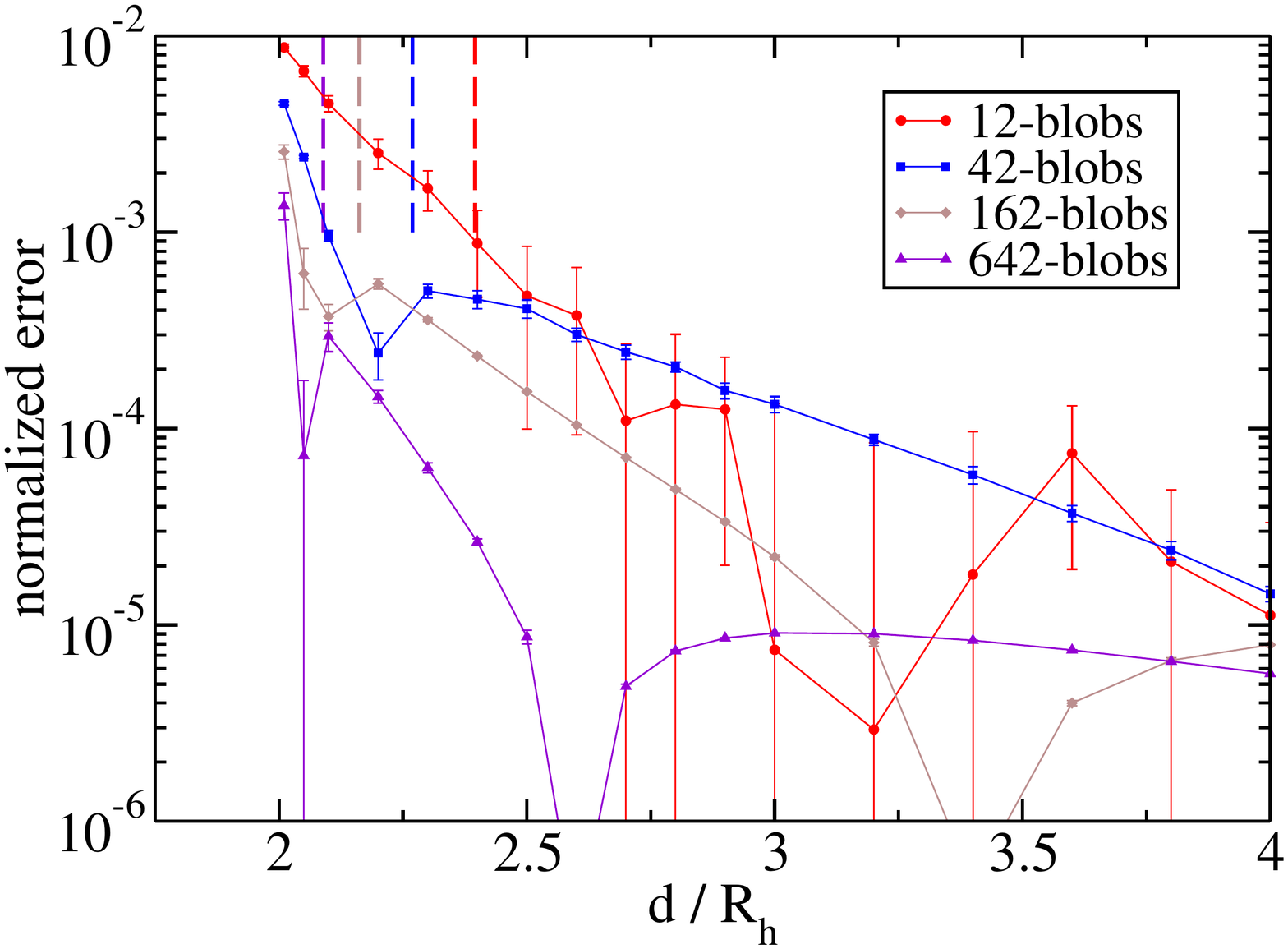}

\caption{\label{fig:lub}Lubrication forces on two identical spheres pulled
toward each other with equal forces, for different resolutions (see
legend). Twice the standard deviation as the two spheres are rotated
relative to one another is shown as an error bar. (Left panel) Dimensionless
normalized hydrodynamic resistance, i.e., the inverse of the hydrodynamic
mobility shown in the top left panel in Fig. \ref{fig:pairMobility}.
(Right panel) Velocity error for one sphere with respect the exact
theoretical result, normalized with the velocity at long distances
($f/6\pi\eta R_{h}$). The distance $d=2\left(R_{g}+a\right)$ at
which blobs start to overlap is marked as a vertical line of the same
color as the corresponding symbols.}
\end{figure*}

We explore the ``lubrication'' forces between spheres at very close
distances in more detail in Fig. \ref{fig:lub}. In the left panel
we show how the hydrodynamic force grows as the gap between the spheres
decreases. The hydrodynamics of the low resolution models start to
deviate from the theory for gaps $\sim R_{h}/2$ or smaller, while
the highest resolution model (2562 blobs) shows a good agreement with
the theory for gaps down to $0.1R_{h}$. The right panel of Fig. \ref{fig:lub}
shows the velocity error for one of the spheres with respect the exact
theoretical result. For all models the error is below $10^{-3}$ until
distances where the blobs forming the spherical shells start to overlap,
which is the intuitive distance above which we expect the rigid multiblob
to act as a good approximation to a sphere.

In our second test, we measure the velocity of one sphere located
at the origin when a second sphere, located at $(x,0,2R_{h})$, is
subject to an external torque applied around the $x$-axis. Since
the Brenner theory is only valid for spheres approaching along the
line of collision we use the expansion of Jeffrey \& Onishi accurate
to order $\mathcal{O}\left(r^{-100}\right)$ \citet{SpheresStokes_JeffreyOnishi}
to compare with our mobility results; this expansion is also used
in Stokesian Dynamics to compute near-field lubrication corrections
for pairs of spheres and can be computed using the libStokes library
of Ichiki \citet{libStokes}. One can see on the lower panels of Fig.
\ref{fig:pairMobility} that the low resolution model (12 blobs) is
similar to a Stokesian dynamics model that includes monopole (forces)
and dipole (torques and stresslet) terms but no lubrication corrections.
As the resolution of the models is increased the results agree better
with the theory, as expected.%
{} Note, however, that the lack of lubrication corrections in our models
prevents a perfect agreement down to contact distances. In the top
right panel of Fig. \ref{fig:pairMobility} we compare the stresslet
computed on the particle at the origin. Again, we observe that the
12 blob model is similar in accuracy (aside from the presence of nonzero
error bars, i.e., variance) to the Stokesian dynamic method without
lubrication corrections, while higher resolutions methods agree better
with the theory, as expected.

\subsection{\label{sub:Squirmer}Squirmer swimming speed}

In this section we confirm the ability of our method to model an active
sphere ``squirmer'' \citet{SquirmersFCM} with a prescribed tangential
surface slip on the surface. This slip $\breve{\V u}_{i}=\breve{\V u}\left(\V r_{i}\right)$
takes the following form in spherical coordinates, 
\[
\breve{\V u}_{r}=0,\quad\breve{\V u}_{\phi}=0,\quad\breve{\V u}_{\theta}=B_{1}\sin\theta
\]
The active translational velocity of the squirmer is well-known to
be the surface average of the surface slip \citet{SquirmerTheory_Stone},
\[
\V u=-\av{\breve{\V u}}=\frac{2}{3}B_{1}\hat{\M z}.
\]

We have numerically computed the swimming speed of a squirmer in an
unbounded domain for different resolutions and compared to the theory.
We obtain that the relative error $\epsilon$ in the swimming speed
is linear in $a/R_{g}$, which is expected. However, the error has
a large prefactor, $\epsilon\approx3.5\, a/R_{g}$, which is not small
for the low-resolution models. Furthermore, observe that linear convergence
with the size of the blobs implies only order one-half convergence
in the number of blobs since the number of blobs required to cover
the surface of the sphere grows quadratically with the sphere radius.

These findings confirm that the rigid multiblob models converge to
the correct swimming speed but the accuracy is not very good. This
is, in fact, not so surprising because we did not include any adjustments
to account for the (potentially large) difference between the effective
hydrodynamic radius $R_{h}$ and the geometric radius. That is, even
though the effective no-slip surface has a radius $R_{h}>R_{g}$,
we imposed the slip (in a locally-averaged way) at the surface of
a sphere of radius $R_{g}$. We will investigate these issues and
potential ways to improve the accuracy with which active slip is imposed
in future work. Here we simply note that rigid multiblob models are
well-suited to qualitative studies of suspensions of many active particles.
If one wishes to accurately model one or a few active particles higher-order
methods such as boundary integral methods are preferable.

\subsection{\label{sub:ConvergenceFMM}Suspension of spheres}

In this section we study the convergence of the preconditioned Krylov
solver for suspensions of many spheres. Our primary goal is to assess
the effectiveness of our block-diagonal preconditioner for different
packing densities (particle-particle distances) and numbers of particles.
In the tests of this section we use spherical shells of 42 blobs subject
to random forces, torques and slips. We form a finite cubic subset
of a simple cubic lattice and place it in an unbounded fluid domain.
We use right preconditioned GMRES without restarts, implemented using
the PETSc \citet{PETSc} library.

\begin{figure*}[tbph]
\includegraphics[width=0.49\textwidth]{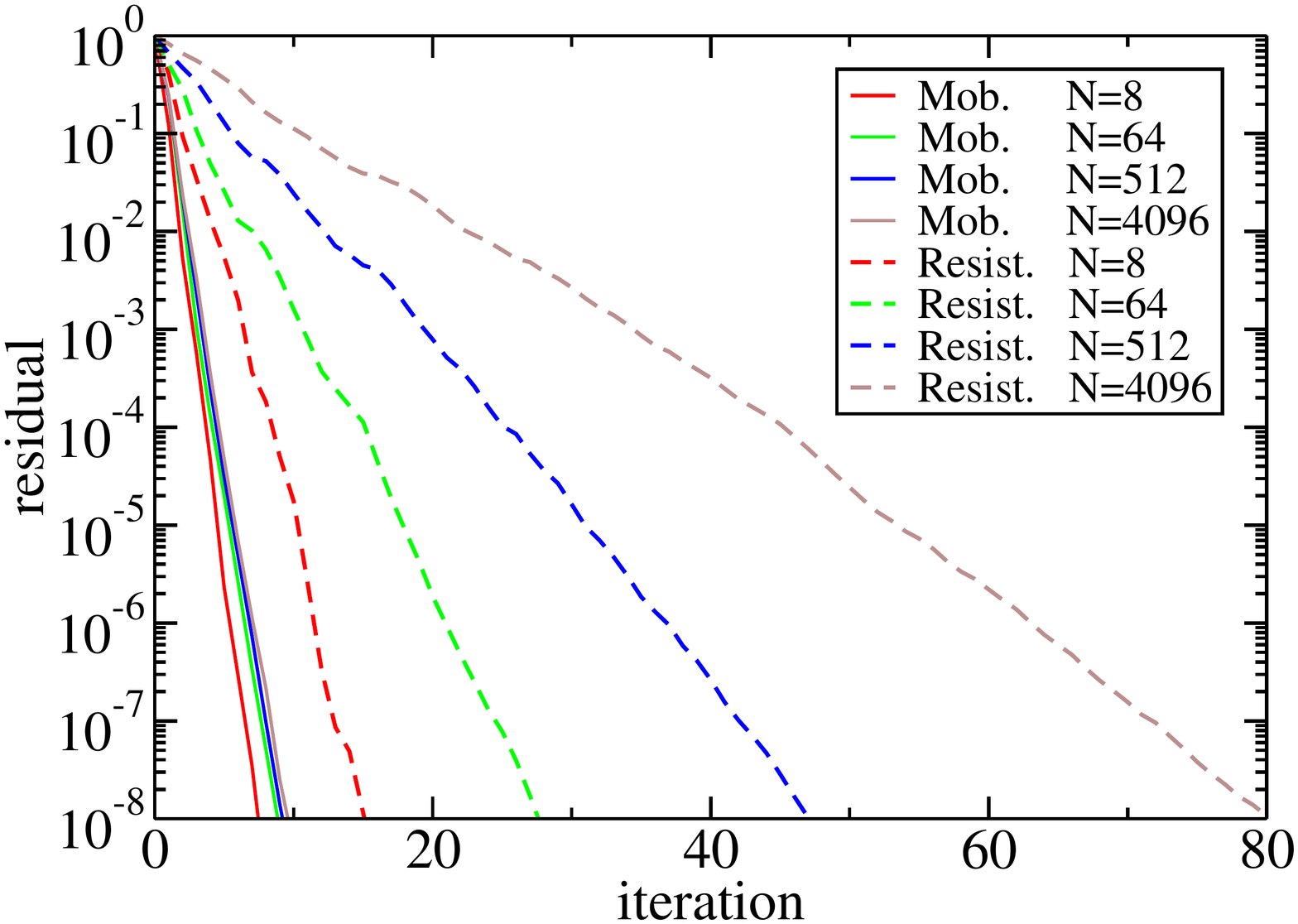}\includegraphics[width=0.49\textwidth]{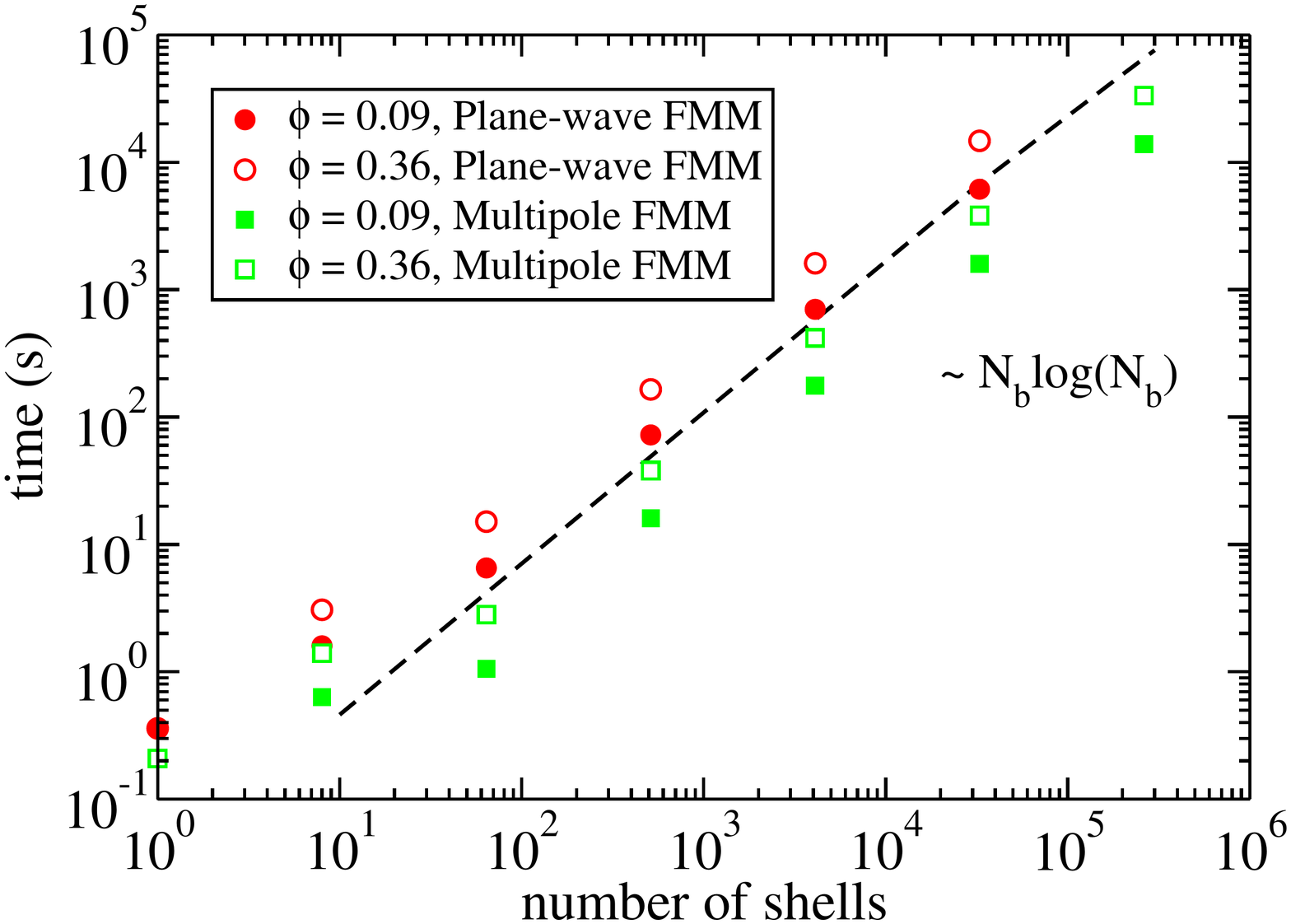}\caption{\label{fig:Convergence-FMM} (Left panel) Convergence of preconditioned
GMRES for the resistance and mobility problems for a finite subset
of a cubic lattice of 42-blob spheres in an unbounded domain, for
different number of particles, keeping the lattice spacing (closest
distance between spheres) at $4R_{g}\sim3.6R_{h}$ (corresponding
to $\phi=0.09$). (Right panel) Wall-clock time to solve the mobility
problem with a tolerance $10^{-8}$ versus the number of spherical
shells for two volume fractions $\phi$, demonstrating the $\mathcal{O}(N_{b}\log(N_{b}))$
asymptotic complexity, where $N_{b}$ is the number of blobs. The
matrix-vector product was computed with relative tolerance $0.5\cdot10^{-9}$
using the FMM method developed in Ref. \citet{RPY_FMM}. We compare
the performance of a parallel multipole-based FMM3DLIB code \citet{OseenBlake_FMM}
with a more efficient but serial plane-wave FMM currently under development
in the group of Leslie Greengard. Parallel runs used 8 cores and serial
runs used a single core of an Intel Xeon 2.40GHz processor.}
\end{figure*}

First, we test the robustness with increasing system size, keeping
the particles well-separated at a distance $4R_{g}\approx3.6R_{h}$,
which corresponds to volume fraction $\phi\approx0.09$. The left
panel of Fig. \ref{fig:Convergence-FMM} shows that the convergence
is uniform and that the number of iterations to reduce the residual
by a given factor depends very weakly on the number of spheres. This
demonstrates the effectiveness of the block-diagonal preconditioner
for the mobility problem. Next we investigate the robustness with
respect to packing density. The right half of Table \ref{tab:Stresslets_3D}
shows the number of iterations to convergence for spheres arranged
in a cubic simple lattice for several packing densities. When particles
are far apart the solver converges fast because the hydrodynamic interactions
between particles are weak and the preconditioner is designed to be
an exact solver for a single body. As the spheres come closer together
the preconditioner is not so effective and the Krylov solver needs
to perform more iterations, as expected. However, even when the particles
are relatively close the solver performs reasonably well. For example,
when the spheres are at a distance of $2.55R_{g}\approx2.3R_{h}$
($\phi\approx0.36$), the solver converges in 23 iterations. Of course,
as the particles come closer and closer, and in particular, as the
blobs on disjoint spheres begin to overlap, we expect to see an increasing
ill-conditioning of the linear system (\ref{eq:saddle_M}) and an
increasing number of iterations. However, the rigid multiblob method
should not be used in this regime, since it does not accurately resolve
lubrication forces at such short distances. In the right panel of
Fig. \ref{fig:Convergence-FMM} we show the wall-clock time to solve
the mobility problem for different number of shells and volume fractions.
Since the number of iterations is essentially independent of the system
size we obtain a quasi-linear scaling by using the FMM to compute
the product between the blob-blob mobility matrix and a vector.

However, for the resistance problem explained in Sec. \ref{sec:RigidMultiblobs},
the left panel of Fig. \ref{fig:Convergence-FMM} shows that the number
of iterations to attain convergence increase with the number of particles
as $N^{1/3}$, i.e., with the linear extent of the system. This is
somewhat better than the $O\left(N^{1/2}\right)$ iterations reported
for Stokesian dynamics by Ichiki \citet{libStokes}, but still much
worse than the mobility problem.

We believe that this difference between resistance and mobility problems
is physical rather than purely numerical. In particular, we expect
that the same behavior will be observed in essentially any other iterative
method, regardless of the specifics of the discretization of the problem
(rigid multiblobs, boundary-integral methods, multipole expansions,
regularized Stokeslets, etc.). To appreciate the difference between
mobility and resistance problems, observe that it is possible to obtain
a low accurate solution to the mobility problem by approximating each
sphere by a single blob and then computing the matrix vector product
$\Mob\V f=\V u$ using an FMM. On the other hand, to solve the resistance
problem the linear system $\Mob\V f=\V u$ has to be solved, which
must account for the collective nature of hydrodynamic interactions.
The difference appears because there is an effective far-field two-body
approximation for the mobility $\BMob$ (equivalently for $\Mob$)
but not for the resistance matrix $\BMob^{-1}$ (or $\Mob^{-1}$),
which is essentially a multibody problem \citet{RPY_Brinkman}.

Mathematically, the difference appears because solving the saddle
point problem (\ref{eq:saddle_M}) is similar to computing the motion
for force- and torque-free particles \citet{IndefinitePreconditioning},
even though forces and torques are applied on the particles. For force-
and torque-free particles, the hydrodynamic fields and thus interactions
with other particles decay faster than $1/r$. Therefore, the effective
interactions that need to be captured by the iterative solver decay
much faster for the mobility problem than for the resistance problem,
making the former much easier. To confirm this intuition, we have
studied (not shown) mixed resistance/mobility problems. When we fix
the angular velocities but leave the linear velocities as free, we
expect to see rapid convergence because the leading order interactions
that the Krylov method needs to capture decay as $1/r^{3}$. Indeed,
we observe numerically that in this case the solver converges almost
as well as for the pure mobility problem. However, when we fix the
linear velocities of the spheres but let them freely rotate, we find
that the solver converges almost as bad as for the pure resistance
problem.

\section{\label{sec:ResultsWall}Results: Single Wall}

In this section we study the accuracy of rigid multiblob models and
the effectiveness of our block diagonal preconditioner for particle
suspensions sedimented near a single no-slip boundary. This is an
important and common occurrence in practice, especially in the field
of active matter, since many active particles have metallic components
and are not density-matched with the solvent, and thus sediment to
the bottom substrate. Some of us studied the diffusive dynamics of
nonspherical particles near a no-slip boundary using a rigid multiblob
approach in Ref. \citet{BrownianMultiBlobs}. However, in that prior
work, we only studied a single body and therefore all of the mobility
matrices were simply formed as dense matrices. Here we explore in
more detail the accuracy of rigid multiblob models and also demonstrate
how to scale rigid multiblob computations to suspensions of thousands
of rigid bodies.

To compute the hydrodynamic interactions between blobs in the presence
of a single wall we use a pairwise approximation to the blobs' mobility
which includes the effects of the wall in the Rotne-Prager tensor
\citet{StokesianDynamics_Wall}. In our implementation, we compute
the product of the blob-blob mobility matrix $\M{\mathcal{M}}$ with
a vector using a direct $\mathcal{O}(N_{b}^{2})$ summation (here
$N_{b}$ is the number of blobs) implemented on a GPU using PyCUDA
\citet{PyCUDA} in double precision; single precision can be used
for lower accuracy requirements. This is an ideal problem for using
GPUs as an accelerator since the computation is trivially parallelized
on shared memory. Furthermore, the communication requirements between
the CPU and GPU are minimal, since only the positions of the blobs
need to be communicated %
\footnote{For suspensions of identical bodies only the positions and orientations
of the rigid bodies (so only up to 7 numbers per body) need to be
communicated to the GPU.%
}. It is possible to implement a fast multipole method (FMM) for the
RP(Y) tensor including wall corrections by using a system of images
together with an FMM for unbounded domains \citet{RPY_FMM,OseenBlake_FMM}.
However, it is important to note that the asymptotically-optimal FMMs
on a CPU (even with multicore acceleration) will only be computationally
more efficient than a direct sum on a GPU for more than about one
hundred thousand blobs (in our testing on current hardware). Therefore,
for many applications a simple GPU implementation is sufficient or
even preferable over an asymptotically scalable implementation. Our
PyCUDA codes are publicly available at \url{https://github.com/stochasticHydroTools/RigidMultiblobsWall}.
Once a Rotne-Prager regularization of the construction of Gimbutas
\emph{et al}. \citet{OseenBlake_FMM} is developed and combined with
an FMM, the asymptotic cost will be reduced to $\mathcal{O}(N_{b}\log N_{b})$
and our computations can be extended to millions of blobs.

In section \ref{sub:SphereWall} we study the accuracy of rigid multiblob
models for modeling a sphere close to a boundary, and in Section \ref{sub:CylinderWall}
we extend this study to a rigid cylinder (rod). In section \ref{sub:ActivePair}
we study the dynamics of a pair of active rods close to a no slip
boundary. In section \ref{sub:ConvergenceWall} we study the performance
of our iterative solver on a suspension of many rods, and demonstrate
that the number of GMRES iterations is essentially independent of
the number of particles just as for suspensions in an unbounded domain
(see Section \ref{sub:ConvergenceFMM}).

\subsection{\label{sub:SphereWall}Sphere}

In this section the mobility $\M{\mu}\equiv\M{\mathcal{N}}$ of a
rigid multiblob sphere whose center is at a distance $H$ from a no-slip
boundary is compared with some theoretical results available in the
literature. We use the shell models of spheres described in Sec. \ref{sec:ResultsUnbounded},
and show the mean and standard deviation of the computed mobility
averaged over a large number of random orientations of the rigid multiblob
relative to the boundary. To denote the specific component of the
mobility matrix we use a subscript $tt$ for translational mobility,
$rr$ for rotational mobility and $tr$ for translation-rotation coupling
mobility, and we use a superscript $\perp$ or $\parallel$ to denote
whether the direction of the force, torque, velocity or angular velocity
is perpendicular or parallel to the wall.

\begin{figure*}[tbph]
\includegraphics[width=0.49\textwidth]{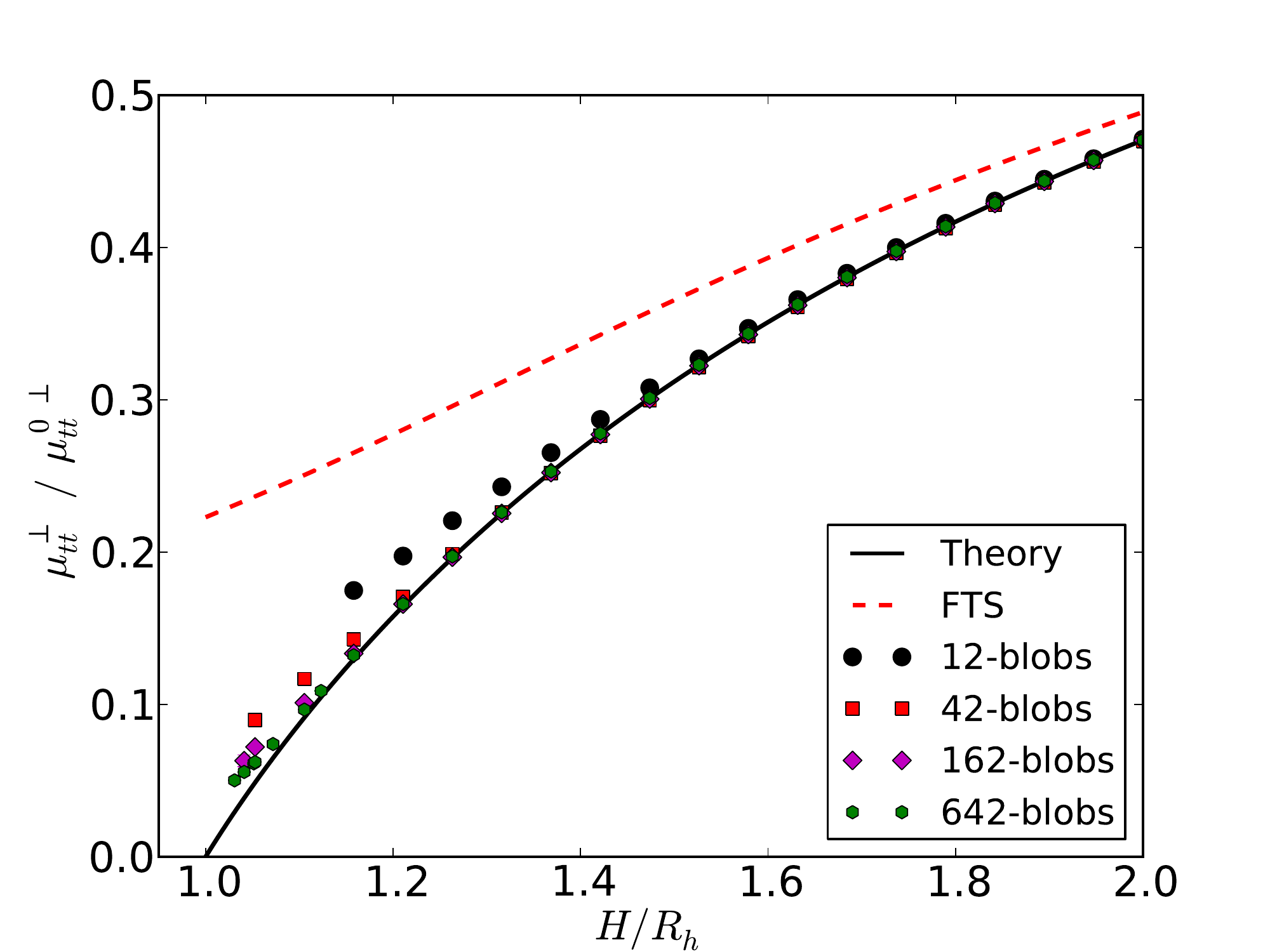}\includegraphics[width=0.49\textwidth]{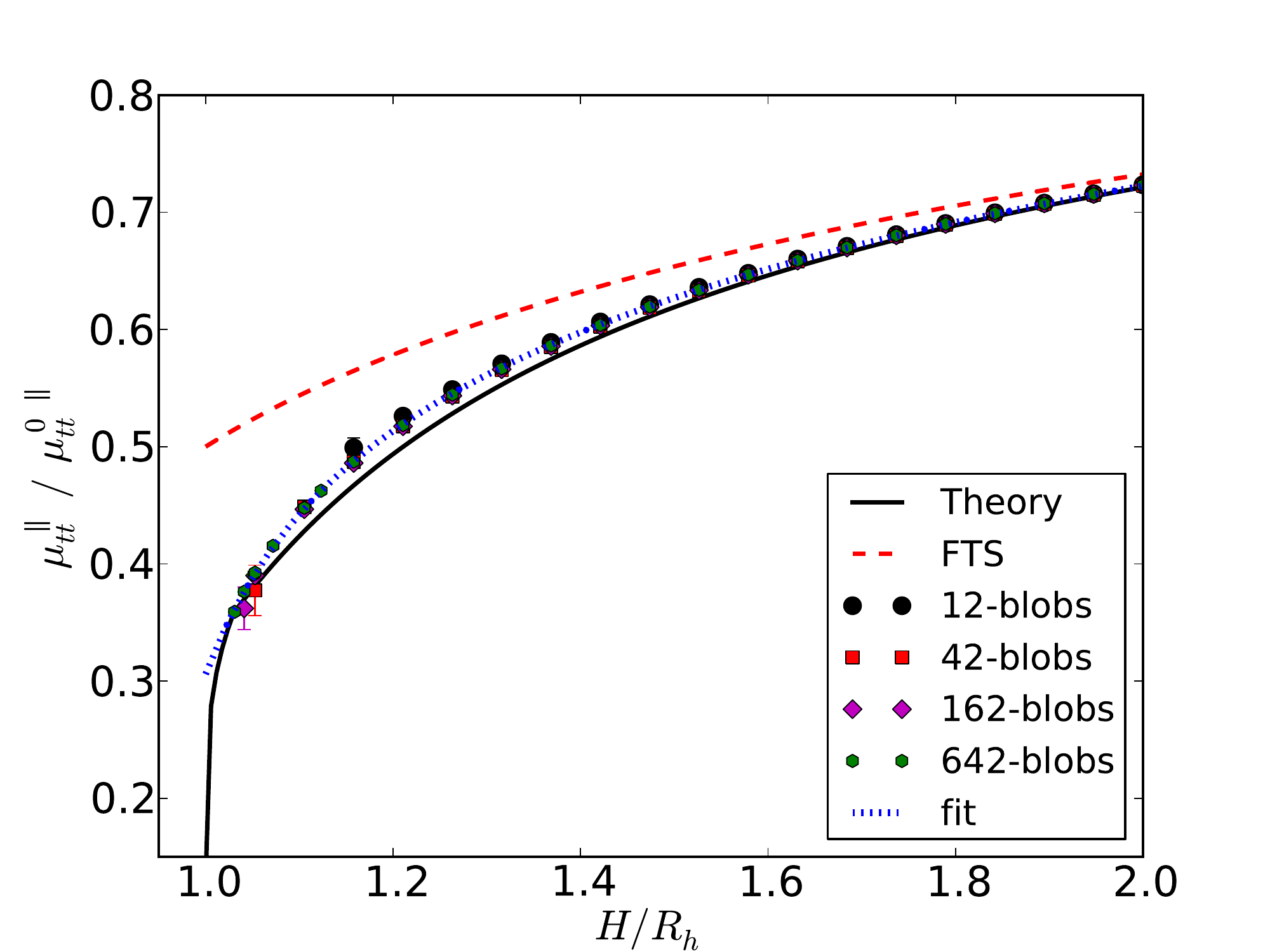}

\includegraphics[width=0.49\textwidth]{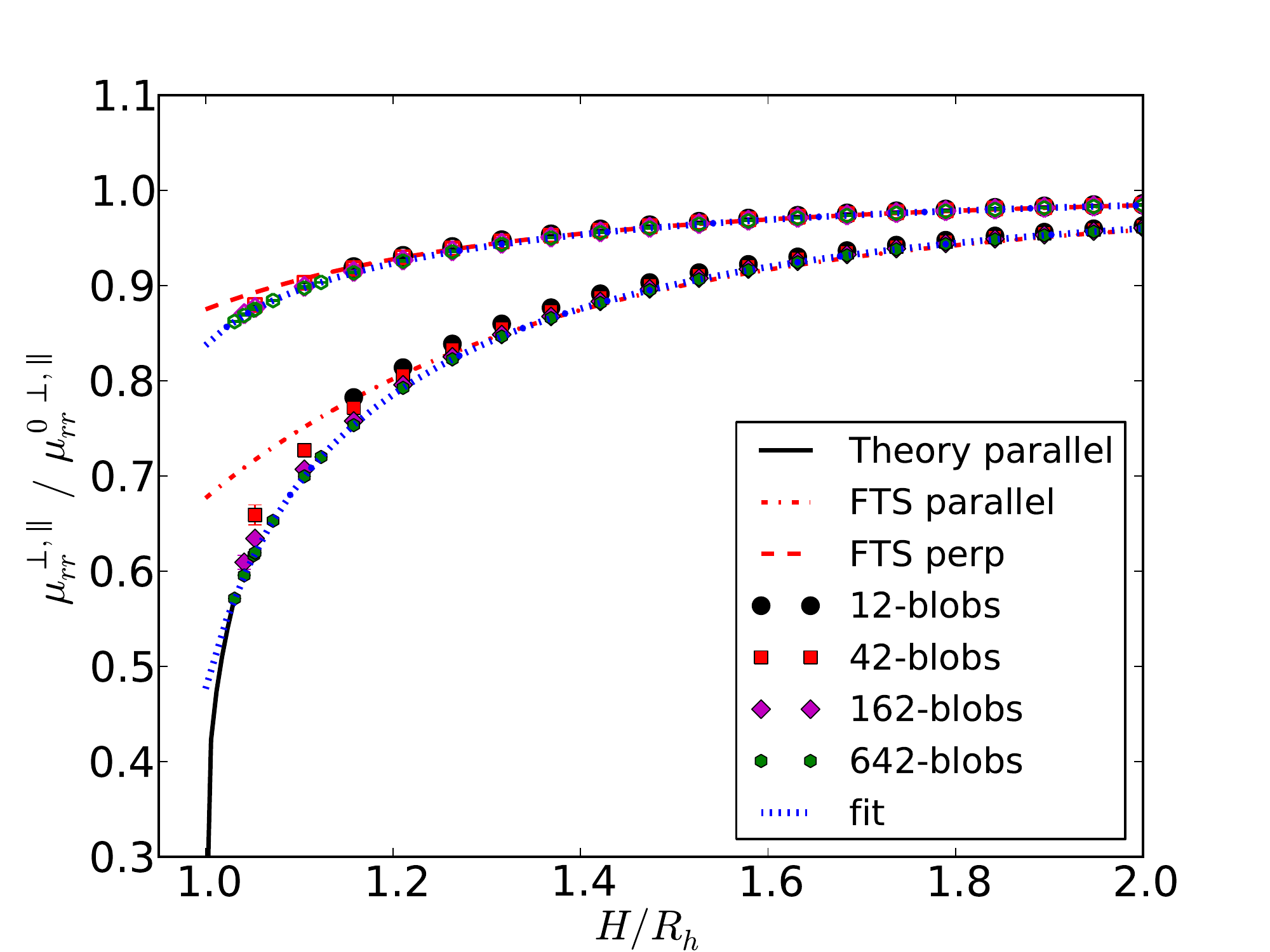}\includegraphics[width=0.49\textwidth]{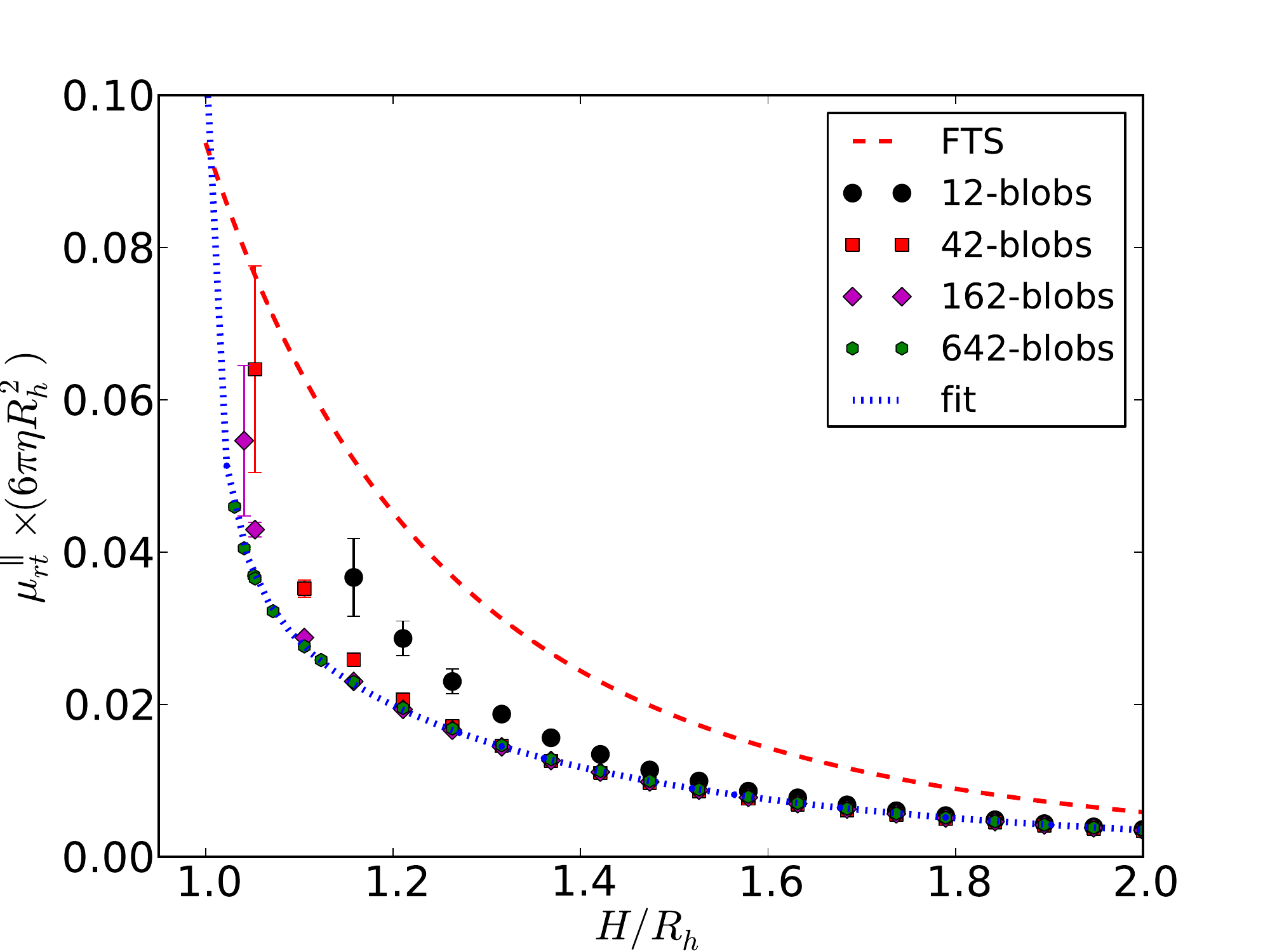}

\caption{\label{fig:Sphere-wall}Selected components of the mobility $\M{\mu}$
matrix for a sphere close to a wall, normalized by the corresponding
bulk (unbounded domain) mobility $\M{\mu}^{0}$ where possible. Comparison
of the rigid multiblob results (symbols) is made to the best available
theoretical results (solid black lines) and the FTS approximation
used in Stokesian dynamics \citet{StokesianDynamics_Wall} (dashed
and dashed-dotted lines). Empirical fits listed in Appendix \ref{app:FitsSphereWall}
are shown with a dotted line. (Top left) Translational mobility $\mu_{tt}^{\perp}$
for a force applied perpendicular to the wall. (Top right) Translational
mobility $\mu_{tt}^{\parallel}$ for a force parallel to the wall.
(Bottom left) Rotational mobility for a torque applied parallel ($\mu_{rr}^{\parallel}$,
filled symbols) or perpendicular ($\mu_{rr}^{\perp}$, empty symbols)
to the wall. (Bottom right) Rotation-translation coupling mobility
$\mu_{tr}^{\parallel}$ for force or torque parallel to the wall.}
\end{figure*}

The top left panel of Fig. \ref{fig:Sphere-wall} presents the translational
mobility of the sphere perpendicular to the wall together with the
exact theory obtained by Brenner \citet{Brenner1961} (see also Eq.
(D2) in \citet{BrownianMultiBlobs} for a simple but accurate approximation).
We also compare to the complete expression for the Rotne-Prager-Blake
tensor derived by Swan and Brady \citet{StokesianDynamics_Wall},
including stresslet corrections, which corresponds to an FTS truncation
(plus degenerate quadrupole corrections) of the multipole hierarchy.
It is evident that a single-blob model of a sphere, just like the
substantially more complicated FTS truncation, does not recover the
strong drop in the mobility (i.e., lubrication) at small distances
to the wall. Rigid multiblob models do substantially better than the
FTS truncation even with only 12 blobs (icosahedral multiblob), and,
as expected, the accuracy is improved with the addition of more blobs.
As in the Sec.\ref{sec:ResultsUnbounded}, the numerical mobility
never goes exactly to zero since we do not add lubrication corrections,
and we expect the rigid multiblob model to only work well when the
blobs do not overlap the boundary itself. In fact, we recall here
that the RP tensor we use here \citet{StokesianDynamics_Wall} does
\emph{not} include near-field corrections when blobs overlap the wall,
and therefore repulsive forces or other mechanisms should be used
to ensure that the rigid multiblob is sufficiently far from the boundary.
We empirically observe that the rotational invariance gets violated
strongly if the gap to the wall is less than $2a/3$, which corresponds
to 7\% of the sphere radius for 12 blobs, and about 2\% of the radius
for the 642 blob model.

In the remaining panels of Fig. \ref{fig:Sphere-wall} we investigate
other components of the mobility. There are no closed-form expressions
(even as infinite sums) for these components that are valid for all
distances to the wall, so we use the best approximations available,
see Appendix D in \citet{BrownianMultiBlobs} for specific formulas.
For the translational mobility parallel to the wall, shown in the
top right panel of the figure, we use a result based on lubrication
theory \citet{NearWallSphereMobility} (see Eq. (D3) in \citet{BrownianMultiBlobs})
when the sphere is very close to the wall ($H<1.03R_{h}$), and an
approximation to order $\mathcal{O}\left((H/R_{h})^{5}\right)$ for
larger distances \citet{ConfinedSphere_Sedimented,BrennerBook} (see
Eq. (D4) in \citet{BrownianMultiBlobs}). It is clear that the rigid
multiblob matches the theory for large distances but that the approximate
theory is not very accurate for $H\lesssim1.5\, R_{h}$ since the
rigid multiblob results are clearly converging to something slightly
different. As for the perpendicular mobility, we see that the icosahedral
model (12 blobs) is substantially more accurate than an FTS truncation. 

Our results for the rotational mobilities, for torque applied either
perpendicular or parallel to the wall, are shown in the bottom left
panel of the figure and agree with the FTS results at large distances.
We see slow but clear convergence of the rigid multiblob results for
the translation-rotation coupling, shown in the bottom right panel
of the figure. This component of the mobility is therefore most difficult
to capture accurately, as is evident from the fact that the FTS truncation
does pretty poorly in this case. Since neither the FTS truncation
nor the asymptotic lubrication results \citet{NearWallSphereMobility}
are sufficiently accurate for comparison to experimental measurements,
we have empirically fitted our highest-resolution results for the
mobilities for which there are no exact theoretical expressions. We
show the fits in Fig. \ref{fig:Sphere-wall} and give details about
the fits in Appendix \ref{app:FitsSphereWall} for the benefit of
other researchers.

\subsection{\label{sub:CylinderWall}Cylinder}

In this section, we consider a cylinder (rod) of length $L=2.12$
and diameter $D=2R=0.325$ of aspect ratio $\alpha=L/D\approx6.35$,
mimicking the metallic rods studied in recent experiments \citet{TripleNanorods_Megan},
for three different levels of resolution. The minimal resolution rigid
multiblob has blobs placed in a row along the axis of the cylinder
(a total of 14 blobs), while in the more resolved models, a hexagon
(86 blobs) or a dodecagon (324 blobs) of blobs is placed along the
circumference of the cylinder to better resolve it, as illustrated
in Fig. \ref{fig:BlobModels}. We study different components of the
mobility matrix $\M{\mu}\equiv\M{\mathcal{N}}$; to specify the direction
of the force, torque, velocity or angular velocity we use a subscript
$\perp$ or $\parallel$ to denote whether the direction is perpendicular
or parallel to the axes of the cylinder, respectively, and a superscript
to denote whether the direction is perpendicular or parallel to the
wall.

\subsubsection{Bulk mobility}

The first question that must be answered when constructing a rigid
multiblob model of a given body is where to place the blobs and how
to choose their hydrodynamic radius, to match the effective hydrodynamic
response of the actual rigid body. Here we generalize the approach
taken in Section \ref{sub:ConvergenceMoments} for spheres to a nonspherical
body and show how to match the (passive) mobility of an actual rigid
cylinder with a rigid multiblob model. Based on the results for spheres,
for resolutions other than the minimally-resolved model we cover the
surface (similarly for the ends of the cylinder) of a cylinder uniformly
with spheres keeping the spacing between blobs both around the circumference
and the length of the cylinder uniform and fixed at $a/s=0.5$.

Because there are no exact analytical results for the mobility of
a cylinder even in an unbounded domain, we estimate the true mobility
of the cylinder $\M{\mu}^{0}$ in an unbounded domain numerically.
Specifically, we place the blobs on the surface of a cylinder of length
$L$ and radius $R$ (i.e., the true geometric surface of the actual
rod we are modeling), and numerically compute the mobility for different
resolutions. We then extrapolate to the limit $a\rightarrow0$ using
the two finest resolutions (86 and 324 blobs) based on our knowledge
that the error is linear in $a$. For the translation-translation
mobility we obtain $\left(\mu_{tt}^{0}\right)_{\parallel}\times4\pi\eta L=3.404$
(compare to $3.295$ from slender-body theory \citet{SlenderBody_Cylinder})
and $\left(\mu_{tt}^{0}\right)_{\perp}\times4\pi\eta L=2.606$ (compare
to $2.619$ from slender-body theory \citet{SlenderBody_Cylinder}),
while for the rotation-rotation mobility we get $\left(\mu_{rr}^{0}\right)_{\perp}\times\pi\eta L^{3}/3=1.212$
(compare to $1.211$ from slender-body theory \citet{IBM_Sphere})
and $\left(\mu_{rr}^{0}\right)_{\parallel}\times\pi\eta L^{3}/3=12.678$,
see the last row in Table \ref{tab:Matched-mobility-coefficients}.

Our goal here is to match the bulk mobility $\M{\mu}^{0}$ of our
rigid multiblob models to that of a true cylinder as close as possible.
To do this, for the surface-resolved models (86 and 324 blobs), we
place the blobs on the surface of a cylinder of the \emph{same} aspect
ratio $\alpha=6.35$ but with the geometric radius $R_{g}$ of this
cylinder allowed to be smaller than the geometric radius of the actual
particle, while keeping the blob spacing $a/s=0.5$. We then numerically
optimize the value of $R_{g}$ to minimize a measure of the error
with respect to (extrapolated) mobility of a true cylinder, to obtain
$R_{g}/R=0.90$ for the 86-blob model and $R_{g}/R=0.95$ for the
324-blob model. For the minimally-resolved model, we empirically tune
both the geometric length (i.e., the distance between the centers
of the two furthest blobs) to $L_{g}=0.914L$ and the blob radius
to $a=1.103R$ while keeping the number of blobs fixed at $N_{b}=L/R+1=14$
as suggested by Bringley and Peskin \citet{IBM_Sphere}. Table \ref{tab:Matched-mobility-coefficients}
shows the resulting infinite-domain mobilities for each resolution
along with the relative error compared to the extrapolated values
for infinite resolution. We see a relative error always less than
$2.5\%$ even for the minimally-resolved model, except for rotation
of the cylinder around its own axis; recall that the minimally-resolved
model cannot support a torque around the axis of the cylinder.

\begin{table}[tbph]
\centering{}%
\begin{tabular}{|c|c|c|c|c|}
\hline 
Resolution &
$\left(\mu_{tt}^{0}\right)_{\parallel}\times4\pi\eta L$ &
$\left(\mu_{tt}^{0}\right)_{\perp}\times4\pi\eta L$ &
$\left(\mu_{rr}^{0}\right)_{\perp}\times\pi\eta L^{3}/3$ &
$\left(\mu_{rr}^{0}\right)_{\parallel}\times\pi\eta L^{3}/3$\tabularnewline
\hline 
\hline 
14 blobs &
$3.422\,(-0.53\%)$ &
$2.612\,(-0.23\%)$ &
$1.216\,(-0.31\%)$ &
NA\tabularnewline
\hline 
86 blobs &
$3.324\,(2.35\%)$ &
$2.541\,(2.48\%)$ &
$1.240\,(-2.34\%)$ &
$11.564\,(8.79\%)$\tabularnewline
\hline 
324 blobs &
$3.360\,(1.29\%)$ &
$2.588\,(0.67\%)$ &
$1.225\,(-1.06\%)$ &
$12.274\,(3.19\%)$\tabularnewline
\hline 
$\infty$ (extrapolated) &
$3.4040$ &
$2.6061$ &
$1.212$ &
$12.678$\tabularnewline
\hline 
\end{tabular}\caption{\label{tab:Matched-mobility-coefficients}Nontrivial elements of the
bulk mobility matrix for empirically-optimized rigid multiblob models
of a cylinder of aspect ratio $6.35$, shown in the three left panels
of Fig. \ref{fig:BlobModels}. The value in the limit of infinite
resolution is extrapolated numerically (see main text) and reported
in the last row. The percentages in parenthesis correspond to the
error relative to the infinite resolution estimates.}
\end{table}

\subsubsection{Mobility for a sedimented rod}

Having determined the geometric parameters for the rigid multiblob
models based on motion in an unbounded domain, we now study the accuracy
of the three different resolutions for a cylinder close to a no-slip
boundary. We assume that the cylinder is parallel to the wall with
the centerline of the rod at a distance $H$ from the no-slip boundary.

\begin{figure*}[tbph]
\begin{centering}
\includegraphics[width=0.5\textwidth]{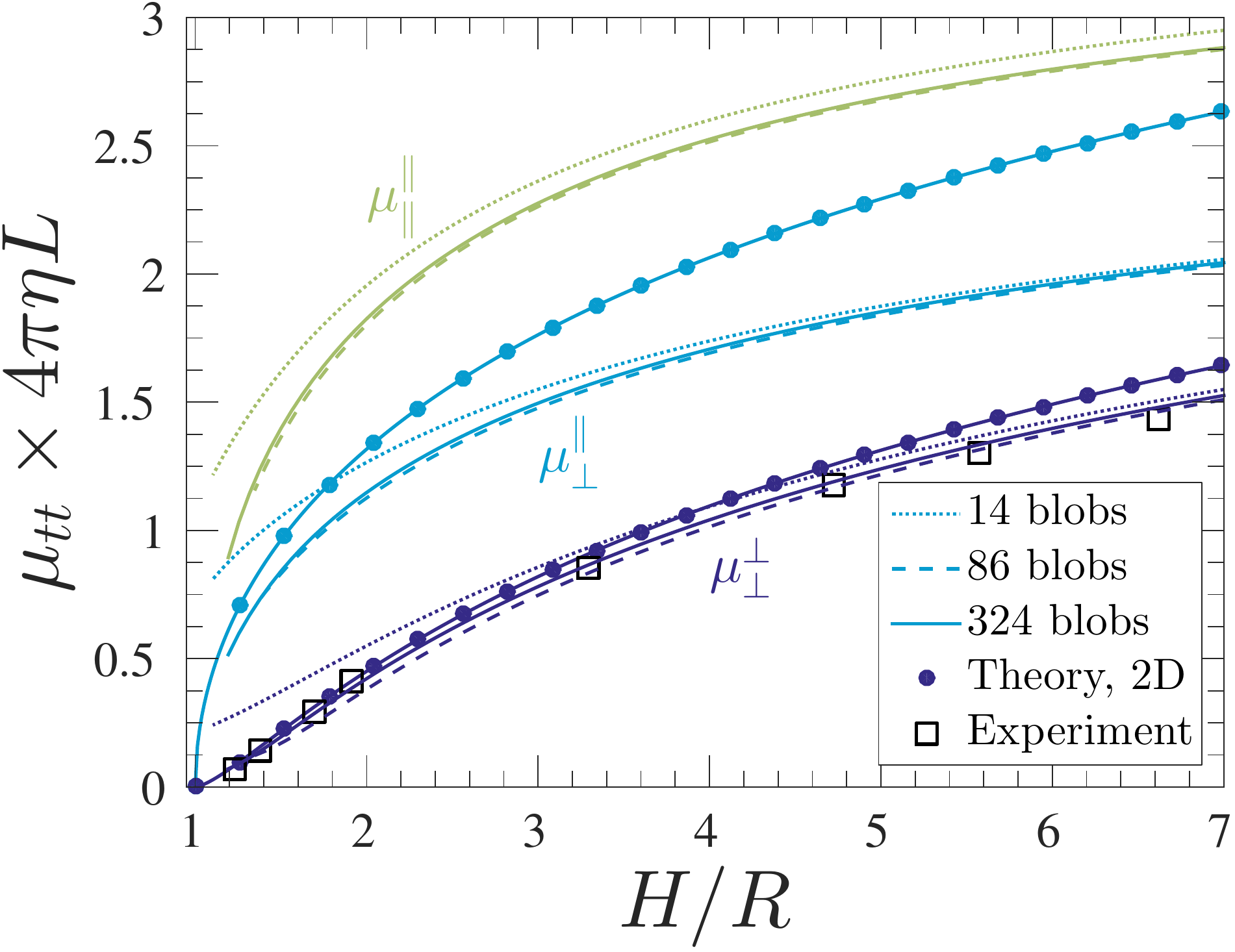}\includegraphics[width=0.5\textwidth]{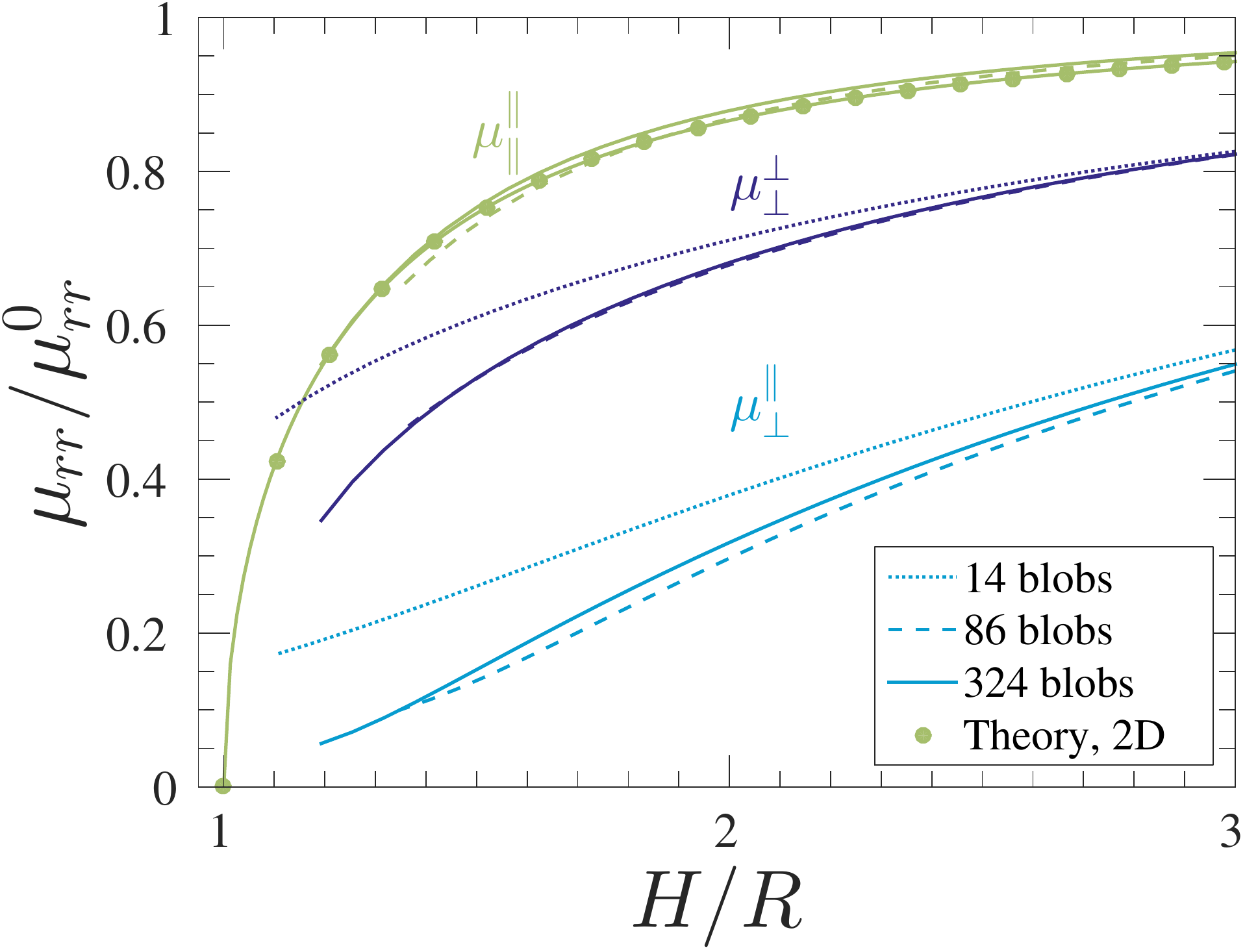}
\par\end{centering}

\centering{}\caption{\label{fig:Mobility-coefficients-Wall}Mobility coefficients for a
cylinder of aspect ratio $\alpha=L/\left(2R\right)=6.35$ when it
is parallel to the wall, as a function of the height of the rod centerline
$H/R$. The superscript of the mobility denotes the direction with
respect to the wall, while the subscript denotes the direction with
respect to the rod axis. (Left panel) Translational mobility coefficients
$\V{\mu}_{tt}$ normalized by $4\pi\eta L$ as in \citet{CylinderWall_Experiments}.
The curves with circles corresponds to the formulas from Jeffrey and
Onishi for an infinite cylinder near a wall \citet{InfiniteCylinderWall}.
The black squares correspond to the experimental measurements of Trahan
and Hussey for a rods with aspect ratio $\alpha=5.05$\textbf{ }\citet{CylinderWall_Experiments}.
(Right panel) Rotational mobility coefficients $\boldsymbol{\mu}_{rr}$
normalized by the corresponding bulk value. The curve with circles
corresponds to an infinite cylinder near a wall \citet{InfiniteCylinderWall}.}
\end{figure*}

Figure \ref{fig:Mobility-coefficients-Wall} compares the computed
mobility coefficients to available theoretical and experimental results.
As could be expected, the decrease in mobility when approaching the
boundary is clearly underestimated with the minimally-resolved model.
The left panel of the figure shows the translational mobilities. For
$\mu_{\perp}^{\perp}$, the higher resolutions are in good agreement
with the experimental measurements of Trahan and Hussey for a sedimenting
rod with aspect ratio $\alpha=5.05$ \citet{CylinderWall_Experiments}.
Our numerical results also match well the theory of Jeffrey and Onishi
\citet{InfiniteCylinderWall} for an infinite cylinder when $H<2R$
. It is important to emphasize that our model is significantly more
accurate than slender-body theory near boundaries; the slender-body
theory results from\textbf{ }\citet{SlenderCylinderWall_2,SlenderCylinderWall_1,SlenderBody_Cylinder}
(not shown here) are reasonably accurate only when $H/R>3.5$ for
aspect ratios $\alpha>9$ \citet{CylinderWall_Experiments}. The rotational
mobilities of the rod are shown in the right panel of Fig. \ref{fig:Mobility-coefficients-Wall}.
For the rotational mobility $\mu_{\parallel}^{\parallel}$, all three
resolutions are in good agreement with the theory of Jeffrey and Onishi
\citet{InfiniteCylinderWall} for an infinite cylinder. For $\mu_{\perp}^{\parallel}$
and $\mu_{\perp}^{\perp}$, the minimally-resolved model shows substantial
errors near the wall, but the two higher resolution models agree with
each other quite well over a broad range of distances.

\subsection{\label{sub:ActivePair}Active Rod Pair}

In this section we apply the rigid multiblob method to a problem of
recent experimental and theoretical interest: the dynamics of a pair
of active ``nanorods'' that exhibit a ``pusher'' or extensile
dipolar flow at large distances. Specifically, we compute the motion
of a dimer of tripartite nanorods, as studied in recent experiments
by Wykes \emph{et al.} \citet{TripleNanorods_Megan}. The rods have
diameter of 0.325$\mu\mbox{m}$ and length of 2.12$\mu\mbox{m}$,
and are in force and torque equilibrium (under the action of gravity
and Van der Waals and electrostatic interaction forces with the boundary)
at some distance from the wall that has not been measured in the experiments.
The rods are constructed of three metal sections, in the arrangement
gold-platinum-gold (Au-Pt-Au) and create a dipolar extensile (pusher)
far-field flow. As such, they do not propel themselves in isolation
but experiments show the formation of dimers of rods that actively
rotate in a direction that is opposite of that predicted by recent
simulations \citet{ActiveFilaments_Adhikari}, which neglect the presence
of the bottom wall. In agreement with the experiments, our simulations
show the formation of a stable rod pair that touch each other tangentially
and rotate (without exhibiting a significant translation) around an
axis perpendicular to the wall in a direction consistent with the
experimental measurements.

The exact details of the active flows near the surface of the rods
have not been measured experimentally and are difficult to predict
analytically because this requires resolving the thin slip boundary
layer (of thickness related to the Debye length) around the rods,
as well as the knowledge of a number of material constants that are
not known accurately. To obtain a qualitative understanding of the
dynamics of the rods we impose an apparent tangential surface slip
velocity on the two gold sections, directed away from platinum center
and having a magnitude of 20$\mu\mbox{m}/\mbox{s}$; no slip is imposed
on the platinum section. Note that both gravity and the active slip
pull the rods toward the wall, so we use an \emph{ad hoc} repulsive
force with the wall to balance the distance between the cylinder centerline
and the wall at one cylinder diameter. Due to electrostatic interactions,
a stacking of the two rods with the gold end of one rod aligned with
the platinum center of the other is observed experimentally \citet{TripleNanorods_Megan};
here we study the flow around such a pair of aligned rods.

\begin{figure*}[tbph]
\begin{centering}
\includegraphics[width=0.33\textwidth]{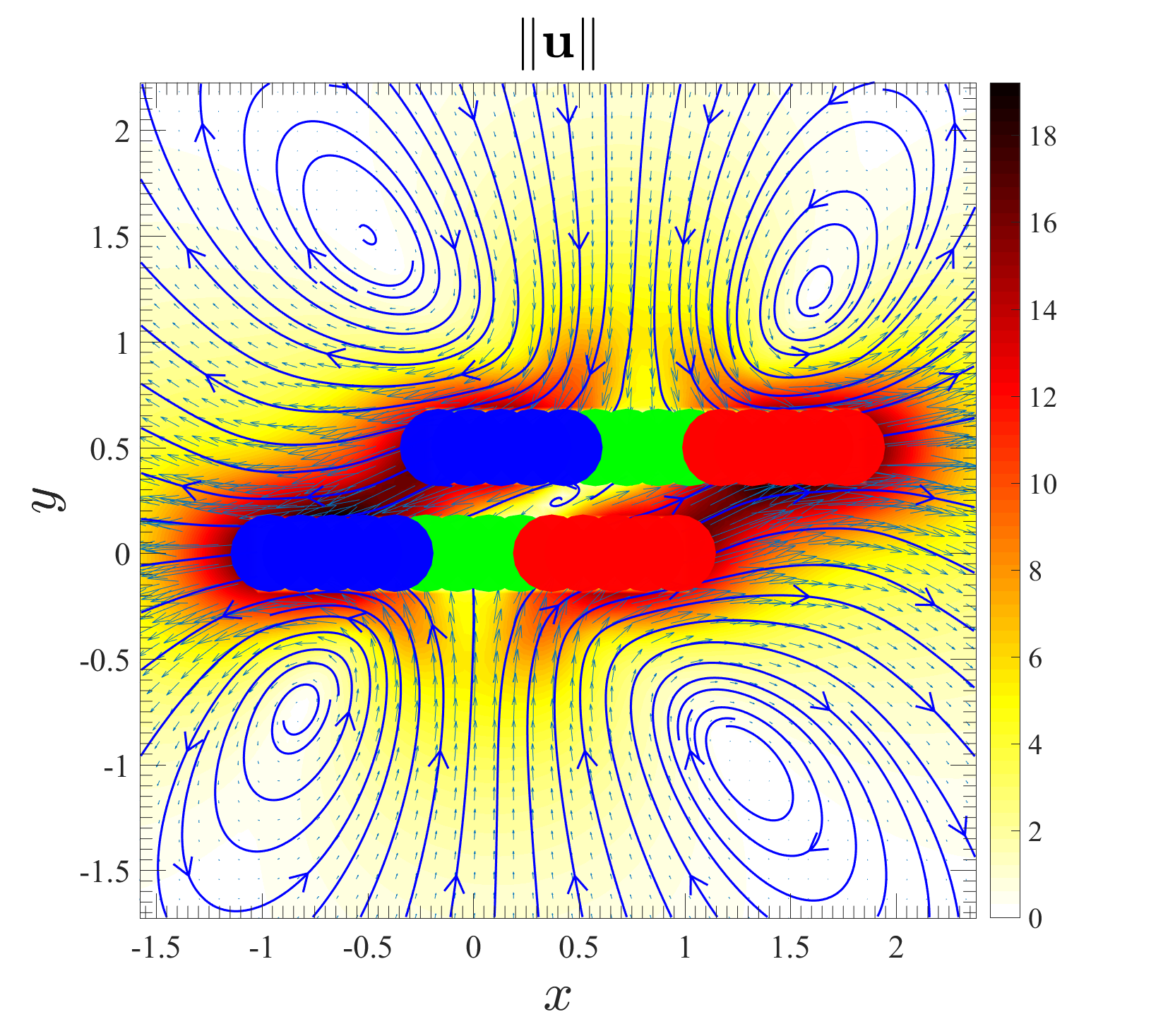}\includegraphics[width=0.33\textwidth]{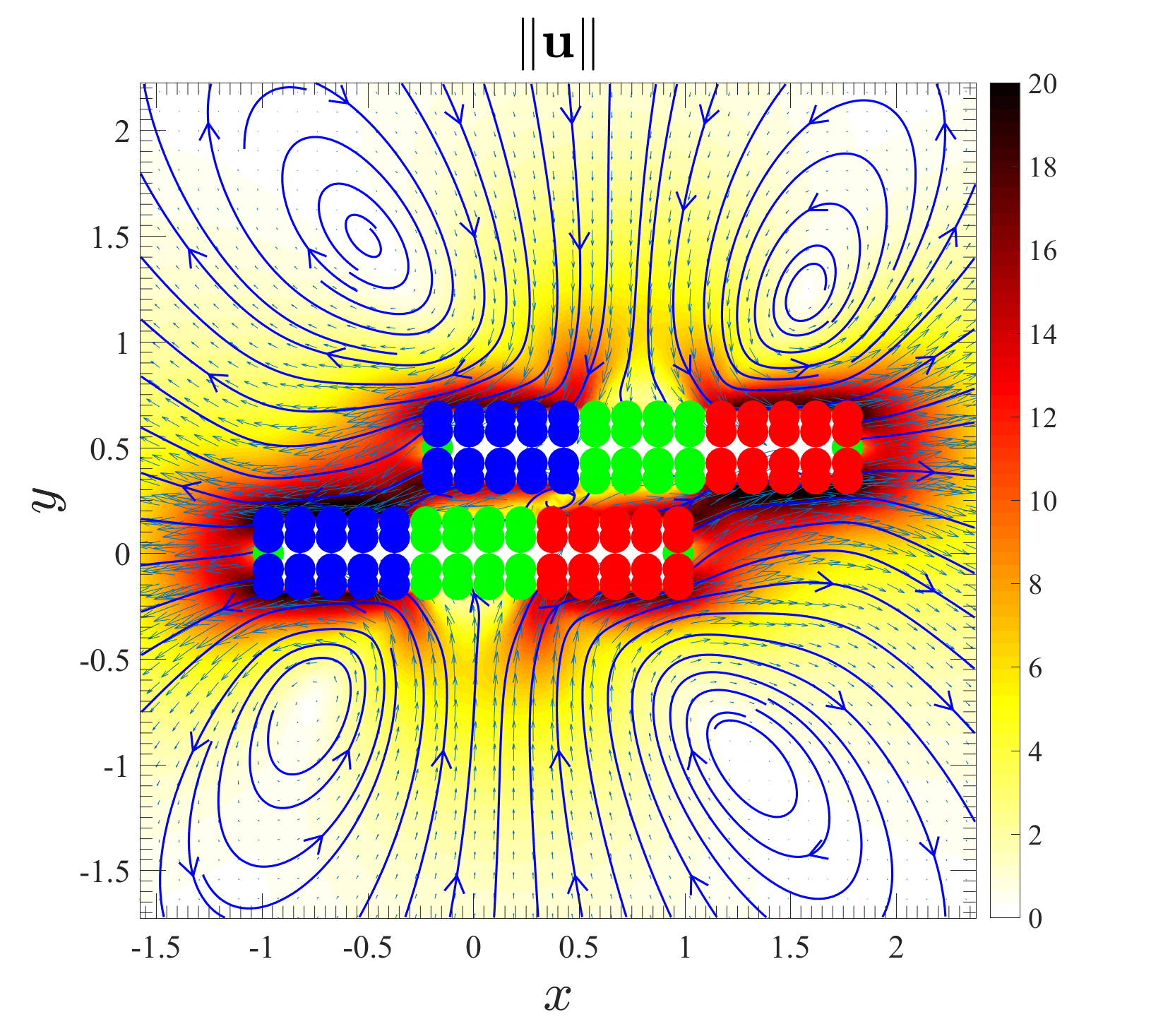}\includegraphics[width=0.33\textwidth]{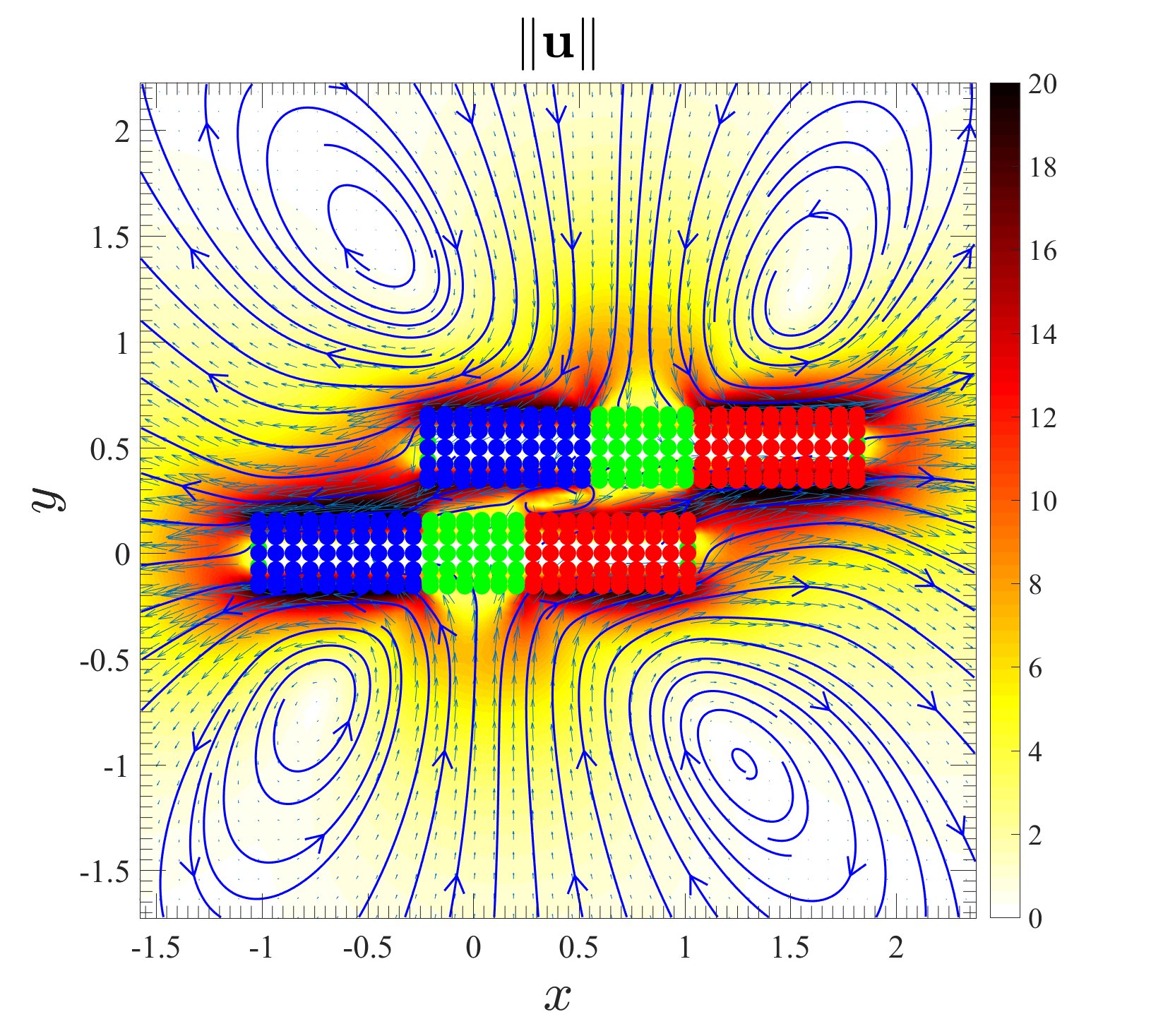}
\par\end{centering}

\caption{\label{fig:ActiveDimer}Active flow around a pair of extensile nanorods
composed of three segments (shown with blue, green and red colors)
sedimented on top of a no-slip boundary (the plane of the image) and
viewed from above for the three different levels of resolution illustrated
in Fig. \ref{fig:BlobModels}: minimally resolved (left), marginally
resolved (middle), and well resolved (right). The colored disks (red,
blue or green) are projections of the blobs, with no-slip conditions
on the green blobs and active slip of magnitude 20$\mu\mbox{m}/\mbox{s}$
on the blue and red blobs, directed away from the green segment. A
cut through the flow field is shown in the\textbf{ }lab frame as a
vector field along with streamlines, with the magnitude of the velocity
shown as a color scale plot.}
\end{figure*}

In Fig. \ref{fig:ActiveDimer} we show the instantaneous flow around
a dimer of active rods, as computed using the procedure described
in Appendix \ref{app:RenderFlow} and seen from above, for three different
resolutions: a minimally-resolved, a moderately-resolved and a well-resolved
model. The imposition of the slip at the surface of the blobs becomes
more and more accurate as the resolution is improved; however, we
see a rather good match between the three flow fields even relatively
close to the rods and wall. Our simulations, which correctly take
into account the physical boundary conditions, estimate the angular
frequency of rotation of the dimer to be approximately $0.64$Hz in
the counter-clockwise direction, consistent with experimental observations
\citet{TripleNanorods_Megan}. The estimated angular velocities are
$0.62,\,0.67$ and $0.63$Hz for each resolution respectively. We
will study the dynamics of active nanorods near a no-slip surface
in more detail in the future; in the next section we demonstrate that
the calculations above can be scaled to suspensions of thousands of
rods.

\subsection{\label{sub:ConvergenceWall}Suspension of rods near a boundary}

In this section we test the efficiency of the preconditionner outlined
in Section \ref{sub:Preconditioner} on a suspension of active rods
sedimented near a wall. We have already seen that the block-diagonal
preconditioner is able to account for the hydrodynamic interactions
among the particles in a modest number of iterations for unbounded
flow. Here we show that this continues to hold even when the wall
strongly dominates the hydrodynamics, and investigate how important
it is for the preconditioner to know about the presence of the boundary.
Namely, recall that in the block-diagonal preconditioner the diagonal
blocks of $\widetilde{\Mob}$ correspond to the blob-blob mobility
for an individual rod \emph{in the presence of the boundary}. Since
$\widetilde{\Mob}$ depends on the configuration of \emph{each} rod
relative to the wall, unlike for an unbounded suspension, all diagonal
blocks need to be factorized anew for each new configuration. However,
it is also possible to use an \emph{approximate }block diagonal preconditioner
that assumes an \emph{unbounded} suspension, i.e., neglects the presence
of the boundary when computing a block-diagonal approximation of the
blob-blob mobility. This seems like a strong approximation to be made
for objects close to a no-slip wall; however, the investigations below
will demonstrate that the Krylov solver can account not only for the
rod-rod interactions, but also for the rod-wall interactions. This
is an important finding because we recall that in the Green's-function-free
method described for confined suspensions in Section \ref{sec:RigidIBAMR},
the boundary conditions are completely ignored in the preconditioner.

In these tests we discretize cylinders of aspect ratio $\alpha=L/D\approx6.4$
either by placing $98$ blobs on the surface of a cylinder of geometric
length $L_{g}=L$ and geometric radius $R_{g}=0.863R$, keeping $a/s=0.5$,
or, by placing $21$ blobs of radius $a=1.02R$ uniformly spaced along
a line segment of length $L$. For testing purposes, we generate random
periodic packings of $N_{r}$ rectangles at a surface packing density
$\phi_{a}$ using a molecular dynamics code \citet{Tetratic_Rectangles}.
We then use these hard-rectangle packings to generate a configuration
of non-overlapping cylinders that are parallel to the wall and at
a constant distance $H$ from the wall; we do not expect to see different
results if some randomness is added to the heights and orientations
of the cylinders relative to the wall, as long as their surfaces remain
sufficiently far from the wall. In our tests we vary the centerline
height $H$ from $H=0.75D$ to $H=2D$, the area fraction $\phi_{a}$
from $0.01$ to $0.6$, and the number of rods $N_{r}$ from $10$
to $10^{4}$; the number of blobs varies in the range from $N_{b}=200$
to about $N_{b}=10^{6}$. For our implementation and hardware (a Tesla
K20 GPU) one GMRES iteration takes around $0.3s$ for $N_{b}=10^{4}$,
$1s$ for $N_{b}\approx2\cdot10^{4}$, $20s$ for $N_{b}\approx10^{5}$,
and $220s$ for $N_{b}\approx3\cdot10^{5}$; we emphasize again that
by using an FMM one can change the scaling from $O\left(N_{b}^{2}\right)$
to $O\left(N_{b}\log N_{b}\right)$ and thus substantially reduce
the computational times for large $N_{b}$, see right panel in Fig.
\ref{fig:Convergence-FMM}. To ensure a nontrivial right hand side
of the linear system when testing the iterative solver, each blob
is prescribed a random slip velocity and a random force, producing
a random force and torque on each cylinder. 

The left half of Table \ref{tab:Efficiency-GMRES} shows the scaling
of the number of GMRES iterations with the area fraction $\phi_{a}$
for a fixed number of rods $N_{r}=1000$, and compares the efficiency
of the preconditioner using the full (wall-corrected) to that using
the approximate (no wall contributions) block-diagonal preconditioner.
We observe that the number of iterations increases slowly with the
area fraction for both resolutions and reaches a maximum of 31 iterations
for $\phi_{a}=0.6$ with the wall-corrected preconditioner. Therefore,
as we already saw in Section \ref{sub:ConvergenceFMM}, the performance
of the preconditioner is not highly sensitive to near-field interactions.
When using the approximate block-diagonal preconditioner without the
wall corrections, the number of iterations is increased, as expected.
However, this increase never exceeds $50\%$, which means that even
a poor approximation of the mobility can be used in the preconditioner
in practice. The right half of Table \ref{tab:Efficiency-GMRES} shows
the scaling of the preconditioner with the number of rods $N_{r}$
for a fixed area fraction $\phi_{a}=0.1$. The number of iterations
rapidly saturates around 20 and becomes independent of the number
of rods for both resolutions and heights. This confirms that the results
obtained for a suspension of spheres in Section \ref{sub:ConvergenceFMM}
apply to confined suspensions as well. Note that for the largest system
sizes studied in the right half of Table \ref{tab:Efficiency-GMRES}
a linear-scaling FMM implementation would likely be substantially
more efficient than the quadratic-scaling GPU implementation employed
here.

\begin{table*}[tbph]
\begin{centering}
\begin{tabular}{|c|c|c|c|}
\hline 
$\phi_{a}$ &
Resolution  &
Wall-corrected &
Unbounded\tabularnewline
\hline 
\hline 
\multirow{2}{*}{0.01} &
21 &
12 &
17\tabularnewline
\cline{2-4} 
 & 98 &
16 &
28\tabularnewline
\hline 
\multirow{2}{*}{0.1} &
21 &
19 &
23\tabularnewline
\cline{2-4} 
 & 98 &
22 &
32\tabularnewline
\hline 
\multirow{2}{*}{0.2} &
21 &
20 &
25\tabularnewline
\cline{2-4} 
 & 98 &
23 &
34\tabularnewline
\hline 
\multirow{2}{*}{0.4} &
21 &
25 &
29\tabularnewline
\cline{2-4} 
 & 98 &
27 &
33\tabularnewline
\hline 
\multirow{2}{*}{0.6} &
21 &
30 &
33\tabularnewline
\cline{2-4} 
 & 98 &
31 &
43\tabularnewline
\hline 
\end{tabular}\hspace{1cm}%
\begin{tabular}{|c|c|c|c|c|c|}
\hline 
\multirow{2}{*}{$N_{r}$} &
\multirow{2}{*}{Resolution } &
\multicolumn{2}{c|}{$H/D=0.75$} &
\multicolumn{2}{c|}{$H/D=2$}\tabularnewline
\cline{3-6} 
 &  & iterations &
time (s) &
iterations &
time (s)\tabularnewline
\hline 
\hline 
\multirow{2}{*}{10} &
21 &
7 &
0.15 &
7 &
0.15\tabularnewline
\cline{2-6} 
 & 98 &
8 &
1.49 &
9 &
1.51\tabularnewline
\hline 
\multirow{2}{*}{100} &
21 &
14 &
1.95 &
13 &
1.52\tabularnewline
\cline{2-6} 
 & 98 &
19 &
18.9 &
18 &
35.6\tabularnewline
\hline 
\multirow{2}{*}{1000} &
21 &
19 &
32.7 &
16 &
29.8\tabularnewline
\cline{2-6} 
 & 98 &
22 &
620 &
20 &
559\tabularnewline
\hline 
\multirow{2}{*}{5000} &
21 &
18 &
520 &
16 &
4,500\tabularnewline
\cline{2-6} 
 & 98 &
23 &
10,200 &
22 &
12,400\tabularnewline
\hline 
\multirow{2}{*}{10000} &
21 &
20 &
2,050 &
17 &
1,430\tabularnewline
\cline{2-6} 
 & 98 &
23 &
39,400 &
21 &
36,300\tabularnewline
\hline 
\end{tabular}
\par\end{centering}

\caption{\label{tab:Efficiency-GMRES}(Left) Number of GMRES iterations required
to reduce the residual by a factor of $10^{8}$ for several surface
packing fractions and two different resolutions (number of blobs per
rod), for $H/D=0.75$ and $N_{r}=1000$ rods. The full block-diagonal
preconditioner, which takes into account the wall corrections for
each body, is compared to the approximate preconditioner, in which
all wall corrections are neglected. (Right) Number of iterations and
wall-clock time (using a direct GPU matrix-vector product on a Tesla
K20 GPU) to solve the mobility problem with tolerance $10^{-8}$ using
the wall-corrected preconditioner at $\phi_{a}=0.1$, for different
number of rods and proximity of the rods to the wall.}
\end{table*}

\section{\label{sec:ResultsConfined}Results: Confined Domains}

In this section, we numerically explore the accuracy and efficiency
of the rigid multiblob immersed boundary (IB) method described in
Section \ref{sec:RigidIBAMR}. This method is suited to fully confined
(bounded) domains, and here we model suspensions of spheres in a periodic
domain, a slit channel (i.e., two parallel walls), and a square (duct)
channel. As discussed in more detail in Section \ref{sec:RigidIBAMR},
for periodic suspensions it is possible to use FFT-based methods \citet{SpectralEwald_Stokes,SD_SpectralEwald,ForceCoupling_Fluctuations,FluctuatingFCM_DC}
to obtain the product of the blob-blob mobility matrix with a vector.
Future work should compare the method developed here with such approaches,
especially for Brownian suspensions.

The effective hydrodynamic radius of an IB blob (also called a ``marker''
or ``IB point'' in the IB literature \citet{IBM_PeskinReview})
can be computed numerically by dragging a single blob with a constant
applied force through a large periodic grid with spacing $h$ and
applying the Hasimoto periodic correction \citet{ISIBM,BrownianBlobs}.
When averaged over many positions of the marker relative to the underlying
grid, for the six-point kernel \citet{New6ptKernel} used here %
\footnote{As summarized in Refs. \citet{ISIBM,BrownianBlobs}, $a\approx1.25h$
for the widely used four-point kernel \citet{IBM_PeskinReview}, and
$a\approx0.91h$ for the three-point kernel \citet{StaggeredIBM}.%
} we obtain $a\approx1.47\, h$. The geometry of the rigid multiblob
models of a sphere used here is the same as in Sections \ref{sec:ResultsUnbounded}
and \ref{sec:ResultsWall}. We also know from Section \ref{sub:Pair-mobility}
that the spacing between the blobs should be around $s\approx2a\approx3h$,
which is somewhat larger than the spacing $s\approx2h$ used in \citet{RigidIBM},
and leads to improved conditioning of the blob-blob mobility matrix.
In fact, we observe that when distinct blobs overlap, the preconditioned
GMRES solver described in Section \ref{sub:Preconditionining-Algorithm}
shows significantly worse performance than when the blobs are not
overlapping (or just touching). We therefore recommend using $s\gtrapprox3h$
for rigid multiblob suspensions at zero Reynolds numbers.

Determining the exact spacing is somewhat of an art and is problem
specific. In the IB approach developed here, the fast multipole method
used in Section \ref{sec:ResultsUnbounded} and the GPU matrix-vector
product used in Section \ref{sec:ResultsWall} are replaced by a geometric
multigrid method, which works best for grid sizes that are powers
of two. Once the exact spacing is determined, the effective hydrodynamic
radius of the rigid multiblob can be determined numerically; we get
very similar results for the IB method to those for an unbounded domain
in Section \ref{sub:ConvergenceMoments}, after setting $a=1.47\, h$.
For confined suspensions, the ratio of the size of the particles to
the domain size is typically fixed to some experimental value, and
this constrains the choice of number of grid cells and spacing between
the blobs. In all of the tests presented here, we have empirically
optimized the appropriate value for the grid size and the spacing
$s$ in the interval $2h$ to $3h$, and report the chosen values.
As explained earlier, it is possible to use split IB kernels to gain
more flexibility in choosing the grid sizes and blob spacing.

In Section \ref{sub:TransInv}, we investigate in more detail the
loss of perfect translational and rotational invariance of the blob-blob
mobility and the mobility of rigid multiblob spheres, and demonstrate
that by using the improved six point kernel \citet{New6ptKernel}
our method is able to minimize the grid artifacts to a significant
extent. Note, however, that there is an additional loss of rotational
invariance for rigid multiblob models that comes from discretizing
the bodies using blobs; this unphysical bias exists even in the absence
of a fluid grid. In Section \ref{sub:LaddTest} we explore in more
detail the accuracy for different resolutions for a periodic suspension
of spheres by comparing to reference results from multipole-based
methods. In Section \ref{sub:SphereSlitChannel} we compute the mobility
of a sphere in a slit channel and compare to existing theories for
a number of resolutions. In Section \ref{sub:Boomerang} we compute
the effective quasi-two-dimensional diffusion coefficient of a boomerang
colloid in a slit channel, and compare to recent experimental measurements
\citet{BoomerangDiffusion,AsymmetricBoomerangs,angleBoomerangs}.
In Section \ref{sub:ConvergenceIBAMR} we optimize the convergence
of the iterative linear solver for suspensions of many bodies, and
demonstrate that the number of GMRES iterations is essentially independent
of the number of particles, even in confined domains such as slit
channels. Finally, we study the sedimentation velocities in a bidisperse
suspension of spheres in Section \ref{sub:BinarySedimentation}, and
compare to recent Stokesian Dynamics simulations \citet{BinarySuspension_SD,SD_SpectralEwald}.
In Appendix \ref{app:Permeable} we study flow around permeable rigid
bodies.

\subsection{\label{sub:TransInv}Translational invariance}

\begin{figure*}[tbph]
\includegraphics[width=0.54\textwidth]{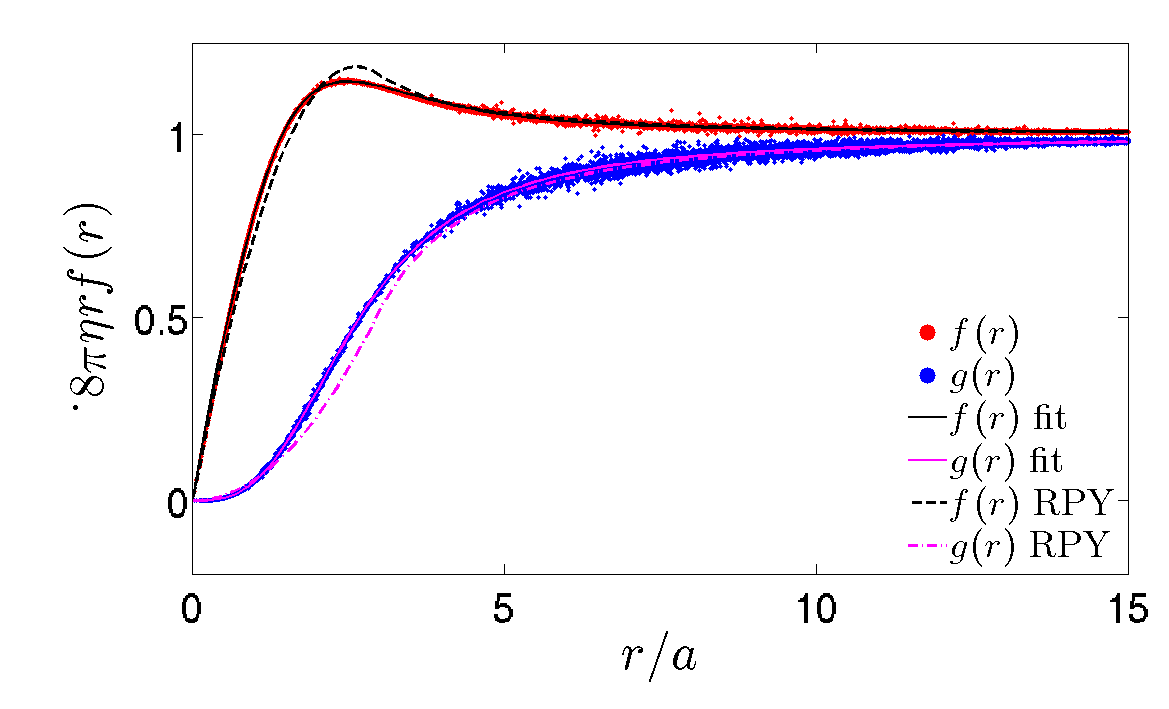}\includegraphics[width=0.44\textwidth]{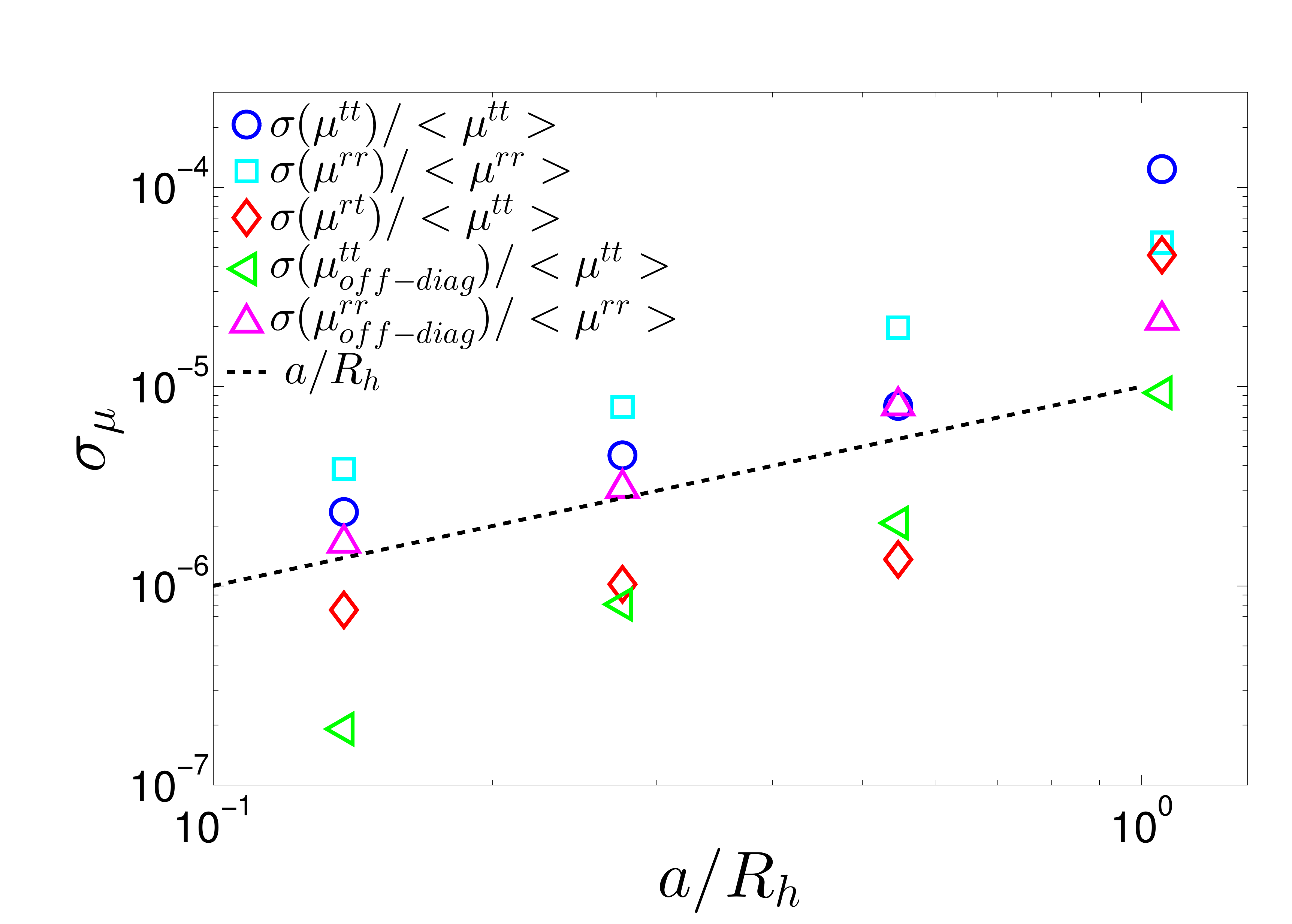}

\caption{\label{fig:TransInv}Translational and rotational invariance of the
rigid multiblob IB method. (Left panel) Empirical values of the blob-blob
mobility functions $f(r)$ and $g(r)$ appearing in (\ref{eq:M_tilde_ij}),
normalized by $8\pi\eta r$ so that they asymptote to unity. An empirical
fit through the data is compared to the RPY tensor for blob radius
$a=1.47h$. Note that the scatter around the fit for $r\gtrsim5a$
is dominated by periodic artifacts due to the finite size of the grid.
(Right panel) Standard deviations of the diagonal (translation-translation,
rotation-rotation) and cross-coupling (rotation-translation) components
of the mobility matrix for spheres discretized with 12, 42, 162 or
642 markers shells (i.e., for decreasing $a/R_{h}$). Also shown is
the typical magnitude of the off-diagonal components of the translation-translation
and rotation-rotation mobility matrices, which should be zero for
a perfect sphere.}
\end{figure*}

As explained in detail in Section \ref{sec:RigidIBAMR}, for sufficiently
large domains the blob-blob mobility computed by the IB method has
the approximate form (\ref{eq:M_tilde_ij}). Deviations from this
formula arise because of the imperfect translational and rotational
invariance due to grid artifacts. The two functions $f(r)$ and $g(r)$
are expected to be similar to those appearing in the RPY tensor (\ref{eq:RPYTensor}).
We obtain the actual form of the functions $f(r)$ and $g(r)$ empirically
by fitting numerical data for the parallel and perpendicular mobilities
of a pair of blobs placed in a large periodic domain, see \citet{RigidIBM}
for more details. The results are shown in the left panel of Fig.
\ref{fig:TransInv}, and are compared to the RPY tensor for spheres
of radius $a=1.47h$. We have empirically fitted the numerical results
for $f(r)$ and $g(r)$ with a fit that has the proper asymptotic
behavior at large and short distances, see Appendix A in \citet{RigidIBM}
for more details; this fit is used in the preconditioner as an analytical
approximation of the diagonal blocks of the blob-blob mobility matrix.
We see that the differences between the fits and the RPY tensor are
rather small, and also confirm the improved translational invariance
of the six-point kernel \citet{New6ptKernel} as evidenced in the
small scatter of the points around the fits. This confirms our expectation
that the rigid multiblob IB method will behave similarly to an RPY-based
method in terms of accuracy.

In the right panel of Fig. \ref{fig:TransInv} we investigate the
translational and rotational invariance of rigid multiblob models
of spheres as a function of the resolution. We randomly move and rotate
a sphere relative to the underlying grid and compute the elements
of the mobility matrix. The mean of these elements defines an effective
translational and rotational radius consistent with the results presented
in Section \ref{sub:ConvergenceMoments} (not shown). In the figure
we show the normalized standard deviation of the elements of the mobility
matrix as a function of the resolution (number of blobs). As expected,
the normalized standard deviation decreases linearly with the size
of the blobs $a$ (equivalently, the inverse of the square root of
$N_{b}$), and is below $10^{-4}$ for all mobility elements even
for the 12-blob (icosahedral) model \citet{MultiblobSprings}.

\subsection{\label{sub:LaddTest}Periodic Suspension of Spheres}

In this section we apply our rigid multiblob method to a benchmark
resistance problem in a periodic suspension of 108 spheres moving
with random linear and angular velocities. This benchmark was developed
by Anthony Ladd, who supplied us with a random and a face-centered
cubic (FCC) configuration of spheres, at a low density of $\phi=0.05$,
as well as a high density of $\phi=0.45$. He also supplied to us
the results for the resulting forces and torques obtained using the
HYDROMULTIPOLE code \citet{HYDROMULTIPOLE}. Note that pairwise lubrication
corrections have been included in the multipole expansion method used
for these calculations \citet{HYDROMULTIPOLE,HydroMultipole_Ladd,HydroMultipole_Ladd_Lubrication};
to our knowledge no method has accounted for three-body lubrication
corrections.

\begin{figure*}[tbph]
\begin{centering}
\includegraphics[width=0.49\textwidth]{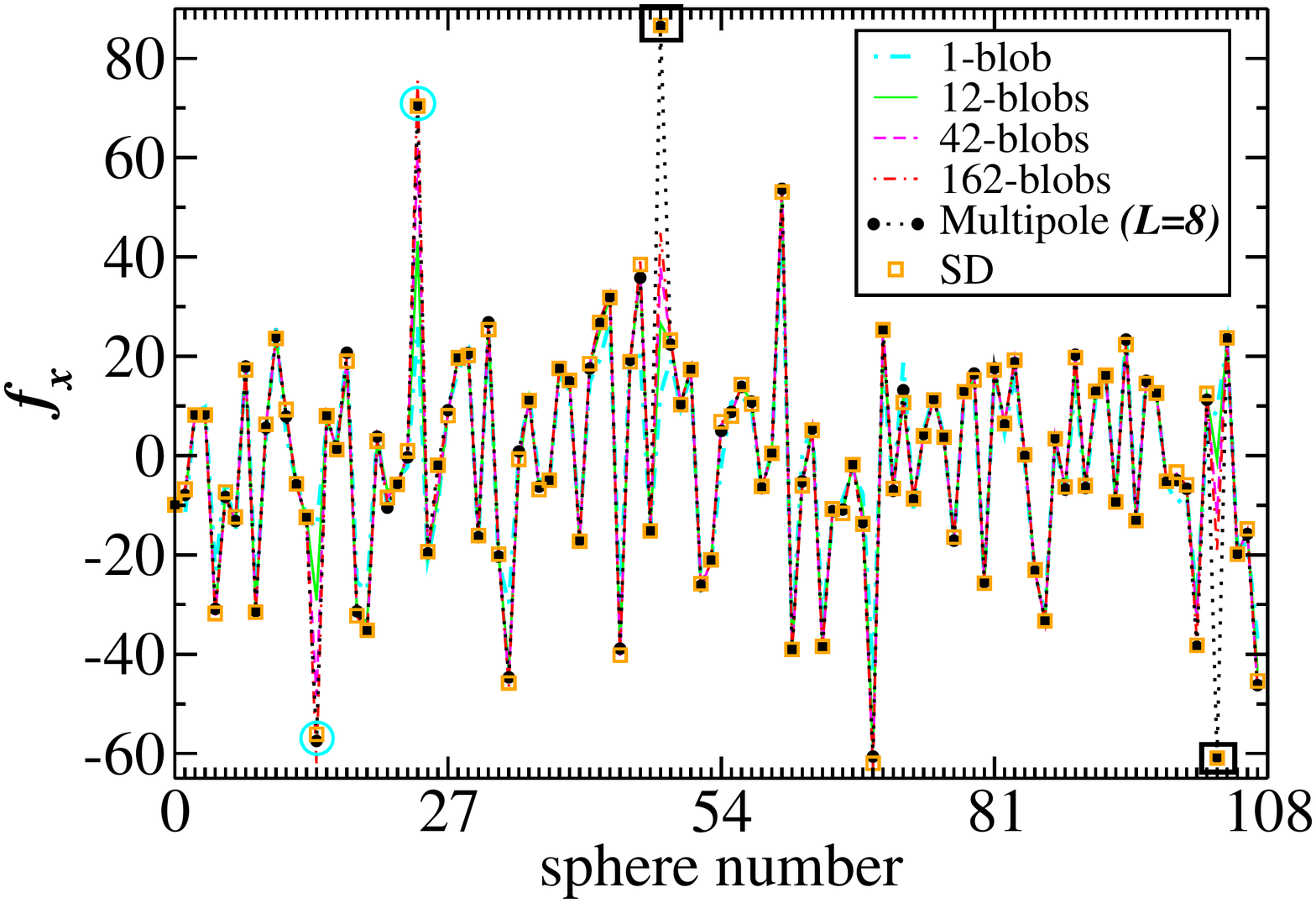}\includegraphics[width=0.49\textwidth]{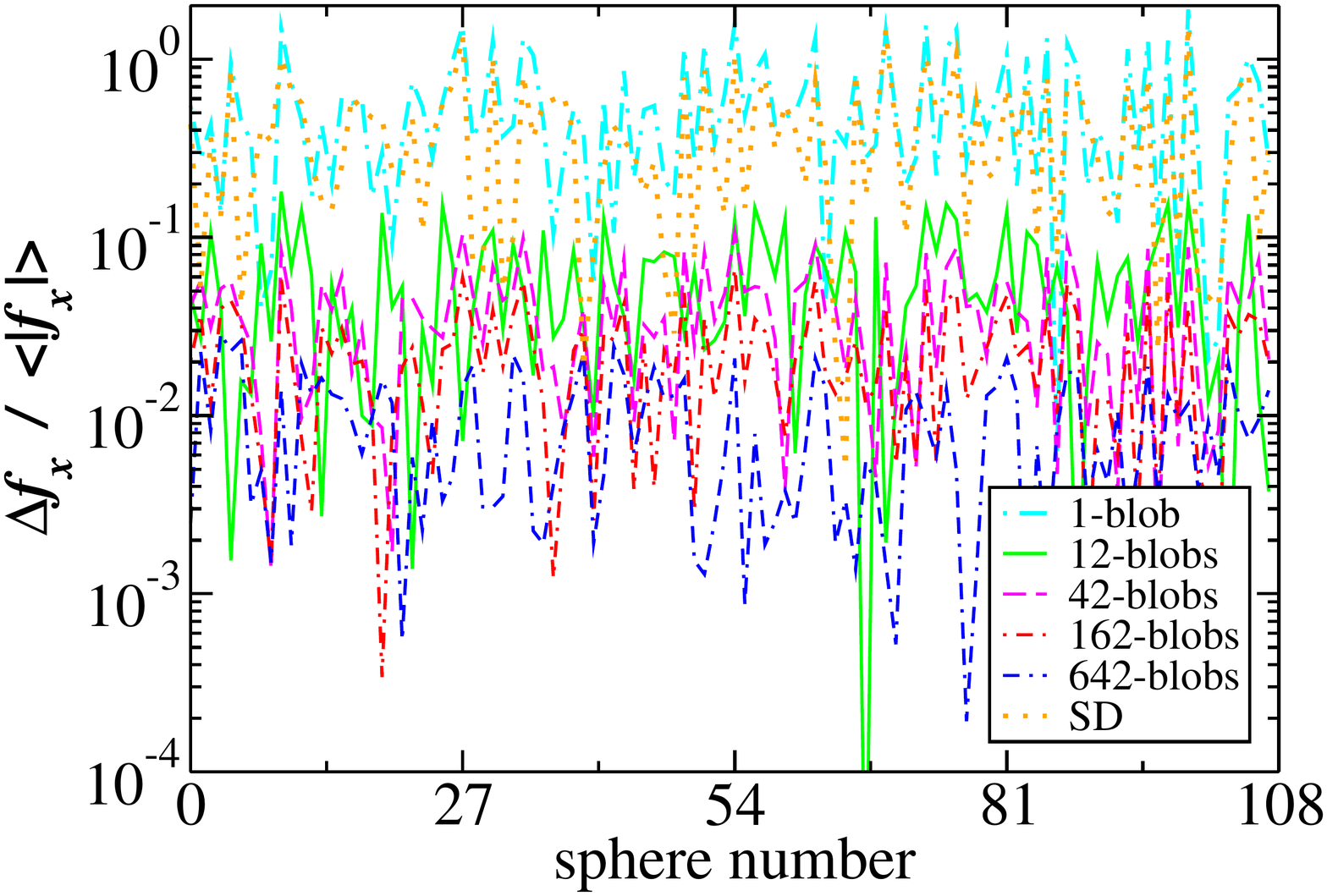}
\par\end{centering}

\caption{\label{fig:LaddTest}Results from the rigid multiblob method applied
to Ladd's benchmark resistance problem (with specified random velocities)
for a periodic suspension of 108 spheres. (Left panel) The $x$-component
of the force on each sphere in a random suspension at a low volume
fraction $\phi=0.05$. For comparison, we show the results of Stokesian
dynamics (SD) \citet{libStokes} and the HYDROMULTIPOLE code \citet{HYDROMULTIPOLE}
with $L=8$ moments retained. Two particles that happen to be at a
distance closer than $2.02$ radii from each other are marked by a
black box and a blue circle, and develop unresolved strong lubrication
forces between them. (Right panel) The normalized error $\abs{f_{x}-f_{x}^{(\text{ref})}}/\av{\abs{f_{x}^{(\text{ref})}}}$
in the $x$-component of the force for an FCC lattice at the high
volume fraction $\phi=0.45$. The HYDROMULTIPOLE code with $L=8$
moments is used as a high-accuracy reference calculation.}
\end{figure*}

The rigid multiblob models used in this study have been chosen to
give a blob spacing close to $s\approx2a\approx3h$, while ensuring
that the number of grid cells is integer given the specified unit
cell length in the benchmark configurations, and to have a specified
effective hydrodynamic radius $R_{h}\approx1$ for five different
resolutions: a single blob per sphere (similar to a truncation with
only one monopole per sphere), 12 blobs (geometric radius $R_{g}=0.7896$,
grid spacing $h=0.2778$), 42 blobs ($R_{g}=0.8899$, $h=0.1667$),
162 blobs ($R_{g}=0.9502$, $h=0.08929$) and 642 blobs ($R_{g}=0.9766$,
$h=0.0463$) per sphere. The results for the $x$ component of the
computed forces on the spheres are illustrated in Fig. \ref{fig:LaddTest};
similar results are observed for other components.

In the left panel of Fig. \ref{fig:LaddTest}, we focus on the low
density suspension ($\phi=0.05$) and compare the forces computed
by the rigid multigrid method with those computed using $L=8$ multipole
moments, as well as the results of the Stokesian dynamics (SD) code
of Ichiki \citet{libStokes} (which roughly corresponds to $L=2$
moments). The overall agreement is quite good, but notice that even
with $162$ blobs per sphere we do not resolve the lubrication force
between particles numbered 48 and 103, marked in the figure, since
this pair of particles have a gap of only $0.024$ radii between them.
This is not surprising given that we do not include pairwise lubrication
corrections in our method; such corrections are included in the reference
results we are comparing to so they produce accurate forces for all
particles. In the rigid multiblob method, we resolve the near field
interactions more and more accurately as we increase the number of
blobs per sphere, but we cannot accurately resolve the hydrodynamic
interactions between pairs of particles with overlapping blobs (see
Fig. \ref{fig:pairMobility}).

At the high packing fraction $\phi=0.45$ there are many pairs of
nearly touching particles in a typical random suspension of hard spheres
in the absence of (electrostatic) repulsive forces, and there is no
hope that the rigid multiblob method can accurately compute the interparticle
forces %
\footnote{The results from HYDROMULTIPOLE suggest that even $L=15$ moments,
which is the maximum that could be afforded with 32GB of memory since
the linear system to be solved is dense and has about $7\cdot10^{4}$
unknowns, convergence is not achieved to sufficiently high accuracy
for the random suspension at $\phi=0.45$.%
}. Therefore, at this density we focus on an FCC lattice configuration,
and compare to the multipole-based code with $L=8$ moments. Here
the closest particle distance is $2.36$ radii and our method is able
to resolve the forces relatively well, especially with more than 12
blobs per sphere, see the right panel in Fig. \ref{fig:LaddTest}.
This is perhaps not surprising; however, the more important point
we wish to make is that the SD results are now not significantly more
accurate than the results obtained from using only a single blob per
sphere. The addition of stresslets and pairwise lubrication does not
appear to help much in resolving the many-body far field hydrodynamic
multiple scattering in this lattice configuration. Using an icosahedral
rigid multiblob already provides an order of magnitude improvement
in the typical error over an FTS truncation, and provides an error
comparable to keeping $L=3$ moments in the HYDROMULTIPOLE method
(not shown), which is also the minimum number of moments necessary
to keep to capture all long-ranged hydrodynamic interactions, as well
as to model active sphere suspensions \citet{BoundaryIntegralGalerkin,IrreducibleActiveFlows_PRL}.

\subsection{\label{sub:SphereSlitChannel}Sphere in a slit channel}

In this section we compute the parallel and perpendicular translational
mobilities of a sphere in a slit channel of thickness $19.2R_{h}$
as a function of the height $H$ of the sphere center above one of
the walls. This problem is of relevance to a number of experiments
involving spherical colloids confined between two glass microscope
slips, and has also been used as a benchmark problem for boundary
integral calculations in \citet{BoundaryIntegral_Wall}. Since the
immersed boundary method used here cannot be used for infinite domains,
we take a domain of dimensions $(76.8,\,19.2,\,76.8)R_{h}$ and apply
no-slip conditions on the $y$ boundaries and periodic conditions
in the other two directions.

There are no manageable theoretical results accurate for all distances
from the wall and all channel dimensions \citet{StokesianDynamics_Slit}.
For the parallel component of the mobility, Faxen obtained exact series
expansions for the mobility at the half and quarter channel locations,
which we use to benchmark our calculations, neglecting the corrections
coming from the use of periodic boundary conditions in the directions
parallel to the walls. For other positions of the sphere we employ
the Modified Coherent Superposition Assumption (MCSA) approximation
given in Eq. (9) in Ref. \citet{MobilitySlitChannel}, with the mobility
relative to a single wall given by the same theoretical lines shown
in Fig. \ref{fig:Sphere-wall}. The rigid multiblob models used in
this study have been chosen to give a blob spacing close to $s\approx2a\approx3h$,
while ensuring that the number of grid cells is integer given the
target channel width relative to the effective hydrodynamic radius
of the sphere $R_{h}$ for all of the resolutions studied: a single
blob as a minimal model of the sphere \citet{BrownianBlobs} (grid
size $128\times28\times128$), 12 blobs (grid size $256\times64\times256$,
geometric radius $R_{g}=2.503h$, giving spacing $s/h=2.63$), and
42 blobs ($512\times128\times512$ grid, $R_{g}=6.047h$, $s/h=3.30$).

\begin{figure}[tbph]
\begin{centering}
\includegraphics[width=0.75\columnwidth]{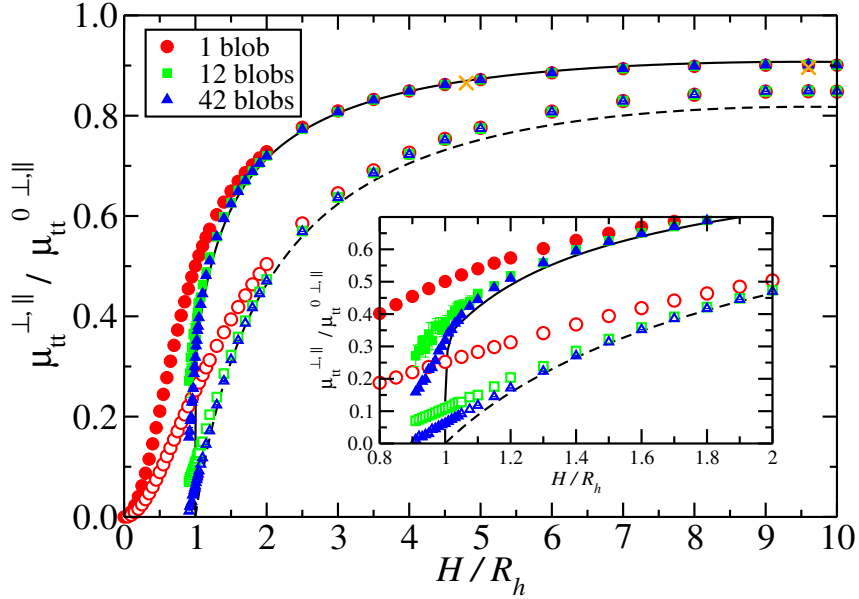}
\par\end{centering}

\caption{\textbf{\label{fig:mob-channel}}Translational mobility of a sphere
in a slit channel of width $19.2R_{h}$ relative to the bulk, for
several resolutions of the rigid multiblobs (see legend), for forces/motion
parallel (filled symbols) and perpendicular (empty symbols) to the
wall. The numerical results are in good agreement with the exact results
of Faxen for distances $H=L/4$ and $H=L/2$ (orange crosses). The
inset shows that close to the wall the results for the $12$ and $42$
blob shells are in reasonable agreement with the approximate MCSA
theory (lines).}
\end{figure}

We have already confirmed that our results are in agreement with Faxen's
theory in the (extrapolated) limit of an infinitely long channel in
our prior work \citet{RigidIBM}. Here we examine the mobility as
the sphere moves through the channel, and in particular, as it comes
close to the wall. The results of our calculations are shown in Fig.
\ref{fig:mob-channel}, and are in good agreement with the approximate
MCSA theory for the parallel mobility. Note that the MCSA theory is
approximate even far from the wall, as seen from the fact that our
results do not match it for the perpendicular mobility even at the
center of the channel. The boundary condition handling described in
Appendix D of \citet{RigidIBM} ensures that for a single blob the
mobility vanishes at the boundary, i.e., for $H=0$, rather than for
$H=R_{h}$ as for a true sphere. The more resolved models, however,
do show a sharp drop in mobility when the sphere nearly touches the
wall. The inset shows that the lubrication interactions are not resolved
very close to the wall, just as we observed for a single wall in Section
\ref{sub:SphereWall}. Nevertheless, we note that unlike the Rotne-Prager-Blake
mobility tensor computed by Swan and Brady \citet{StokesianDynamics_Wall}
and used in Section \ref{sub:SphereWall}, the mobility computed by
the grid-based Stokes solver is physically realistic even when the
blobs overlap the wall. This has two important implications. First,
the rigid multiblob IB method does not run into singularities for
$H_{\text{min}}<H<R_{h}$, where $H_{\text{min}}$ is the distance
at which the center of some blob first leaves the physical domain.
This is particularly beneficial in Brownian simulations, where stochastic
motion can push the sphere to slightly overlap the wall \citet{BrownianBlobs,BrownianMultiBlobs}.
Furthermore, the immersed boundary results in the inset of Fig. \ref{fig:mob-channel}
are substantially more rotationally-invariant as the sphere approaches
the wall than the corresponding results in the top two panels of Fig.
\ref{fig:Sphere-wall}; the error bars in the mobility due to discretization
artifacts are very small for the immersed boundary method even for
$H<R_{h}$.

\subsection{\label{sub:Boomerang}Boomerang in a slit channel}

In this section we study the diffusion of a boomerang colloidal particle
in a narrow slit channel, as recently studied experimentally and theoretically
\citet{BoomerangDiffusion,AsymmetricBoomerangs,angleBoomerangs}.
The boomerangs are confined to essentially remain in the plane parallel
to the wall by the tight confinement, and thus perform quasi two-dimensional
diffusion. We previously studied the diffusion of such a boomerang
colloid sedimented against a single no-slip boundary in \citet{BrownianMultiBlobs},
using a strong gravitational force to keep the boomerang in quasi
two dimensions. However, the colloids in the actual experiments are
almost neutrally buoyant and a slit channel is used to confine them
to two dimensions. Here we use our rigid multiblob IB method to determine
the effective two dimensional diffusion coefficients of a single boomerang
in slit confinement.

As discussed in detail in \citet{BrownianMultiBlobs}, it is, in principle,
necessary to perform long Brownian simulations to determine the long-time
diffusion tensor $\M D$ of nonspherical particles. However, if the
mean square displacement (MSD) is linear in time, the long and short-time
time diffusion coefficients are equal and can be obtained from the
Stokes-Einstein relation 
\begin{equation}
\M D=k_{B}T\left\langle \BMob\right\rangle =k_{B}T\,\bar{\M{\mu}},\label{eq:SE_short}
\end{equation}
where $\bar{\M{\mu}}$ is the average mobility over configurations
following the Gibbs-Boltzmann (GB) distribution. Therefore, the diffusion
coefficient can be computed by generating samples from the GB distribution
of boomerang configurations, and then solving a mobility problem for
each configuration and averaging over the samples. For quasi two-dimensional
diffusion the MSD can be made nearly linear by a careful choice of
the \emph{tracking point} \citet{BrownianMultiBlobs,BoomerangDiffusion,AsymmetricBoomerangs},
which is the point whose translation is measured and around which
torques are expressed \citet{bernal1980transport}. Chakrabarty \emph{et
al.} \citet{BoomerangDiffusion} have shown that for particles diffusing
in two dimensions, the optimal choice of tracking point is the center
of hydrodynamics stress (CoH), the location of which can be computed
from the bulk mobility tensor \citet{BrownianMultiBlobs}. 

To compare with the experimental results of Chakrabarty \emph{et al}.
we compute the diffusion coefficient of a single boomerang colloidal
particle between two walls. In the experiments \citet{BoomerangDiffusion},
colloidal particles with boomerang shape diffuse in a channel of width
$\sim2\mu m$. The boomerang particles, produced by photolithography,
have two arms of length $2.1\mu m$, thickness $0.51\mu m$ and width
$0.55\mu m$ forming a right angle. In our computations, we use no-slip
boundary conditions on the walls of the channel and periodic boundary
conditions in the directions parallel to them. We construct two rigid
multiblob models of such boomerangs (see \citet{BrownianMultiBlobs}
for geometric details): a minimally resolved model with 15 blobs (grid
size $128\times9\times128$, blob spacing $s/h=1.36$) that is essentially
a bent version of the cylinder model shown in the left-most panel
of Fig. \ref{fig:BlobModels}, and a moderately resolved model with
120 blobs (grid size $256\times18\times256$ and blob spacing $s/h=2.22$),
shown in the right-most panel of Fig. \ref{fig:BlobModels}.

We assume a hard-core potential between each of the blobs and the
walls, and average the mobility over 100 samples generated from the
Gibbs-Boltzmann distribution using an accept-reject Monte Carlo procedure
\citet{BrownianMultiBlobs}. In the experiments, there is likely an
additional electrostatic repulsion from the wall; we have checked
that adding a short-ranged Yukawa-type repulsion with the walls does
not change our results significantly %
\footnote{In fact, our computations (not shown) indicate that a rather accurate
estimate of the average mobility can be computed quickly by simply
evaluating the mobility of a boomerang lying exactly on the center
plane of the channel.%
}. Following \citet{BoomerangDiffusion}, we report the diffusion coefficients
computed using (\ref{eq:SE_short}) in the continuous body frame (CoB)
\citet{BoomerangDiffusion} attached to the colloidal particle, such
that the axis $X_{1}$ goes along the line that bisect the boomerang
angles and the axis $X_{2}$ is orthogonal to $X_{1}$ (see Fig. 1
in \citet{BoomerangDiffusion}). The diffusion coefficient for the
boomerang particle are given in Table \ref{tab:boomerangDiffusion}.
We see that the computed location of the CoH is in good agreement
with experimental estimates. However, both translational and the rotational
in-plane diffusion coefficients computed in the simulations are twice
larger than those measured experimentally, for both resolutions.

\begin{table}[tbph]
\begin{tabular}{|c|c|c|c|}
\hline 
\multirow{2}{*}{} &
\multirow{2}{*}{Experiments} &
\multicolumn{2}{c|}{ratio = (Experiment / Simulation)}\tabularnewline
\cline{3-4} 
 &  & 15 blobs &
120 blobs\tabularnewline
\hline 
$d$ &
$1.16\;(\mu m)$ &
1.06 &
1.06\tabularnewline
\hline 
$D_{11}$ &
$0.049\;(\mu m^{2}/s)$ &
0.55 &
0.47\tabularnewline
\hline 
$D_{22}$ &
$0.058\;(\mu m^{2}/s)$ &
0.50 &
0.46\tabularnewline
\hline 
$D_{\theta}$ &
$0.044\;(\mbox{rad}^{2}/s)$ &
0.46 &
0.46\tabularnewline
\hline 
\end{tabular} 

\caption{\textbf{\label{tab:boomerangDiffusion}}Comparison of experimentally-measured
diffusion coefficients for a boomerang particle in a slit channel
to numerical estimates obtained from the rigid multiblob IB method.
The tracking point is chosen to be the CoH, which is a point on the
boomerang bisector line at a distance $d$ (first row) from the crossing
point of the two boomerang arms. We report the translational diffusion
coefficients $D_{11}$ and $D_{22}$ in the Continuous Body Frame
(CBF) of reference as in Ref. \citet{BoomerangDiffusion}, averaged
over 100 samples from the Gibbs-Boltzmann distribution of particle
configurations, for two different resolutions.}
\end{table}

To investigate this large mismatch between simulations and experiments,
we explore further the difference between the right angle boomerangs
used in the two distinct experiments \citet{BoomerangDiffusion} (arms
of length $2.1\mu m$ and width $0.55\mu m$) and \citet{angleBoomerangs}
(arms of length $2.33\mu m$ and width $0.7\mu m$), both for a reported
channel width of $2\mu m$. The boomerangs in \citet{angleBoomerangs}
are reported to be about 10\% larger than those used in \citet{BoomerangDiffusion};
however, the reported diffusion coefficients are reported to be about
25\% larger. This is inconsistent with a purely hydrodynamic model,
since the larger particles should have smaller bulk diffusion coefficients
and are more confined. Therefore, the larger particles must have a
translational diffusion coefficient that is \emph{more} than $1.1$
times smaller for translation, and \emph{more} than $1.1^{3}\approx1.33$
times smaller for rotation, if particle size is the only difference
between the two experiments. Indeed, in our simulations the translational
diffusion coefficient is about $1.25$ times smaller for the larger
particles, and the rotational one is $1.57$ times smaller. This suggest
that there are some unreported experimental effects that are not taken
into account in the simulations, such as a potentially non-uniform
channel thickness or polydispersity in the particles. More careful
future investigations are required to understand the origin of the
difference between simulations and experiments reported in Table \ref{tab:boomerangDiffusion}.

\subsection{\label{sub:ConvergenceIBAMR}GMRES convergence}

In this section we investigate the performance of the preconditioner
described in Sec. \ref{sub:Preconditionining-Algorithm} and determine
an optimal value for $N_{s}^{(1)}$ and $N_{s}^{(2)}$, the number
of iterations in the first and second approximate Stokes solves in
the preconditioner. As a Krylov solver, here we use the restarted
right-preconditioned GMRES method, but we have also observed good
performance with the short-term recurrence BiCGStab method, which
typically requires a few more iterations than GMRES but has smaller
memory requirements. It is important to emphasize that the exact number
of iterations depends strongly on the geometry of the rigid multiblobs,
notably, the spacing between the blobs. The performance depends even
more strongly on the efficacy of the geometric multigrid preconditioner
for the Poisson equation, which in the implementation used here is
strongly degraded for grids that have a nearly prime number of cells
in each dimension, and for grids of large aspect ratios (even if all
directions are powers of two). Our focus here is on investigating
the trends in the number of GMRES iterations with the various parameters
in the preconditioner and the system size.

The computational cost of the solver is dominated by the application
of the full preconditioner, whose complexity depends on nontrivial
ways on its different steps and on the parameters of the simulations.
However, in most cases the cost is dominated by the multigrid cycles
for the Poisson equation, and each application of the projection preconditioner
for the Stokes equation (\ref{P_Stokes}) involves $d+1=4$ scalar
V cycles. Here we use preconditioned Richardson iteration as an iterative
solver for the (unconstrained) Stokes equations %
\footnote{Richardson iteration is not effective as a stand-alone Stokes solver,
but, as already explained, it is more efficient in this constrained
context because we only need a rather approximate Stokes solver for
the unconstrained fluid equations.%
}. Therefore, the total number of scalar multigrid cycles per GMRES
iteration is $4(N_{s}^{(1)}+N_{s}^{(2)})$ and we can use $N_{s}^{(1)}+N_{s}^{(2)}$
as a proxy for the computational cost. It should be noted, however,
that this is only an approximation and in practice it may be better
to allow a small increase on the total number of Stokes preconditioner
applications if it reduces significantly the number of outer GMRES
iterations.

We study the solver convergence for a random bidisperse suspensions
of hard spheres with aspect ratio $R_{h,1}/R_{h,2}=1/2$ at different
concentrations and system sizes and geometries. The parameters of
the rigid multiblob models are identical to those reported in Section
\ref{sub:SphereSlitChannel}. We investigate a suspension at a moderate
volume fraction $\phi=0.15$ ($\phi_{1}=\phi_{2}=0.075$ for the two
components), as well as a suspension at a high volume fraction $\phi=0.45$
($\phi_{1}=\phi_{2}=0.225$). We investigate three different types
of boundary conditions, a \emph{periodic} suspension in a cubic domain,
a suspension in a \emph{slit channel} with periodic boundaries in
the directions parallel to the wall, and a \emph{square channel} with
periodic boundaries in the direction of the channel axis. The configurations
of hard spheres were generated using a Monte Carlo algorithm with
hard core exclusion radius equal to the effective hydrodynamic radius
of the spheres.

\begin{figure}[tbph]
\includegraphics[width=0.49\textwidth]{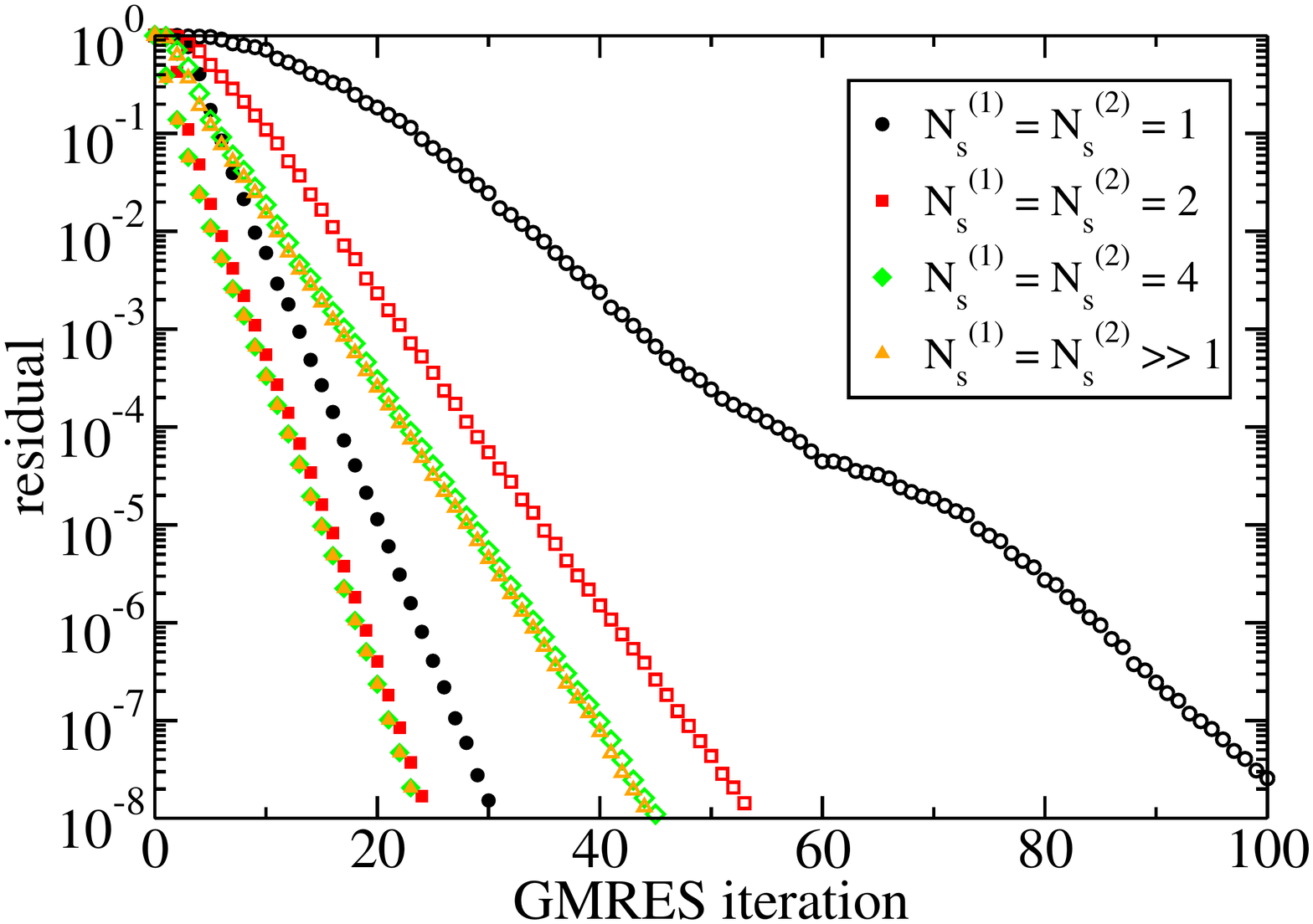}\includegraphics[width=0.49\textwidth]{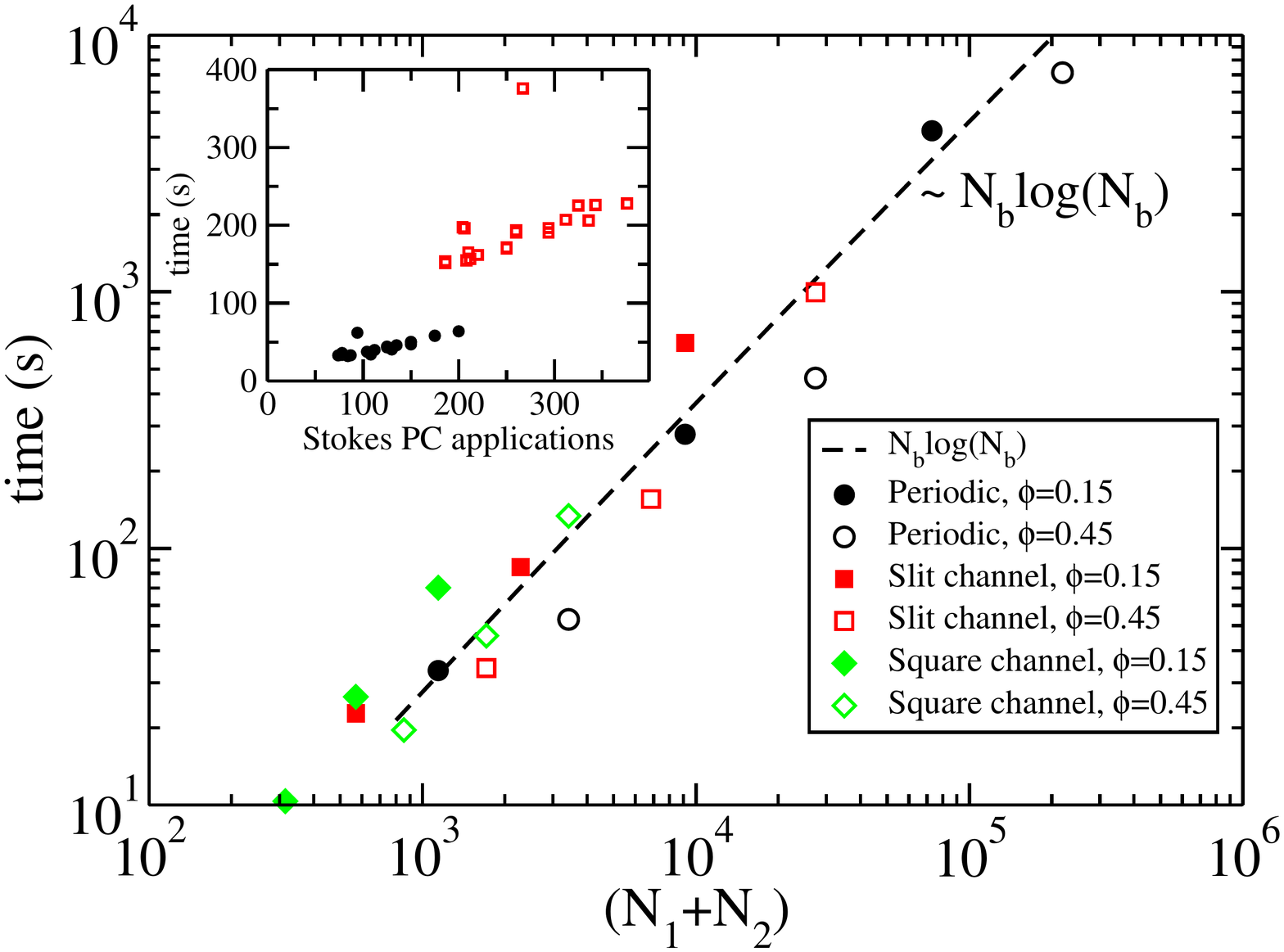}

\caption{\label{fig:ibamr-convergence}(Left panel) Convergence of GMRES with
restart frequency of 60 iterations for a bidisperse suspension of
spheres in a cubic periodic domain (filled symbols, grid size $128^{3}$,
$N_{s}=1014+127=1141$ spheres), and in a slit channel of dimensions
$4L\times L\times4L$ (open symbols, grid size $256\times64\times256$,
$N_{s}=6083+760=6843$ spheres). All spheres are subject to random
forces, torques and slips. (Left panel) Normalized residual versus
number of iterations of the outer solver for different number of iterations
in the first ($N_{s}^{(1)}$) and second ($N_{s}^{(2)}$) Stokes subsolves
inside the preconditioner. For comparison, we also solve the Stokes
subproblems to high accuracy using an inner iteration of GMRES, marked
$N_{s}^{(1)}=N_{s}^{(2)}\gg1$ in the legend. (Right panel) Total
computing time to solve the mobility problem as a function of the
number of spheres $N_{s}=N_{1}+N_{2}$ using an implementation based
on the IBAMR library and eight cores of an Intel Xeon (E5-2665, 2.4GHz)
processor. The inset shows the time to solve the linear system versus
the total number of Stokes preconditioner applications $(N_{s}^{(1)}+N_{s}^{(2)})\, N_{\text{iter}}$.}
\end{figure}

In the left panel of Fig. \ref{fig:ibamr-convergence} we show the
relative residual versus the number of iterations of the outer solver
for different values of $N_{s}^{(1)}$ and $N_{s}^{(2)}$. Because
of all of the approximations in the analytical blob-blob mobility
matrix, the convergence does not improve with increasing $N_{s}^{(1,2)}$
beyond some point. Therefore, it is not necessary to perform nearly
exact Stokes solves (e.g., complete FFTs in periodic domains) inside
the preconditioner; a few (spectrally equivalent) cycles of multigrid
are sufficient. In the inset in the right panel of Fig. \ref{fig:ibamr-convergence}
we show that the total number of applications of the Stokes preconditioner
$N_{s}=\left(N_{s}^{(1)}+N_{s}^{(2)}\right)N_{\text{iter}}$ is a
reasonable proxy for the computational time, where $N_{\text{iter}}$
is the number of GMRES iterations %
\footnote{The actual cost has the form $aN_{s}+bN_{\text{iter}}$ where $b$
grows with the number of blobs, therefore, two outliers are observed
for $N_{s}^{(1)}=1$ and $N_{s}^{(2)}=0$ since these require a large
number of GMRES iterations to converge.%
}.

\textbf{}

Table \ref{tab:IBAMR-convergence} shows the number of GMRES iterations
\footnote{The GMRES iteration is restarted every 60 iteration except for the
largest system ($512^{3}$ fluid cells and about $2\cdot10^{5}$ particles)
which uses a restart frequency of 20 to reduce memory requirements.%
} required to reduce the residual by a factor of $10^{8}$ for a variety
of system sizes, keeping $N_{s}^{(1)}=2$ and $N_{s}^{(2)}=1$. As
seen in the table, the convergence for periodic domains, just as for
the methods based on Green's functions studied in Sections \ref{sub:ConvergenceFMM}
and \ref{sub:ConvergenceWall}, is independent on the system size
and only depends weakly on the packing density. The largest system
has a grid of $512^{3}$ cells and almost $3.4$ million blobs on
$2.2\cdot10^{5}$ spherical shells packed to a rather high density
$\phi=0.45$, yet the GMRES iteration converges after only 44 iterations.
For confined systems the solver requires more iterations, as expected
because the boundary conditions are not taken into account in either
the Stokes solver preconditioner nor the block-diagonal mobility approximation.
At first sight, it may appear that the number of iterations grows
with the system size for non-periodic domains. This increase, however,
comes not because of the increase in system size but rather because
the aspect ratio of the domain grows and the multigrid algorithm used
in our implementation becomes less effective. This can be confirmed
by noting that the number of iterations grows very weakly with system
size if we keep the domain aspect ratio fixed; for a square channel
and $\phi=0.45$ we require $58$ iterations for a grid with $128\times64\times64$
cells, 60 iterations for $256\times128\times128$ cells, and 65 iterations
for $512\times256\times256$ cells (compared to 119 for $512\times64\times64$
cells). This demonstrates that our preconditioner robustly handles
large system sizes even in the presence of physical boundaries. We
believe the performance of the solver for high-aspect ratio domains
can be greatly improved with a new multigrid implementation capable
of dealing with highly non-cubic domains at the coarsest levels of
the multigrid hierarchy. The right panel of Fig. \ref{fig:ibamr-convergence}
shows the total computing time as a function of system size and demonstrates
the near linear scaling of the method at fixed computing power, at
least for cubic periodic domains, for which the multigrid implementation
used in the IBAMR library is nearly optimal.

\begin{table}[tbph]
\begin{tabular}{|c|c|c|c|c|c|c|c|c|c|c|c|c|}
\hline 
\multirow{2}{*}{$\phi$} &
\multicolumn{4}{c|}{Periodic} &
\multicolumn{4}{c|}{Slit channel} &
\multicolumn{4}{c|}{Square channel}\tabularnewline
\cline{2-13} 
 & cells &
$N_{1}$ &
$N_{2}$ &
$N_{G}$ &
cells &
$N_{1}$ &
$N_{2}$ &
$N_{G}$ &
cells &
$N_{1}$ &
$N_{2}$ &
$N_{G}$\tabularnewline
\hline 
\multirow{3}{*}{0.15} &
$128\times128\times128$ &
1014 &
127 &
28 &
$128\times64\times128$ &
507 &
63 &
34 &
$128\times64\times64$ &
283 &
32 &
37\tabularnewline
\cline{2-13} 
 & $256\times256\times256$ &
8111 &
1014 &
29 &
$256\times64\times256$ &
$2028$ &
$253$ &
42 &
$256\times64\times64$ &
507 &
63 &
47\tabularnewline
\cline{2-13} 
 & $512\times512\times512$ &
64885 &
8111 &
29 &
$512\times64\times512$ &
$8111$ &
$1014$ &
63 &
$512\times64\times64$ &
1014 &
127 &
65\tabularnewline
\hline 
\multirow{3}{*}{0.45} &
$128\times128\times128$ &
3041 &
380 &
42 &
$128\times64\times128$ &
1521 &
190 &
52 &
$128\times64\times64$ &
760 &
95 &
58\tabularnewline
\cline{2-13} 
 & $256\times256\times256$ &
24332 &
3041 &
43 &
$256\times64\times256$ &
6083 &
760 &
62 &
$256\times64\times64$ &
1521 &
190 &
76\tabularnewline
\cline{2-13} 
 & $512\times512\times512$ &
194656 &
24332 &
44 &
$512\times64\times512$ &
24332 &
3041 &
96 &
$512\times64\times64$ &
3041 &
380 &
119\tabularnewline
\hline 
\end{tabular}

\caption{\textbf{\label{tab:IBAMR-convergence}}GMRES convergence results for
a bidisperse suspension in periodic and confined domains. A random
configuration of $N_{1}$ hard spheres of radius $R_{h}=1$ (12 blobs)
and $N_{2}$ hard spheres of radius $R_{h}=2$ (42 blobs) is generated,
and random forces, torques and slips are applied on all of the particles.
We report the number of GMRES iterations $N_{G}$ needed to reduce
the residual by a factor $10^{8}$ for the mobility problem for a
variety of system sizes and boundary conditions.}
\end{table}

\subsection{\label{sub:BinarySedimentation}Sedimentation velocity in a binary
sphere suspension}

In our last test we use our rigid multiblob IB method to compute the
mean and variance of the instantaneous sedimentation velocity in a
random binary suspension of hard spheres, as done using Stokesian
Dynamics (SD) by Wang and Brady \citet{BinarySuspension_SD,SD_SpectralEwald}.
The binary suspension has two components, $\alpha=1$ and $\alpha=2$,
with equal volume fractions $\phi_{1}=\phi_{2}=\phi/2$ and size ratio
$R_{2}/R_{1}=2$. The two types of particles are assumed to be much
denser than the solvent and to have the same density, so that the
ratio of the gravitational forces is set to $F_{2}/F_{1}=8$. Here
we average the sedimentation velocity statistics over an ensemble
of sphere packings that are sampled from the equilibrium distribution
in the absence of gravity. To generate configurations of spheres,
we use the Lubachevsky-Stillinger packing algorithm \citet{Event_Driven_HE}
to create an initial packing of spheres, and then use equilibrium
event-driven hard sphere MD to equilibrate the packings. We then apply
gravitational forces on all spheres and solve the mobility problem
to compute the instantaneous sedimentation velocities $U_{s,\alpha}$
for each species. As described in more detail in \citet{RigidIBM},
the total gravitational force on the spheres must be balanced by an
equal and opposite uniform force density in the fluid because of the
use of periodic boundary conditions.

The rigid multiblob models used in this study have either 12 blobs
($R_{g}=0.6643$, $R_{h}\approx1$, $s/h=2.052$) for the smaller
species and 42 blobs for the larger species ($R_{g}=1.7714$, $R_{h}\approx2$,
$s/h=2.843$), or, for improved resolution, 42 blobs for the smaller
species ($R_{g}=0.8692$, $R_{h}\approx1$, $s/h=2.553$) and $162$
blobs for the larger species ($R_{g}=1.8935$, $R_{h}\approx2$, $s/h=2.808$).
To correct for finite system size effects, for each volume fraction
$\phi$ we run simulations for three grid resolutions, specifically,
we use grids of size $64^{3}$, $128^{3}$, and $256^{3}$ cells for
the smaller resolution ($12-42$ blobs per sphere), and $128^{3}$,
$256^{3}$, $512^{3}$ for the higher resolution ($42-162$ blobs).
The average sedimentation velocity was extrapolated to the infinite
system size limit by assuming that the finite-size corrections scale
as $1/N^{-1/3}$, where $N$ is the total number of particles, instead
of assuming a specific analytical form for the corrections \citet{BinarySuspension_SD,SedimentationMonodisperse_Ladd}.
The largest example in our simulations is for $\phi=0.5$ with a $512^{3}$
grid, for $N_{1}=51,200$ and $N_{2}=6,400$ spheres, corresponding
to a total of about 10 million Lagrangian (i.e., blob/body) degrees
of freedom (DoFs), and about half a billion Eulerian (i.e., fluid)
DoFs.

\begin{figure*}[tbph]
\includegraphics[width=0.49\textwidth]{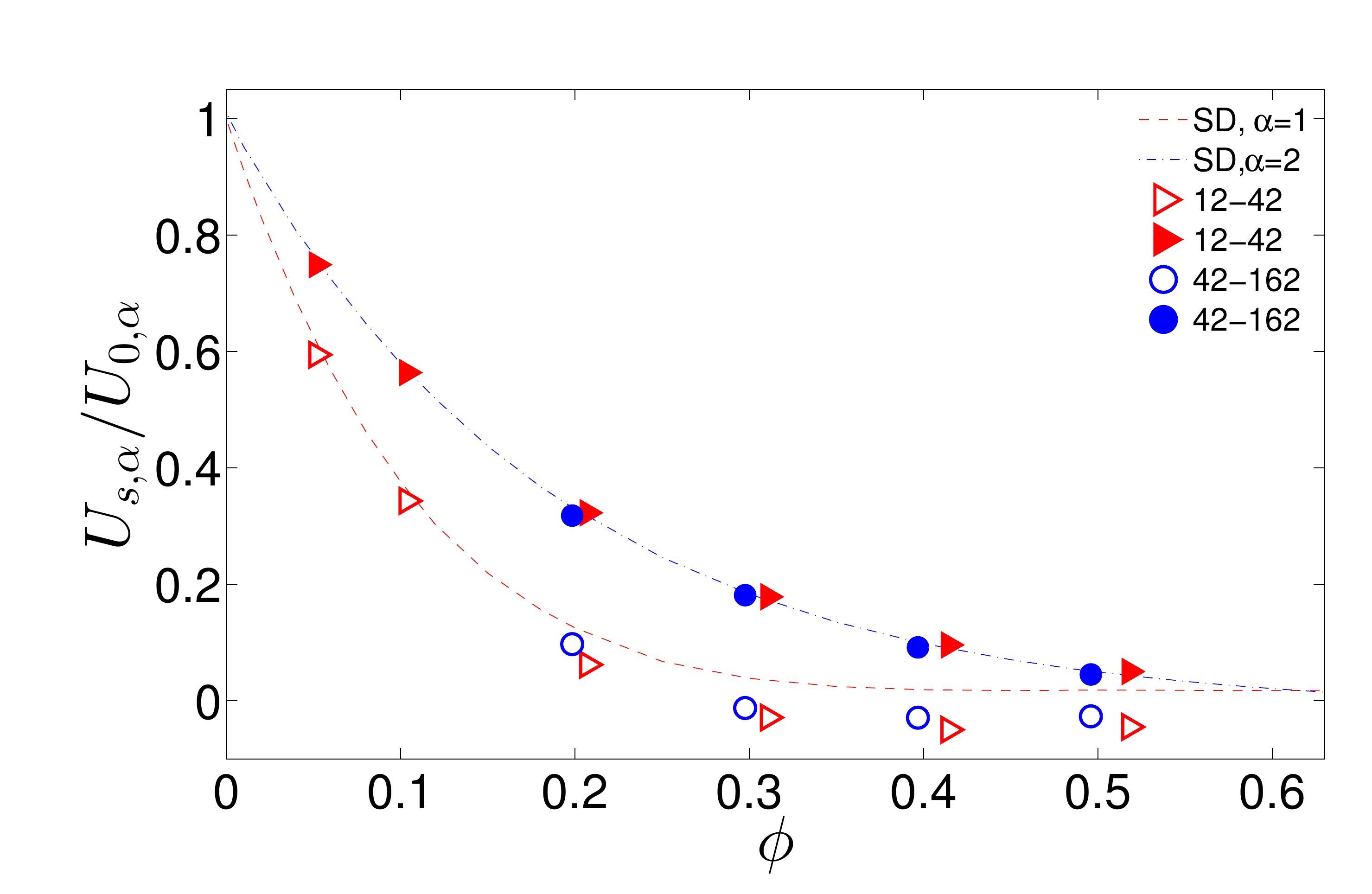}\includegraphics[width=0.49\textwidth]{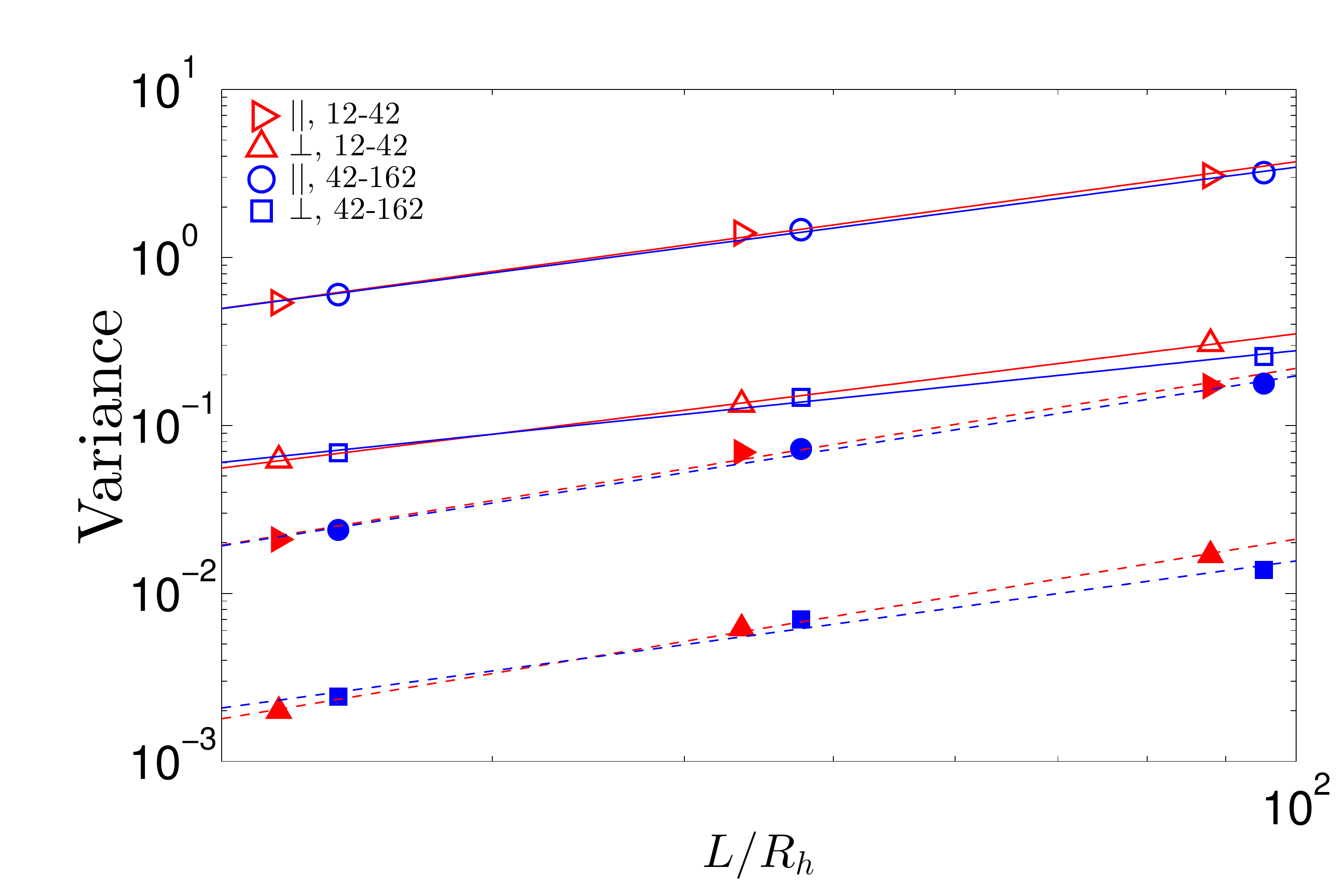}

\caption{\label{fig:BinarySediment}Instantaneous sedimentation rates of the
two species, $\alpha=1$ (empty symbols) and $\alpha=2$ (filled symbols),
in a binary suspension of hard spheres, for two different resolutions
(see legend). (Left panel) Average vertical sedimentation velocity
normalized by $U_{0,\alpha}=F_{\alpha}/(6\pi\eta R_{\alpha})$ as
a function of the total volume fraction $\phi$. The data from recent
Stokesian dynamics simulations \citet{BinarySuspension_SD} is shown
as lines. (Right panel) Normalized variance $\D U_{\alpha}^{2}=\av{\delta U_{s,\alpha}^{2}}/U_{0,\alpha}^{2}$
of the sedimentation velocity parallel and perpendicular to gravity
for $\phi\approx0.2$. Linear fits to the data are shown as dashed
lines.}
\end{figure*}

The left panel of Fig. \ref{fig:BinarySediment} compares our results
for the mean sedimentation velocity of the different species with
results obtained using traditional (i.e., non-accelerated) SD without
pairwise lubrication corrections \citet{BinarySuspension_SD}. It
is well known that a standard FTS truncation is not particularly accurate
for sedimentation because of the importance of a nontrivial mean quadrupole
\citet{SedimentationMonodisperse_SD}, therefore, the SD simulations
include a mean-field estimate of the quadrupole contribution, see
Eq. (2.29) in \citet{SD_Suspensions}. Such a correction is not included
in the accelerated SD method developed in \citet{SD_SpectralEwald},
and this leads to a strong over-estimation of the sedimentation velocity
at larger densities and even a reversal of the trend toward increasing
sedimentation rate \citet{SD_SpectralEwald}. Our results show a consistently
decreasing sedimentation rate with increasing density, and are in
good agreement between the two resolutions, except that the agreement
is only qualitative at the higher densities for the smaller spheres
(thus indicating a lack of convergence in our numerical results).
Our results are consistent with the SD results for the larger particles
over the range of densities studied here. However, for the smaller
particles we find a smaller sedimentation rate and even a negative
rate, which arises due the strong backflow created around the larger
particles. As discussed in Section \ref{sub:LaddTest}, lubrication
forces can be very important at densities as large as $\phi=0.5$,
although they are often assumed to play little role in sedimentation
due to lack of relative motion among the particles, and are therefore
not included in the SD simulations. Neverthess, it may be that lubrication
forces play a role for dense binary suspensions due to the relative
motion of the small spheres around the large spheres. We therefore
believe that the binary sedimentation problem should be revisited
by more accurate methods or experiments.

The right panel of Fig. \ref{fig:BinarySediment} shows the normalized
variance of the instantaneous sedimentation velocities for the two
species at $\phi=0.2$ as a function of system size. Consistent with
theory and simulation for random suspensions of monodisperse suspensions
\citet{SedimentationMonodisperse_Ladd}, we find that the variance
grows linearly with system size, consistently between the two resolutions.
This unphysical growth has been the subject of a long-standing controversy
in the literature, which cannot be resolved by our static (i.e., instantaneous)
computations. Namely, it has been noted that the structure of the
suspension changes during sedimentation \citet{SedimentationMonodisperse_Ladd2},
although not enough to suppress the variance growth in existing Lattice
Boltzmann (LB) simulations \citet{SedimentationMonodisperse_Ladd}.
More recent LB studies have suggested that boundaries, polydispersity
and stratification all play roles in the sedimentation of a realistic
suspension \citet{SedimentationMonodisperse_Ladd3}.

\section{Conclusions}

In this paper we described a numerical method for simulating non-Brownian
Stokesian suspensions of passive and active rigid particles of essentially
arbitrary shape in either unconfined, partially confined, and fully
confined geometries. Following a number of prior works, we discretized
rigid bodies using a collection of minimally-resolved spherical blobs
to move as a rigid body. A key contribution of our work was the development
of preconditioned iterative solvers for the potentially large linear
system of equations for the unknown Lagrange multipliers $\V{\lambda}$
and rigid-body motions $\V U$. We demonstrated that an effective
and scalable approach is to solve the saddle-point problem for both
$\V{\lambda}$ and $\V U$ using a block-diagonal preconditioner that
ignores hydrodynamic interactions of distinct bodies, or even interactions
between the bodies and the boundary.

The hydrodynamic interactions between the blobs are captured using
the Rotne-Prager-Yamakawa (RPY) tensor tailored to the specific geometry
(boundary conditions). For unbounded domains, we used a fast multipole
method to compute the product of the blob-blob mobility $\Mob$ and
a force vector. For a single no-slip boundary, we used a GPU to directly
sum the Rotne-Prager-Blake tensor over all pairs of blobs; FMM methods
for half-space Stokes flow have recently been developed \citet{OseenBlake_FMM}
and could be used to scale these computations to millions of blobs.
We showed empirically that the number of GMRES iterations required
to solve for $\V{\lambda}$ and $\V U$ is bounded independent of
the number of bodies, and grows only weakly with increasing packing
density. This paves the way for the development of linear-scaling
methods for solving the mobility problem in moderately dense suspensions
of hundreds of thousands of particles. At the same time, we find that
solving the resistance problem is substantially more difficult since
the number of iterations grows approximately linearly with the linear
dimensions of the system.

For more complex boundary conditions such as fully confined domains,
there is no simple analytical approximation to the RPY tensor \citet{RPY_Periodic_Shear}.
While it is possible to construct fast methods for computing the product
$\Mob\V{\lambda}$ in \emph{specific} geometries, e.g., using Ewald
summation for periodic domains \citet{RigidMultiblobs_Swan}, this
requires knowing the Green's function analytically, and more importantly,
requires a new method and code for each specific combination of boundary
conditions. As an alternative, in this work we developed a rigid multiblob
method for periodic suspensions or suspensions confined in slit and
square channels that uses a grid-based Stokes solver to compute the
action of the (regularized) Green's function ``on the fly'' \citet{BrownianBlobs,BrownianDynamics_OrderN,BD_IBM_Graham}.
Specifically, we extended a recently-developed rigid-body immersed
boundary method \citet{RigidIBM} to suspensions of freely-moving
passive or active rigid particles at zero Reynolds number. We demonstrate
that GMRES applied to the coupled fluid plus rigid body equations
converges in a bounded number of iterations independent of the system
size, with a weak growth of the number of iterations with the packing
density, and a moderate growth with increased confinement. Unlike
in methods based on Green's functions, each Krylov iteration in our
approach only requires a few cycles of a geometric multigrid solver
for the Poisson equation, and an application of the block-diagonal
preconditioner for the blob-blob mobility.

We used our methods to compute the mobility of a cylinder near a no-slip
boundary and found good agreement with experimental measurements.
We also demonstrated that a pair of active pusher tripartite nanorods
sedimented near a boundary form dimers that rotate in a direction
consistent with recent experimental measurements \citet{TripleNanorods_Megan}.
Our numerical results for the effective planar diffusion coefficient
of a boomerang colloid confined to a narrow slit channel were not
in agreement with recent experimental measurements \citet{BoomerangDiffusion,angleBoomerangs}
by a factor of two. In the future we will carry out more careful and
systematic quantitative comparisons between simulations and experiments
for confined passive and active colloids.

It is worthwhile to point out some specific differences between our
approach and existing methods. We focus in this comparison on methods
based on Green's functions. For confined domain, our Green's-function-free
method described in Section \ref{sec:RigidIBAMR} is quite different
from most existing methods. The equations (\ref{eq:rigidSystem})
appear, perhaps in somewhat modified form, in a number of works \citet{HYDROPRO,HYDROPRO_Globular,RotationalBD_Torre,HYDROLIB,SphereConglomerate,RigidBody_SD,Multiblob_RPY_Rotation}.
The key distinguishing feature of our work is the use of iterative
methods as a way to scale these computations to suspensions of thousands
of bodies. While preconditioned iterative solvers have been used in
the recent work of Swan and Wang \citet{RigidMultiblobs_Swan}, we
believe the preconditioned saddle-point approach developed here is
notably superior both in efficiency \emph{and} simplicity.

The rigid multiblob approach is quite similar to the method of regularized
Stokeslets developed by Cortez \emph{et al}. \citet{RegularizedStokeslets,RegularizedStokeslets_2D,RegularizedStokeslets_Walls,RegularizedStokeslets_Periodic}.
This method is usually presented as a regularized first-kind boundary
integral formulation \citet{RegularizedStokeslets} for solving (\ref{eq:first_kind}).
The method has been made more accurate by using higher-order quadratures
\citet{RigidRegularizedStokeslets,RegularizedStokesletsPhoretic},
and has very recently been generalized to second-kind formulations
that account for active slip \citet{RegularizedStokesletsPhoretic}.
However, these works do not consider preconditioners and existing
regularized Stokeslet methods do not scale well to many-body suspensions.
We note that a first-kind formulation preconditioned by a block-diagonal
preconditioner as we do in this work is spectrally equivalent to a
second-kind formulation for well-separated bodies %
\footnote{We thank Leslie Greengard for sharing this observation with us.%
}. In \citet{RegularizedStokesletsPhoretic} double layer terms (i.e.,
second-kind boundary integrals) are included to account for the active
slip. As we argued in Appendix \ref{sub:BoundaryIntegral}, this is
not necessary if one is not interested in surface tractions, and therefore
we prefer our simpler regularized first-kind formulation. Another
key difference between our approach and the method of regularized
Stokeslets is that the mobility used in regularized Stokeslets methods
is different from the RPY tensor; most importantly, it is \emph{not}
symmetric. Notably, Cortez \emph{et al}. apply the regularization
on the forces (sources) but not also on the velocities (targets),
which approximately corresponds to omitting $\left(\M I+a^{2}/6\,\grad_{\V r}^{2}\right)$
in (\ref{eq:MobilityFaxen}). Using a non-symmetric blob-blob mobility
is not physical, for example, incorporating thermal fluctuations becomes
impossible since this requires the square root of the mobility.

Our work is very closely related to that of Swan, Brady \emph{et al.}
\citet{StokesianDynamics_Wall}. The following are key differences.
First, we use only the RPY form of the mobility matrix, that is, we
only have a force (monopole) degree of freedom at each blob, as done
in more recent work by Swan and Wang \citet{RigidMultiblobs_Swan}.
This can be seen as a direct but regularized discretization of (\ref{eq:first_kind}),
where the unknown is the surface ``traction''. By contrast, Swan
et al. use Stokesian Dynamics (SD) to represent the blobs as ``spheres'',
more precisely, to associate with each blob a force, torque and stresslet
(FTS) %
\footnote{Note that a degenerate quadrupole correction corresponding to the
Faxen operators in (\ref{eq:MobilityFaxen}) is also included in the
RPY tensor even for ``monopoles.''%
}; more multipoles have been included in other works based on multipole
expansions \citet{SphereConglomerate,ActiveFilaments_Adhikari,ConstrainedFCMChains}.
This makes the number of degrees of freedom (DoF) per blob at least
$3+3+5=11$ in three dimensions, instead of just $3$ as in our formulation.
In recent work \citet{Multiblob_RPY_Rotation}, rotational degrees
of freedom (angular velocity and constraint torques) have been added
to the blobs without including stresslets (i.e., an FT truncation),
which doubles the number of DOFs per blob relative to the approach
we follow (6 DOFs instead of 3). Our investigations have shown this
to lead to insufficient improvement in accuracy to justify the doubling
of the number of DOFs. For active suspensions, in the formulation
of \citet{ActiveFilaments_Adhikari,BoundaryIntegralGalerkin,Galerkin_Wall_Spheres},
active slip is imposed on the surface of the beads composing the rigid
body, i.e., each bead is active \emph{individually}. By contrast,
our blobs do not really have a well-defined surface, and in our formulation
active slip is imposed on the surface of the body and not on blobs
individually, consistent with a discretization of (\ref{eq:first_kind}).
Our approach \emph{only} requires a way to compute the (action of
the) RPY blob-blob mobility matrix, and is therefore much simpler
to use in practice and adopt to different boundary conditions. As
we explained in Section \ref{sec:RigidIBAMR}, the RPY tensor can
be approximated using grid-based solvers quite straightforwardly using
immersed boundary methods, but going to higher orders requires additional
differentiability (smoothness) \citet{ForceCoupling_Stokes} than
afforded by simple immersed boundary methods.

Another key difference between the rigid multiblob method and traditional
Stokesian dynamics is that we do \emph{not} include lubrication (near-field)
corrections in addition to the far-field RPY mobility. If it is necessary
to resolve near-field interactions between particles, for example,
to study the rheology of concentrated suspensions, one can increase
the resolution by using more blobs per rigid body. For sufficiently
dense suspensions, very close contacts become numerous and in practice
lubrication forces need to included as a correction to the FTS expansion.
We choose, however, not to include \emph{uncontrolled }pairwise lubrication
approximations for several reasons. First, we believe that it is important
to control the approximations so that accuracy can be confirmed by
comparing different levels of resolution. Second, it is difficult
to generalize pairwise lubrication corrections to dense suspensions
of rigid particles of \emph{arbitrary} shape.

We carefully studied the accuracy of the rigid multiblob approach
on a variety of standard problems for spherical particles. We demonstrated
that, once the effective hydrodynamic radius of the rigid multiblobs
is matched to the target sphere radius, even a 12-blob (icosahedral)
model of a sphere \citet{MultiblobSprings} provides a substantial
improvement over the widely-used force-torque-stresslet (FTS) truncation
of the multipole hierarchy, especially near boundaries. However, we
note that the rigid multiblob models are not rotationally invariant
and this leads to notable discretization artifacts as blobs on distinct
bodies begin to overlap. Furthermore, our method does not include
pairwise lubrication corrections for nearby pairs of spheres (for
reasons discussed in the body of the paper), and can therefore only
accurately resolve the hydrodynamic interactions between objects if
the blobs on the two bodies do not overlap each other. It remains
as a grand challenge for future work to construct a \emph{scalable}
method that applies to particles of complex shape with complex boundary
conditions \emph{and} resolves lubrication interactions among nearly
touching particles with \emph{controllable} accuracy.

There are a number of possible extensions of the computational method
described here. An important direction of work is to compute a tractable
formulation of the RPY-Blake tensor for a single no-slip boundary
that ensures an SPD mobility matrix even when blobs overlap the wall,
which is important for the inclusion of thermal fluctuations (Brownian
motion). While a general SPD formulation of RPY in confined domains
has been developed in \citet{RPY_Periodic_Shear}, that formulation
does not apply a regularization when the blobs overlap the wall %
\footnote{In fact, the overlapping correction derived in \citet{RPY_Periodic_Shear}
is \emph{independent} of the boundary conditions.%
}. Such a regularization is important physically, in particular, we
believe it is crucial that the velocity of a blob go to zero smoothly
as its position approaches the boundary. This prevents unphysical
motion of blobs along the no-slip boundary, or even worse, blobs leaving
the domain. Observe that the alternative formulation of the blob-blob
mobility (\ref{eq:GreensMobility}), together with the modification
near no-slip boundaries first proposed by Yeo and Maxey \citet{ForceCoupling_Channel}
and generalized to other boundary conditions in Appendix D in \citet{RigidIBM},
is SPD for all configurations \emph{and} vanishes as a blob approaches
a boundary. If the integral in (\ref{eq:GreensMobility}) can be performed
analytically for some choice of the kernel $\delta_{a}$, this would
give a simple formula for a Rotne-Prager-Yamakawa-Blake tensor that
can be used in practical simulations. Another approach to constructing
a regularization is to use the simple image construction proposed
recently by Gimbutas \emph{et al.} \citet{OseenBlake_FMM} and combine
with the free-space RPY tensor.

The rigid multiblob formulation can be seen as a low-order regularized
quadrature rule for the first-kind integral equation (\ref{eq:first_kind}).
It is natural to consider using higher-order quadrature rules. This
has been done in the context of the method of regularized Stokeslets
in \citet{RigidRegularizedStokeslets,RegularizedStokesletsPhoretic},
and has been done in the context of immersed boundary methods in \citet{IBFE}.
Specifically, Griffith and Luo have proposed an alternative IB approach
that models the deformations and stresses of immersed elastic body
using a finite element (FE) representation \citet{IBFE}. In their
IB/FE approach, the degrees of freedom associated with $\V{\lambda}$
are represented in a finite-element basis set, and the interaction
between the fluid grid and body mesh is handled by placing IB markers
at the numerical quadrature points of the FE mesh. When such an approach
is generalized to rigid bodies, it simply amounts to \emph{filtering}
the mobility operator (\ref{eq:mob_def}), 
\[
\M{\mathcal{M}}_{FE}=\M{\Psi}\left(\M{\mathcal{J}}\M{\mathcal{L}}^{-1}\M{\mathcal{S}}\right)\M{\Psi}^{T}=\M{\Psi}\Mob\M{\Psi}^{T},
\]
where $\M{\Psi}$ is a matrix that contains quadrature weights as
well as geometric information about the FE mesh. Future work should
explore whether this approach provides a significant improvement in
accuracy or efficiency over the simple rigid multiblob approach presented
here, and to compare this to the methods described in \citet{RigidRegularizedStokeslets,RegularizedStokesletsPhoretic}.

In the method used here, we used a regular (staggered) grid and therefore
no-slip boundary conditions can only be imposed on the boundaries
of a rectangular prism. Domains of complex shapes, such as (patterned)
microfluidic channels, can be handled in two ways. The first way is
to construct the boundaries out of rigidly fixed blobs \citet{BD_IBM_Graham}.
While this is flexible and straightforward, it requires solving a
combined mobility-resistance problem that our investigations suggest
cannot be solved scalably using existing methodologies. An alternative
and promising approach is to use an FEM method to solve the Stokes
equations on a boundary-fitted unstructured tetrahedral grid \citet{SELM_FEM},
and combine this with the rigid-body immersed boundary ideas presented
here. Even if a rectangular grid is appropriate, our regular-grid
method requires very large grids for very low densities or inhomogeneous
suspensions, such as, for example, a suspension of particles sedimented
near a bottom wall in a slit channel where the top wall needs to be
taken into account as well. A substantial challenge for future work
is to develop a stable discretization of the steady Stokes equations
on an adaptively-refined (e.g., block structured) staggered grid;
this has been accomplished for unsteady flow \citet{IBAMR} but the
steady Stokes equations pose several notable challenges.

We believe that a number of the preconditioning ideas developed in
this work can also be applied to other related methods, such as methods
based on boundary integral formulations. Some of these methods can
provide a notable improvement in accuracy over the low-accuracy rigid
multiblob method, and with a suitable preconditioner they can potentially
be scaled to suspensions of tens of thousands of particles. For certain
simple confined geometries, such as periodic boundaries or semi-infinite
slit channels, it is possible to develop fast methods for applying
the RPY and related tensors based on FMM or FFT methods. This may
be preferable to the immersed boundary approach followed here, which
requires a dense grid of spacing \emph{smaller} than the hydrodynamic
radius of the blobs $a$. By contrast, the spectral Ewald method \citet{SpectralEwald_Stokes}
completely decouples the spacing of the FFT grid from $a$, while
controlling the accuracy. We believe that is important for the community
working on Stokes suspensions to develop benchmark problems and compare
different methods in terms of both accuracy and efficiency, to identify
which methods are most appropriate under which conditions and accuracy
requirements.

To account for thermal fluctuations (Brownian motion), one adds a
fluctuating component $\left(k_{B}T\eta\right)^{\frac{1}{2}}\left(\M{\mathcal{Z}}\left(\V r,t\right)+\M{\mathcal{Z}}^{T}\left(\V r,t\right)\right)$
to the fluid stress $\M{\sigma}$ in (\ref{eq:steady_fluct_Stokes})
\citet{BrownianBlobs,ForceCoupling_Fluctuations,SELM,FluctuatingFCM_DC}
in the spirit of fluctuating hydrodynamics \citet{VACF_Langevin,LangevinDynamics_Theory,LLNS_Staggered},
where $\M{\mathcal{Z}}\left(\V r,t\right)$ denotes a white-noise
random Gaussian tensor field with uncorrelated components. This adds
a fluctuating component to the rigid-body velocity and leads to the
overdamped Langevin equation
\begin{equation}
\left[\begin{array}{c}
\V u\\
\V{\omega}
\end{array}\right]=\left[\begin{array}{c}
d\V q/dt\\
d\V{\theta}/dt
\end{array}\right]=\M{\mathcal{N}}\left[\begin{array}{c}
\V f\\
\V{\tau}
\end{array}\right]-\breve{\M{\mathcal{N}}}\breve{\V u}+\left(2k_{B}T\,\M{\mathcal{N}}\right)^{\frac{1}{2}}\diamond\M{\mathcal{W}}\left(t\right).\label{eq:OverdampedVelocity}
\end{equation}
where $\M{\mathcal{W}}\left(t\right)$ denotes a collection of independent
white-noise scalar processes and $\diamond$ denotes a suitable (kinetic)
stochastic product \citet{KineticStochasticIntegral_Ottinger,BrownianMultiBlobs}.
In this work we did not consider the generation of the fluctuating
component of the rigid-body motion $\left(2k_{B}T\,\M{\mathcal{N}}\right)^{\frac{1}{2}}\V W$,
where $\V W$ is a collection of standard normal variates. This is
an important missing component for suspensions in an unbounded or
half-space domain. For the immersed boundary method described in Section
\ref{sec:RigidIBAMR}, computing the random motion of rigid multiblobs
is straightforward and can be accomplished at essentially \emph{no
additional cost} by simply including the stochastic stress on the
right-hand side in the fluid equations. The difficulty, which we will
address in future work, is to develop a temporal integration scheme
for (\ref{eq:OverdampedVelocity}) that correctly accounts for the
stochastic drift terms without incurring significant additional costs
compared to non-Brownian suspensions \citet{BrownianBlobs,FluctuatingFCM_DC,BrownianMultiBlobs}.
\begin{acknowledgments}
We have benefited greatly from extended discussions with Leslie Greengard,
James Swan, Anthony Ladd, Michael Shelley, Megan Davies Wykes, Anna-Karin
Tornberg and Ronojoy Adhikari. We thank Shidong Jiang for sharing
with us their RPY FMM code, Manas Rachh and Leslie Greengard for sharing
with us the latest plane-wave FMMLIB3D codes, James Swan for sending
us Mathematica code to evaluate pairwise sphere mobilities to high
accuracy, Anthony Ladd for sharing with us his numerical results on
random sphere suspensions, Maciej D\l{}ugosz for sharing with us results
from his code, and Jonn Brady for sharing with us the results of SD
for binary sphere suspensions.\textbf{ }A. Donev and B. Delmotte were
supported in part by the National Science Foundation under award DMS-1418706.
A. Donev and F. Balboa were supported in part by the U.S. Department
of Energy Office of Science, Office of Advanced Scientific Computing
Research, Applied Mathematics program under Award Number DE-SC0008271.
B. Delmotte was supported partially by the Materials Research Science
and Engineering Center (MRSEC) program of the National Science Foundation
under Award Number DMR-1420073. B. E. Griffith and A. P. S. Bhalla
were supported in part by the National Science Foundation under awards
DMS-1460368, ACI-1460334, and ACI-1450327. We thank the NVIDIA Academic
Partnership program for providing GPU hardware for performing some
of the simulations reported here.
\end{acknowledgments}
\appendix

\section*{Appendix}

\section{\label{sec:ContinuumFormulation}Continuum Formulation}

The basic problem we consider is the motion of a number of rigid bodies
suspended in a Stokesian fluid. For simplicity, consider a single
body $\Omega$ rotating with angular velocity $\V{\omega}$ around
a \emph{tracking point} (origin) that is translating with linear velocity
$\V u$, under the combined influence of an external force $\V f$
and torque $\V{\tau}$; the generalization to many bodies is straightforward.
Without loss of generality let us assume that the fixed (lab) and
body coordinate frames are identical at the point in time under consideration.
Outside the body we have the steady Stokes equations for the fluid
velocity $\V v\left(\V r\right)$ and the pressure $\pi\left(\V r\right)$,
\begin{align}
-\grad\cdot\M{\sigma} & =\grad\pi-\eta\grad^{2}\V v=0\label{eq:steady_fluct_Stokes}\\
\grad\cdot\V v & =0,\nonumber 
\end{align}
along with some suitable boundary conditions at infinity or the boundary
of a domain $\mathcal{D}\supset\Omega$. The \emph{no slip} boundary
condition on the surface of the body is
\begin{equation}
\V v\left(\V q\right)=\V u+\V{\omega}\times\V q+\breve{\V u}\left(\V q\right)\mbox{ for all }\V q\in\partial\Omega,\label{eq:n_slip_surface}
\end{equation}
where $\breve{\V u}$ is a specified \emph{apparent slip} velocity
due to active boundary layers on the surface of the rigid body. Here
$\V u$ and $\V{\omega}$ are Lagrange multipliers for the force and
torque balance conditions,
\begin{equation}
\int_{\partial\Omega}\V{\lambda}\left(\V q\right)d\V q=\V f\quad\mbox{and}\quad\int_{\partial\Omega}\left[\V q\times\V{\lambda}\left(\V q\right)\right]d\V q=\V{\tau},\label{eq:ft_balance}
\end{equation}
where $\V{\lambda}\left(\V q\right)$ is the normal component of the
stress on the outside of the surface of the body, i.e., the traction
\[
\V{\lambda}\left(\V q\right)=\M{\sigma}\cdot\V n\left(\V q\right),
\]
where $\V n$ is the surface normal and the fluid stress is
\begin{equation}
\M{\sigma}=-\pi\M I+\eta\left(\grad\V v+\grad^{T}\V v\right).\label{eq:stress_def}
\end{equation}
The solution of the above system of equations is, by linearity, an
affine mapping of the form (\ref{eq:N_def_cont}).

\subsection{\label{sub:BoundaryIntegral}Boundary integral re-formulation}

Observe that in the Stokes regime, the details of what happens inside
the body do not actually matter for the motion of the body and its
hydrodynamic interactions with other bodies or boundaries. For instance,
a fluid-filled ``bacterium'' with a rigid membrane and a solid particle
of the same shape will move identically for the same surface slip
and total force and torque. Similarly, to an outside observer, a bacterium
covered with a layer of cilia on the outside will be indistinguishable
from a bacterium that also has a layer of cilia on the inside of its
membrane. Therefore, it is possible to extend the fluid equation (\ref{eq:steady_fluct_Stokes})
to the whole domain and pretend that there is fluid inside the body
moving with a velocity that is continuous across the boundary of the
body. For a strictly rigid body motion on the surface, the fluid inside
will move as a rigid body and thus be free of strain \citet{RegularizedStokeslets}.
If there is slip on the surface, when we extend the flow inside we
are assuming that the velocity is continuous at the boundary so that
the same slip is present on the inside of the body surface. This will
drive \emph{internal} active flows inside the body in addition to
the \emph{external} active flow outside. Once we extend the fluid
equation everywhere we can write down an equivalent \emph{first-kind}
boundary integral equation for Stokes flow \citet{BoundaryIntegral_Pozrikidis,RegularizedStokeslets}
\begin{equation}
\V v\left(\V q\right)=\V u+\V{\omega}\times\V q+\breve{\V u}\left(\V q\right)=\eta^{-1}\int_{\partial\Omega}\Greens\left(\V q,\V q^{\prime}\right)\tilde{\V{\lambda}}\left(\V q^{\prime}\right)d\V q^{\prime}\mbox{ for all }\V q\in\partial\Omega,\label{eq:first_kind}
\end{equation}
which along with the force and torque balance condition (\ref{eq:ft_balance})
defines a linear system of equations to be solved for the single-layer
potential $\tilde{\V{\lambda}}\left(\V q\right)$ and the velocities
$\V u$ and $\omega$. Here $\Greens\left(\V q,\V q^{\prime}\right)$
is the Green's function for steady Stokes flow with unit viscosity
and with the specified boundary conditions on the domain boundary
$\partial\mathcal{D}$. 

In this work we will require that the total volume of fluid is preserved
by the slip, i.e., there is no source or sink for the flow inside
the particle (as would be the case for swelling bodies),
\begin{equation}
\int_{\partial\Omega}\breve{\V u}\left(\V q\right)\cdot\V n\left(\V q\right)d\V q=0,\label{eq:slip_solvability}
\end{equation}
which is always true for tangential slip. This condition is required
to be able to extend the flow inside the body and still keep it incompressible
everywhere in the domain. This condition is related to a known issue
with first-kind boundary integral formulations having a nontrivial
null space, or, equivalently, the single layer operator having an
incomplete range \citet{BoundaryIntegral_Pozrikidis}. Removing the
restriction (\ref{eq:slip_solvability}) requires switching to a second-kind
or a mixed first-second kind integral equation \citet{RegularizedStokesletsPhoretic,BoundaryIntegralGalerkin}.

The single-layer potential $\tilde{\V{\lambda}}\equiv\V{\lambda}=\M{\sigma}\cdot\V n$
if there is no slip, i.e., if $\breve{\V u}=0$, which is a property
that relies closely on the fact that the specified velocity on the
surface of the body is a rigid-body velocity, see the book of Pozrikidis
\citet{BoundaryIntegral_Pozrikidis} for details but also \citet{RegularizedStokeslets}
for a simple and relevant derivation using a regularized (non-singular)
Green's function. If there is slip, then $\V{\lambda}$ does not have
a direct physical interpretation as a surface traction, rather, it
is the jump in the stress when going across the body surface from
the ``interior'' flow to the ``exterior'' flow. If one wants to
determine the actual traction in the presence of nontrivial slip a
second kind integral formulation ought to be used, which includes
an additional term on the right hand side of (\ref{eq:first_kind})
involving $\breve{\V u}$ \citet{BoundaryIntegral_Pozrikidis,StokesSecondKind_Shelley,BoundaryIntegralGalerkin}.
The fact that the same force and torque balance condition (\ref{eq:ft_balance})
applies even though $\V{\lambda}$ is not the physical traction follows
from the fact that the fictitious fluid inside the body is not accelerating;
equivalently, one observes that a double-layer density does not contribute
to the total force and torque on the body since it is a dipole rather
than a monopole density. As discussed at length by Cortez \emph{et
al}. \citet{RegularizedStokeslets}, both the method of regularized
Stokeslets and the rigid multiblob method presented here can be seen
as a particularly straightforward technique for solving a suitably
regularized version of (\ref{eq:first_kind}) \citet{RegularizedStokesletsPhoretic,RigidRegularizedStokeslets}.

\section{\label{app:RenderFlow}Computing flow fields}

Observe that, unlike the immersed boundary method, the Green's-function-based
rigid multiblob method described in Section \ref{sec:RigidMultiblobs}
does not compute the actual flow (velocity and pressure) around the
bodies. Rendering flow fields is useful in a number of applications
for visualizing the flow around passive and active rigid bodies. There
are a number of different ways to define a flow field around a multiblob,
here we follow the following procedure that reuses existing code and
produces smooth non-singular flow fields everywhere, including inside
the blobs. The input to the calculation are the constraint forces
$\V{\lambda}$, and the output is a fluid velocity $\V v\left(\V r\right)$
evaluated at an arbitrary position in the domain. 

The basic idea is to estimate the velocity that a freely-moving tracer
blob of a given size $a^{\prime}\ll a$ would have, where $a^{\prime}$
is a desired resolution scale for the flow that could be chosen to
match the size of actual tracer particles used in an experiment. We
replace each of the $N_{b}$ blobs with $N_{b}^{\prime}$ smaller
blobs of radius $a^{\prime}$, i.e., we treat each blob $i$ as a
sphere of radius $a$ and discretize it using smaller blobs. We divide
the constraint force on blob $i$ \emph{uniformly} (this is consistent
with the approximation used to construct the RPY tensor \citet{RPY_Shear_Wall})
among the small blobs, $\V{\lambda}_{j}^{\prime}=\V{\lambda}_{i}/N_{b}^{\prime}$,
where $j=1,\dots,N_{b}^{\prime}$. The velocity field is then \emph{defined}
at an arbitrary point in space via $\V v\left(\V r\right)=\sum_{j}\Mob^{\prime}\left(\V r-\V r_{j}\right)\V{\lambda}_{j}^{\prime}$,
where $\Mob^{\prime}$ is the blob-blob mobility for blobs of radius
$a^{\prime}$. Observe that the above sum can be evaluated using the
existing matrix vector product, but now applied to the collection
of $N_{b}N_{b}^{\prime}$ small blobs.

\section{\label{app:Permeable}Permeable Bodies: Brinkman Equations}

When the suspended rigid bodies are made of a porous material and
thus partially permeable to the fluid, one can model the flow inside
the particles using the (Debye-Bueche-)Brinkman \citet{Brinman_Original,RPY_Brinkman}
equation, as done for suspensions of permeable spheres using multipole
expansion methods in \citet{HYDROMULTIPOLE_Permeable}. In this Appendix
we demonstrate analytically and numerically how a small change in
the formulation can be used to allow for a finite permeability of
the particles with minimal changes to the algorithm and implementation.

For particles with permeability (porosity) $\kappa$ (possibly different
for different bodies), the velocity equation extends to the whole
domain including the interior of the bodies, and takes the form of
Brinkman's equation,

\begin{equation}
\grad\pi=\eta\grad^{2}\V v-\sum_{p}\frac{\eta}{\kappa_{p}}\left[\V v-\left(\V u_{p}+\V{\omega}_{p}\times\left(\V r-\V q_{p}\right)\right)\right]\,\M 1_{p},\label{eq:Brinkman}
\end{equation}
where $\M 1_{p}\left(\V r\right)$ is the characteristic function
of body $p$, with the condition that both the velocity and the stress
are continuous across the particle-fluid interface. Note that the
rigid body case corresponds to the limit $\kappa\rightarrow0$ and
is a singular limit in which the stress becomes discontinuous.

For permeable bodies, we fill the interior of the bodies with blobs
as well, rather than just covering the surface with blobs as we did
for impermeable bodies. Such filled rigid multiblob models can be
constructed, for example, by covering the body with an unstructured
tetrahedral grid with good uniformity properties and placing blobs
at the nodes (vertices) of the grid. One also needs to assign a volume
$\D V_{i}$ to each blob; this can be done by assigning one quarter
of the volume of each tetrahedron to each of its four vertices. Once
a \emph{filled }rigid multiblob model of the body is constructed,
the only change to the formulation (\ref{eq:semi_continuum}) is to
make the effective slip on blob $i$ on body $p$ proportional to
the fluid-blob force,
\begin{equation}
\slip_{i}=-\frac{\kappa_{p}}{\eta\D V_{i}}\V{\lambda}_{i}.\label{eq:permeable_slip}
\end{equation}
Note that this makes the system (\ref{eq:semi_continuum}) strictly
easier to solve than the case of impermeable bodies; the system is
no longer a saddle-point problem for $\kappa>0$. For the Green's
function based methods described in Section \ref{sec:RigidIBAMR},
accounting for (\ref{eq:permeable_slip}) simply adds $\kappa_{p}/\left(\eta\D V_{i}\right)$
to the diagonal elements $\mathcal{M}_{ii}$. For the Stokes solver
based methods described in Section \ref{sec:RigidIBAMR}, all that
is required is to set $\M{\Omega}$ to be a diagonal matrix with $\Omega_{ii}=\kappa_{p}/\left(\eta\D V_{i}\right)$
for blob $i\in\mathcal{B}_{p}$ in (\ref{eq:constrained_Stokes}).

To demonstrate that (\ref{eq:permeable_slip}) is consistent with
the Brinkman equations (\ref{eq:Brinkman}), we focus on the semi-continuum
formulation (\ref{eq:semi_continuum}). Solve the third equation in
(\ref{eq:semi_continuum}) for $\V{\lambda}_{i}$ %
{} (note that this is only possible for nonzero permeability) and substitute
the result in the first equation in (\ref{eq:semi_continuum}) to
obtain
\begin{equation}
\grad\pi=\eta\grad^{2}\V v-\sum_{p}\frac{\eta}{\kappa_{p}}\sum_{i\in\mathcal{B}_{p}}\D V_{i}\left[\int\delta_{a}\left(\V r_{i}-\V r^{\prime}\right)\V v\left(\V r^{\prime},t\right)\, d\V r^{\prime}-\left(\V u_{p}+\V{\omega}_{p}\times\left(\V r_{i}-\V q_{p}\right)\right)\right]\delta_{a}\left(\V r_{i}-\V r\right).\label{eq:semidisc_Brinkman}
\end{equation}
In the limit in which the number of blobs goes to infinity and the
regularized delta function $\delta_{a}$ becomes a true delta function,
the sum over $i\in\mathcal{B}_{p}$ converges to 
\begin{align*}
\int_{\Omega_{p}}\left[\int\delta\left(\V r^{\prime\prime}-\V r^{\prime}\right)\V v\left(\V r^{\prime},t\right)\, d\V r^{\prime}-\left(\V u_{p}+\V{\omega}_{p}\times\left(\V r^{\prime\prime}-\V q_{p}\right)\right)\right]\delta\left(\V r^{\prime\prime}-\V r\right)d\V r^{\prime\prime} & \rightarrow,\\
\left[\V v\left(\V r\right)-\left(\V u_{p}+\V{\omega}_{p}\times\left(\V r-\V q_{p}\right)\right)\right]\,\M 1_{p}\left(\V r\right)
\end{align*}
and therefore the fluid equation (\ref{eq:semidisc_Brinkman}) is
a regularized semi-discretization of the Brinkman equation (\ref{eq:Brinkman}).

\subsection{Numerical Results}

In this section we confirm the consistency of (\ref{eq:permeable_slip})
with the Brinkman equations by comparing to analytical results. We
also assess the accuracy of the method for different resolutions.
Here we use the immersed boundary formulation, but we expect similar
results to apply to methods based on analytical Green's functions.

\begin{figure*}[tbph]
\begin{centering}
\includegraphics[width=0.49\textwidth]{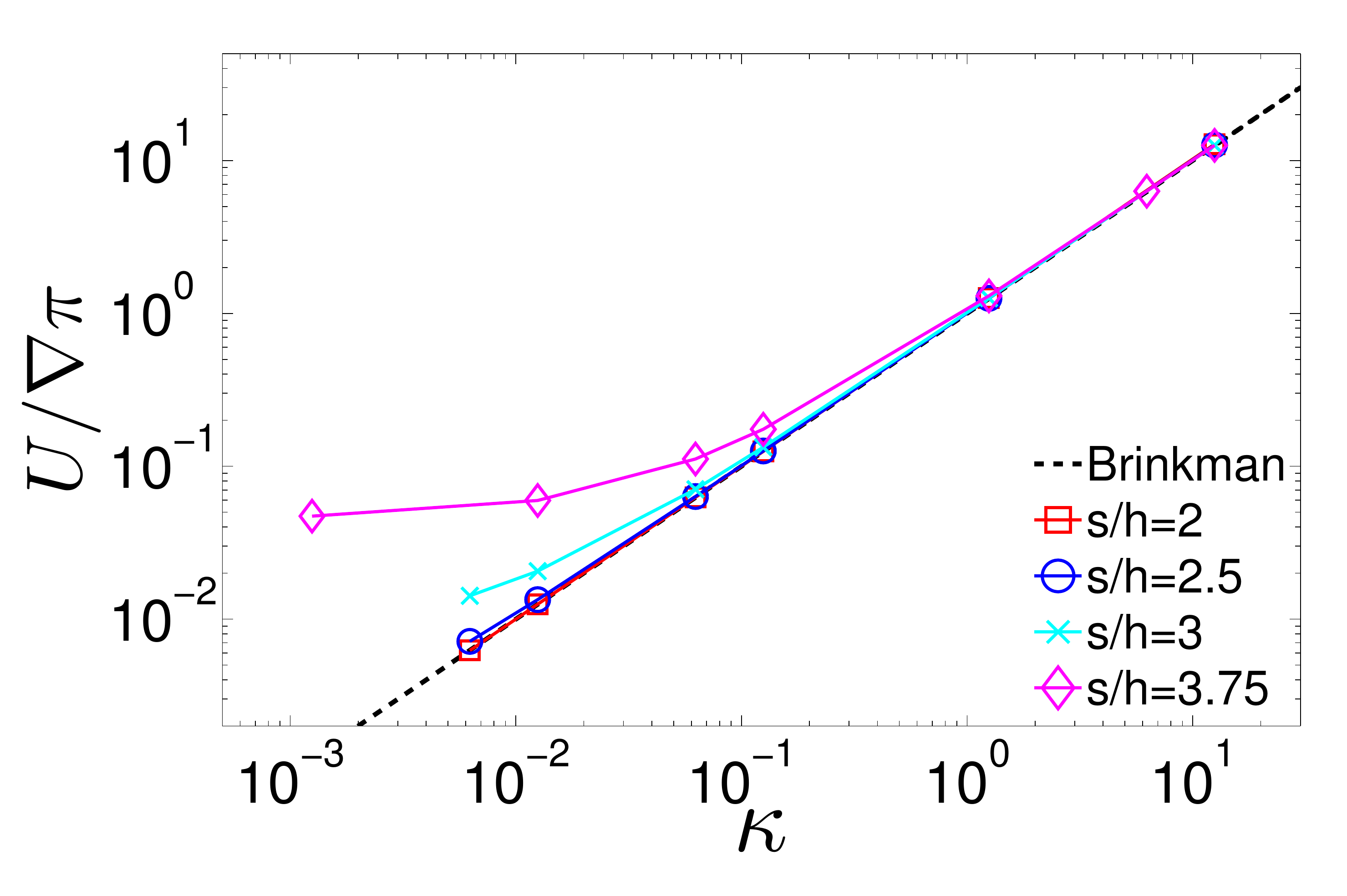}\includegraphics[width=0.49\textwidth]{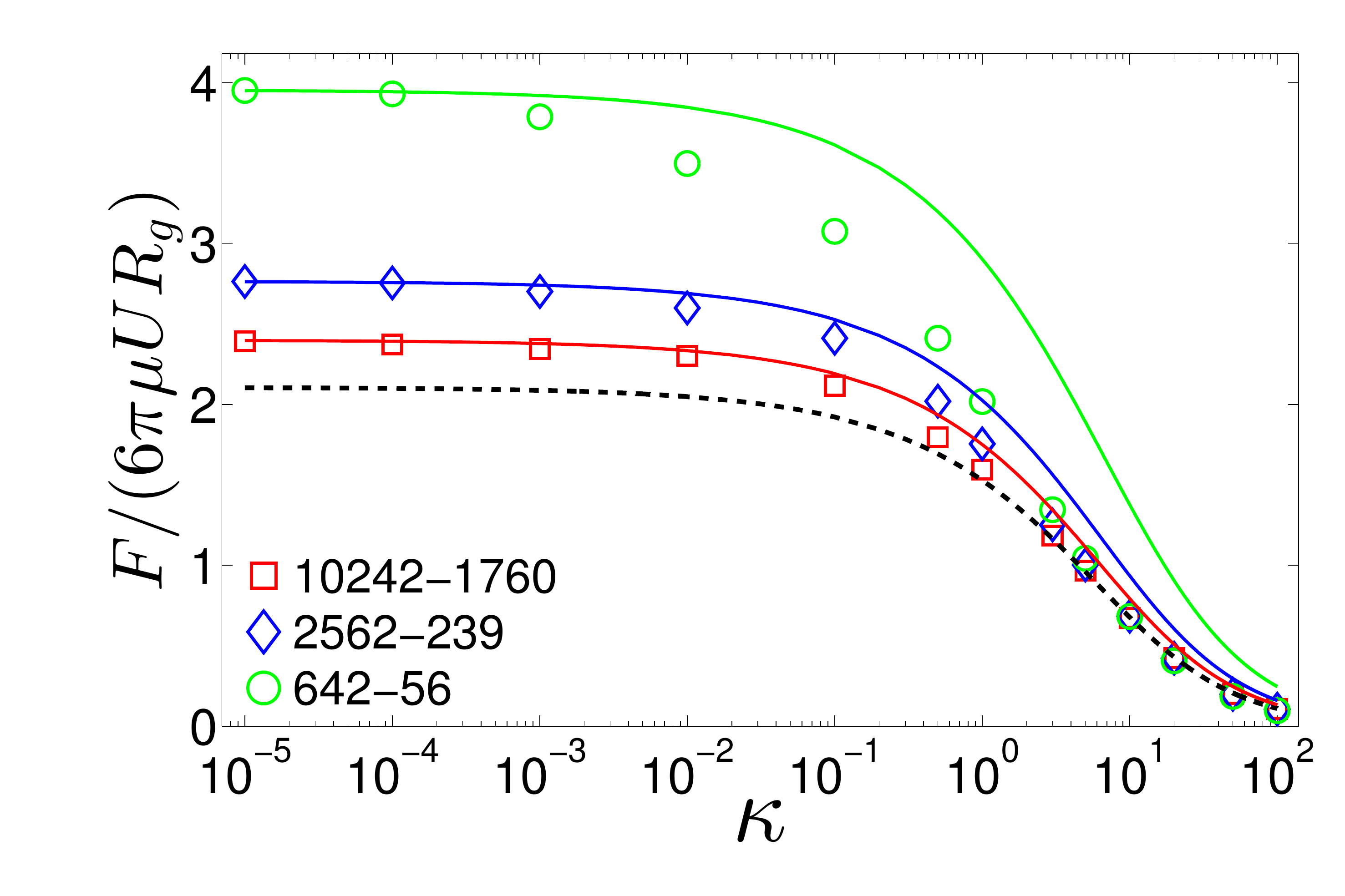}
\par\end{centering}

\caption{\label{fig:Permeable}(Left panel) Numerically measured permeability
of a slab as a function of the input permeability for different blob-blob
spacings $s$. (Right panel) Measured drag (symbols) on a permeable
sphere moving inside a fixed impermeable sphere, as a function of
the input permeability, for three different resolutions, indicated
as the number of blobs on the outer shell (642, 2562 or 10242 blobs)
and inner sphere (56, 239 or 1760 blobs, respectively), see legend.
The theoretical result based on the geometric radii of the spheres
is shown with a dashed black line, while the theoretical result based
on the effective hydrodynamic radii in the impermeable limit (which
vary with resolution) is shown with a solid line of the same color
as the corresponding symbols.}
\end{figure*}

\subsubsection{Permeable Slab}

First, we compute the flow through a permeable slab to numerically
estimate the effective permeability of a rigid multiblob for several
spacings between the blobs. We compose a slab of thickness $5s$ from
blobs placed on a cubic lattice with spacing $s$ (i.e., the slab
has five layers of blobs), and place the slab in the middle of a cubic
domain of side $L$. We impose no-slip for the tangential stress (traction)
on all boundaries of the domain using the technique developed by Griffith
\citet{NonProjection_Griffith}. For the normal component, on the
sides of the domain perpendicular to the slab we impose no-slip, and
we impose a pressure jump of magnitude $\D{\pi}$ across the boundaries
parallel to the slab. We measure the velocity $U$ of the resulting
nearly uniform flow (leak) through the slab as the velocity at the
centerline of the slab a quarter of the domain from the left boundary.

At steady state we expect a uniform flow inside the slab with magnitude
determined from the Brinkmann equation,
\begin{equation}
\nabla\pi=\frac{\D{\pi}}{L}=-\frac{\eta}{\kappa}U,\label{eq:Brinkman-1}
\end{equation}
where $\kappa$ is the permeability (porosity). In the left panel
of Fig. \ref{fig:Permeable} we compare this to the numerical observations.
We see that for a variety of spacings between the blobs we get the
correct permeability for large target values of $\kappa$. However,
as we make the the slab less and less permeable and approach the (singular)
impermeable limit, we start to see a small but measurable leak in
the rigid multiblob results. This leak is larger the larger the spacing
between the blobs is, consistent with the intuition that leak occurs
between the blobs. This suggests that for permeable bodies it is better
to reduce the spacing between the markers to $s\approx2h$ as suggested
in \citet{RigidIBM}. Note that the conditioning of the blob-blob
mobility matrix is significantly improved for permeable bodies compared
to impermeable bodies, so that this reduction in the spacing does
not lead to conditioning problems except for very small values of
$\kappa$.

\subsubsection{Permeable Sphere}

Next we examine the translational mobility of a permeable sphere of
radius $a$. The drag force on a permeable sphere of radius $R$ moving
through an unbounded domain with velocity $U$ is given in \citet{BrinkmanSphereAsymptotics,BrinkmanSphere},
\[
\left(6\pi\eta UR\right)^{-1}F=\frac{G}{1+3G/\left(2\sigma^{2}\right)}=1-\frac{1}{\sigma}+O\left(\frac{1}{\sigma^{2}}\right),
\]
where $G=1-\sigma^{-1}\tanh\sigma$ and $\sigma=\sqrt{a^{2}/\kappa}$.
To eliminate finite-size effects, and following our prior work \citet{RigidIBM}
for impermeable spheres, we consider here a permeable sphere inside
an impermeable spherical shell, that is, we impose a no-slip boundary
condition on a spherical shell of radius $b=a/\lambda$ that is concentric
to the permeable sphere.%

The equations we need to solve are the Stokes equations in the region
between the spherical shells and the Brinkman equation (\ref{eq:Brinkman})
inside the permeable sphere, with no-slip boundary conditions on the
outer shell and continuity of both velocity \emph{and} stress on the
boundary of the permeable sphere. The drag force can be shown to be
\begin{equation}
\alpha\left(6\pi\eta UR\right)^{-1}F=5\lambda^{5}+G(\sigma)\left[1-\lambda^{5}\left(6+\frac{15}{\sigma^{2}}\right)\right],\label{eq:permeable_in_shell}
\end{equation}
where 
\begin{eqnarray}
\alpha & = & 1-5\lambda^{3}+\lambda^{5}\left(9+\frac{15}{2\sigma^{2}}\right)-5\lambda^{6}\\
 &  & +G(\sigma)\Big[\frac{3}{2\sigma^{2}}-\frac{9}{4}\lambda+\frac{15}{2}\lambda^{3}\left(1+\frac{1}{\sigma^{2}}\right)\nonumber \\
 &  & -\frac{9}{2}\lambda^{5}\left(\frac{5}{2}+\frac{7}{\sigma^{2}}+\frac{5}{\sigma^{4}}\right)+\lambda^{6}\left(6+\frac{15}{\sigma^{2}}\right)\Big].\nonumber 
\end{eqnarray}
The solution to this problem in the limit of an impermeable sphere
has been computed by Brenner \citet{BrennerBook} (see Appendix C
in \citet{RigidIBM} for the full solution), 
\begin{equation}
\left(6\pi\eta UR\right)^{-1}F=\frac{1-\lambda^{5}}{1-\frac{9}{4}\lambda+\frac{5}{2}\lambda^{3}-\frac{9}{4}\lambda^{5}+\lambda^{6}}.\label{eq:drag_inner}
\end{equation}

The outside impermeable fixed shell is constructed as a rigid multiblob
using the same recursive triangulation as before. Recall that the
inner sphere has to be uniformly filled with blobs for $\kappa>0$;
we construct a filled sphere model with typical spacing between nearest-neighbor
blobs of $s\approx2h$ using a tetrahedral mesh generator, starting
from a uniform surface triangulation. In the right panel of Fig. \ref{fig:Permeable}
we show the drag on the inner sphere compared to the theory (\ref{eq:permeable_in_shell}),
for several different resolutions. We observe that for large permeabilities
there is an excellent agreement with the theory based on the\emph{
geometric} radii of the inner and outer spheres, even for rather modest
resolutions. But for small permeabilities, we see deviations from
the theory. This is not unexpected, since in the limit of zero permeability
we must converge back to the rigid sphere case, and we know that in
this case the drag is determined by the larger effective \emph{hydrodynamic}
and not the geometric radius. Of course, as the resolution is refined
we get convergence of the geometric and hydrodynamic radius, but convergence
is very slow.

Our numerical observations are consistent with physical intuition.
For large permeability, the flow is smooth and there is no jump in
the pressure (and velocity derivatives) across the surface of the
body, making the rigid multiblob models relatively accurate even for
modest resolutions. However, for impermeable bodies the flow develops
a boundary layer near the surface of the inner sphere and the pressure
and velocity are no longer sufficiently smooth and the accuracy is
lowered. We were able to account for the smearing of the no-slip condition
for an impermeable (passive) sphere by adjusting the hydrodynamic
radius $R_{h}>R_{g}$, but this adjustment cannot be done uniformly
for all permeabilities. This is similar to the situation for active
spheres discussed in Section \ref{sub:Squirmer}, and highlights the
inherent accuracy limitations of regularized methods, including both
the rigid multiblob method and the method of regularized Stokeslets.

\section{\label{app:FitsSphereWall}Empirical mobility of a sphere near a
wall}

A number of theoretical predictions are available for the mobility
of a sphere close to a wall \citet{BrennerBook} (see Appendix D in
\citet{BrownianMultiBlobs} for a summary). However, except for the
translational mobility perpendicular to the wall, for which Brenner
computed an exact infinite sum \citet{Brenner1961} (see \citet{SingleWallPerpMobility}
for an approximate rational fit), the theoretical results are based
on asymptotic expansions and have a limited range of validity. Since
the dynamics of spherical colloids near a no-slip boundary is relevant
to a number of experimental studies, we give here empirical fits to
the mobility computed in Section \ref{sub:SphereWall} using a rigid
multiblob with 642 blobs (our highest resolution).

We have fitted the mobilities shown in the panels of Fig. \ref{fig:Sphere-wall}
with a rational function of the form 
\[
\frac{\mu(x=H/R_{h})}{\mu^{0}}=\delta+\left(\frac{1}{x}\right)^{\alpha}\,\frac{f_{n,2}x{}^{2}+f_{n,1}x+f_{n,0}}{x^{2}+f_{d,1}x+f_{d,0}},
\]
where $\mu^{0}$ is the bulk mobility and $\delta$, $\alpha$ and
$f_{n,2}$ have been chosen to ensure the correct leading-order asymptotic
scaling for large distances to the wall $H$ and the rest of the constants
are fitting parameters. The values of all the coefficients are given
in Table \ref{tab:FittingSphereWall}.

\begin{table}[tbph]
\begin{tabular}{|c|c|c|c|c|c|c|c|c|c|}
\hline 
Mobility &
$\mu^{0}$ &
$\delta$ &
$\alpha$ &
$f_{n,2}$ &
$f_{n,1}$ &
$f_{n,0}$ &
$f_{d,1}$ &
$f_{d,0}$ &
max. relative error\tabularnewline
\hline 
\hline 
$\mu_{tt}^{\parallel}$ &
$6\pi\eta R_{h}$ &
1 &
1 &
$-\frac{9}{16}$ &
0.826024 &
-0.311607 &
-1.4297 &
0.498974 &
$5.6\cdot10^{-3}$\tabularnewline
\hline 
$\mu_{rr}^{\parallel}$ &
$8\pi\eta R_{\tau}^{3}$ &
1 &
3 &
$-\frac{5}{16}$ &
0.15118 &
0.0830598 &
-0.443529 &
-0.406958 &
$4.9\cdot10^{-4}$\tabularnewline
\hline 
$\mu_{rr}^{\perp}$ &
$8\pi\eta R_{\tau}^{3}$ &
1 &
3 &
$-\frac{1}{8}$ &
0.122506 &
-0.0105777 &
-0.953632 &
0.0339739 &
$7.2\cdot10^{-5}$\tabularnewline
\hline 
$\mu_{rt}^{\parallel}$ &
$6\pi\eta R_{h}^{2}$ &
0 &
4 &
$\frac{3}{32}$ &
-0.142813 &
0.0508471 &
-0.528495 &
-0.454638 &
$6.6\cdot10^{-3}$\tabularnewline
\hline 
\end{tabular}

\caption{\label{tab:FittingSphereWall}Fitting parameters for the mobility
of a sphere close to a wall obtained using the numerical mobility
of the 642-blob model. In the last column we report the maximum relative
error between the numerical mobility and the fit in the interval $(1.03,10)R_{h}$.}
\end{table}


\end{document}